\documentclass[aps,rmp,reprint,amsmath,amssymb,graphicx,longbibliography]{revtex4-1}

\usepackage{bm}
\usepackage{graphicx}
\usepackage{color}

\usepackage[normalem]{ulem}

\newcommand{\be}{\begin{equation}
}
\newcommand{\ee}{\end{equation}
}

\newcommand{\projsm}[2]{\vert#1\rangle\langle#2\vert}
\newcommand{\ket}[1]{\left\vert#1\right\rangle}
\newcommand{\pare}[1]{\left(#1\right)}
\newcommand{\proj}[2]{\left\vert#1\rangle\langle#2\right\vert}

\begin{document}

\title{Ultrastrong coupling regimes of light-matter interaction}

\author{P.~Forn-D\'iaz}\email{pforndiaz@ifae.es}
\affiliation{Institut de F\'isica d'Altes Energies (IFAE), The Barcelona Institute of Science and Technology
(BIST), Bellaterra (Barcelona) 08193, Spain}
\affiliation{Barcelona Supercomputing Center - CNS, Barcelona 08034, Spain}
\author{L.~Lamata}
\affiliation{Department of Physical Chemistry, University of the Basque Country UPV/EHU, E-48080 Bilbao, Spain}
\author{E.~Rico} 
\affiliation{Department of Physical Chemistry, University of the Basque Country UPV/EHU, 48080 Bilbao, Spain}
\affiliation{IKERBASQUE, Basque Foundation for Science, 48013 Bilbao, Spain}
\author{J. Kono}
\affiliation{Department of Electrical and Computer Engineering, Rice University, Houston, Texas 77005, USA}
\affiliation{Department of Physics and Astronomy, Rice University, Houston, Texas 77005, USA}
\affiliation{Department of Materials Science and NanoEngineering, Rice University, Houston, Texas 77005, USA}
\author{E.~Solano}
\affiliation{Department of Physical Chemistry, University of the Basque Country UPV/EHU, E-48080 Bilbao, Spain} 
\affiliation{IKERBASQUE, Basque Foundation for Science, E-48013 Bilbao, Spain}
\affiliation{Department of Physics, Shanghai University, 200444 Shanghai, China}

\date{\today{}}

\begin{abstract}
Recent experiments have demonstrated that light and matter can mix together to an extreme degree,
and previously uncharted regimes of light-matter interactions are currently being explored in a variety
of settings. The so-called ultrastrong coupling (USC) regime is established when the light-matter
interaction energy is a comparable fraction of the bare frequencies of the uncoupled systems.
Furthermore, when the interaction strengths become larger than the bare frequencies, the deep-strong
coupling (DSC) regime emerges. This article reviews advances in the field of the USC and DSC
regimes, in particular, for light modes confined in cavities interacting with two-level systems. An
overview is first provided on the theoretical progress since the origins from the semiclassical Rabi
model until recent developments of the quantum Rabi model. Next, several key experimental results
from a variety of quantum platforms are described, including superconducting circuits, semiconductor
quantum wells, and other hybrid quantum systems. Finally, anticipated applications are
highlighted utilizing USC and DSC regimes, including novel quantum optical phenomena, quantum
simulation, and quantum computation.

\end{abstract}

\vspace*{1.5cm}

\maketitle

\tableofcontents{}

\section{Introduction}\label{sec:1}

\begin{figure*}[!hbt]
\centering
\includegraphics[width = 1.5\columnwidth]{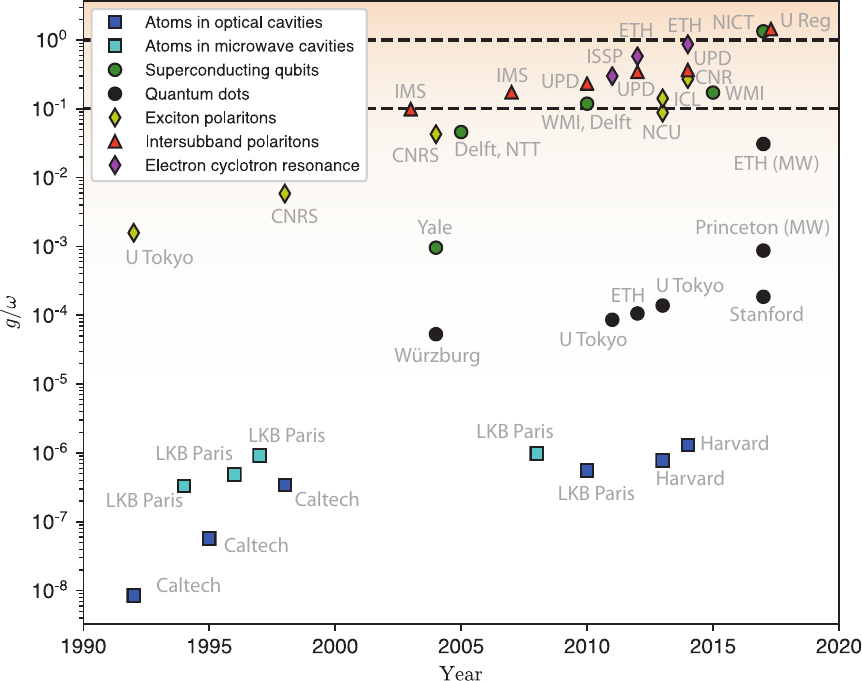}
\caption{\label{fig:g-w_r} (Color online) Evolution in cavity QED of the highest value of $g/\omega$, with $\omega$ the cavity frequency, as a function of time for different physical platforms. The dotted lines at  $g/\omega\simeq0.1$ and $g\simeq\omega$ mark the beginning of the USC and DSC regimes, respectively. References for the data, chronological: atoms in optical cavities \cite{thompson1992}, \cite{TurchettePRL95}, \cite{hood1998}, \cite{colombe2007} \cite{thompson2013}, \cite{Tiecke2014}; atoms in microwave cavities \cite{Brune1994}, \cite{Brune1996}, \cite{Maitre1997}, \cite{Brune2008}; superconducting qubits \cite{wallraff2004}, \cite{chiorescu2004}, \cite{johansson2006}, \cite{niemczyk2010}, \cite{forn-diaz2010}, \cite{baust2016}, \cite{yoshihara2017a}; quantum dots \cite{Reithmaier2004}, \cite{Reinhard2012}, \cite{Takamiya2013}, \cite{Kelaita2017}, \cite{Mi2017}, \cite{Stockklauser2017}; exciton polaritons \cite{WeisbuchetAl92PRL}, \cite{BlochetAl98APL}, \cite{BellessaetAl04PRL}, \cite{WeietAl13OE}, \cite{Kena-CohenetAl13AOM}, \cite{GambinoetAl14ACS}; intersubband polaritons \cite{DupontetAl03PRB}, \cite{DupontetAl07PRB}, \cite{TodorovetAl10PRL}, \cite{DelteiletAl12PRL}, \cite{AskenazietAl14NJP}; and electron cyclotron resonance \cite{MuravevetAl11PRB}, \cite{ScalarietAl12Science}, \cite{MaissenetAl14PRB}, \cite{BayeretAl17NL}.}
\end{figure*}

The Rabi model~\cite{PhysRev.49.324, Rabi1937} arguably describes the simplest class of light-matter interactions, namely, the dipolar coupling between a two-level quantum system (qubit) and a classical radiation field mode. This semiclassical model has a fully quantum counterpart, where the electromagnetic radiation is specified by a single-mode quantum field, yielding the so-called quantum Rabi model (QRM)~\cite{Braak2011}. The QRM describes with accuracy the dynamical and static properties of a wide variety of physical systems, such as quantum optics and solid-state settings. Moreover, a variety of protocols in modern quantum information theory~\cite{nielsen_chuang} employ the QRM as a fundamental building block, with plausible applications in quantum technologies, including, e.g., universal two-qubit gates \cite{Schmidt-Kaler2003, Chow2012, Barends2014}, nondestructive readout \cite{Schuster2005}, quantum state transfer \cite{Majer2007, Richerme2014}, ultrafast quantum gates~\cite{Romero2012}, quantum error correction~\cite{Corcoles2015, kyaw2015prb}, and remote entanglement generation~\cite{PhysRevLett.113.093602, Ritter2012, Campagne2018}. In consequence, the QRM is extremely important in both applied and theoretical physics.

Historically, and for nonrelativistic energies, light and matter have been studied at the fundamental level using single atoms interacting with the electromagnetic mode of an optical~\cite{Kimble98} or a microwave cavity (Raimond et al., 2001), a field known as cavity quantum electrodynamics (cavity QED). The standard cavity QED experiments are usually constrained to light-matter couplings orders of magnitude smaller than the natural frequencies of the noninteracting contributions. Therefore, these experiments take place in the realm of the well-known Jaynes-Cummings (JC) model~\cite{jaynes1963}, which can be obtained by performing the rotating-wave approximation (RWA) on the QRM~\cite{Braak2011}. However, the exploration of cavity QED physics in atomic systems could only be initiated once the light-matter interaction strength was engineered comparable to~\cite{Meschede1985, Rempe1987} or larger~\cite{thompson1992} than all decay rates of the system. This regime of coupling, known as the strong coupling (SC) regime, is necessary to observe coherent quantum dynamics between light and matter, leading to the study of fundamental single atom--single photon processes~\cite{Haroche2013}, and, most importantly, developing the different architectures on which most existing quantum computing technologies are based. Thus, the JC model has represented a theoretical and experimental milestone in the history of light-matter interactions and quantum optics.

During the past decade, a novel coupling regime of the QRM has been theoretically investigated in which the coupling strength is a sizable fraction of the natural frequencies of the noninteracting parts~\cite{CiutietAl05PRB, LiberatoetAl07PRL, bourassa,Beaudoin2011,daniel_prx,Pedernales15,TodorovetAl10PRL}, and experimentally achieved in several quantum systems~\cite{TodorovetAl10PRL,forn-diaz2010,niemczyk2010,ScalarietAl12Science,AnapparaetAl09PRB,GunteretAl09Nature,MuravevetAl11PRB,schwartz2011,GeiseretAl12PRL,PhysRevApplied.2.054002,ZhangetAl16NP,chen2017,braumuller2016,Lv2018, LietAl18NP}. In this ultrastrong coupling (USC) regime, the RWA is not valid anymore, while the counterrotating terms produce novel, unexpected physical phenomena \cite{CiutietAl05PRB} as well as applications in quantum information~\cite{Romero2012,kyaw2015prb,PhysRevLett.113.093602}. In the regime in which the counterrotating terms can still be analyzed with perturbation theory~\cite{TodorovetAl10PRL,forn-diaz2010,niemczyk2010,ScalarietAl12Science,AnapparaetAl09PRB,GunteretAl09Nature,MuravevetAl11PRB,schwartz2011,GeiseretAl12PRL,PhysRevApplied.2.054002,ZhangetAl16NP,chen2017}, the QRM can be described by the Bloch-Siegert (BS) Hamiltonian~\cite{Cohen1973, Beaudoin2011, klimov2009}. On the other hand, some experiments have recently reached the nonperturbative USC regime~\cite{MaissenetAl14PRB,forn-diaz2017,yoshihara2017a,BayeretAl17NL}, where the coupling strength exceeds the natural frequencies of the noninteracting parts, and the full-fledged QRM has to be considered. Under these conditions, a new regime of light-matter interaction emerges, with absolutely different physics than the USC regime. In this deep strong coupling (DSC) regime~\cite{casanova2010}, an approximate solution can reasonably describe some aspects of the QRM. In fact, recently, the DSC regime has been experimentally achieved with a superconducting circuit~\cite{yoshihara2017a} and in a two-dimensional electron gas coupled with terahertz metamaterial resonators~\cite{BayeretAl17NL}. 

\begin{figure*}[!hbt]
\centering
\includegraphics[width = 1.5\columnwidth]{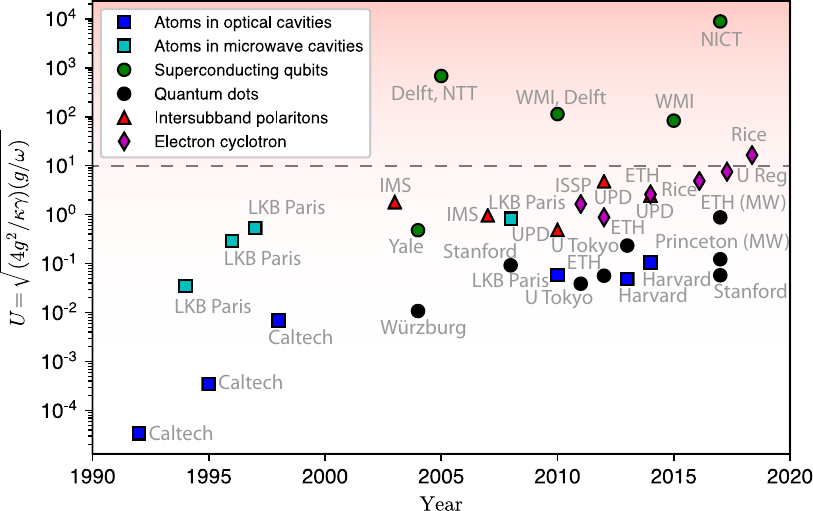}
\caption{\label{fig:U} (Color online) Evolution in time in cavity QED of the highest value of the parameter $U=(Cg/\omega)^{1/2}$ for different physical platforms from the same experimental points in Fig.~\ref{fig:g-w_r}. $C=4g^2/\kappa\gamma$ is the cooperativity, with $\kappa$ and $\gamma$ being the cavity and qubit loss rates, respectively. $U$ is an indicator of combined coupling strength and quantum coherence. References in addition to those in Fig.~\ref{fig:g-w_r}: quantum dots \cite{Faraon2008}; and cyclotron resonance~\cite{ZhangetAl16NP}, \cite{LietAl18NP}.}
\end{figure*}

Figure~\ref{fig:g-w_r} presents the evolution over time of the highest reported coupling strength $g$ normalized to the frequency of light of a confined mode $\omega$, in all fields exploring light-matter interactions. Clearly, experimental ultrastrong couplings are a recent advent over the past decade, mostly as a consequence of the interdisciplinary influence each area has had on the others. Figure~\ref{fig:U} shows the evolution over time of the parameter $U$, which we propose as a novel figure of merit in the USC regime. $U$ corresponds to the geometric mean between reduced coupling $g/\omega$ and the cooperativity factor used in atomic systems, $C=4g^2/\kappa\gamma$, with $\kappa$ and $\gamma$ representing the cavity and atomic losses, respectively. $U$ is therefore a measure of coherence in ultrastrongly coupled systems and, as observed in experiments, when its value largely exceeds unity $U\gg1$ it is possible to access the exotic physics of the USC regime without the blurring effects of dissipation. Otherwise, one could enter the USC regime without satisfying the usual definition of the strong coupling regime, i.e., $g > \gamma, \kappa$ \cite{DeLiberato2017}. From the data collected in Fig.~\ref{fig:U}, the superconducting qubits have entered well into the coherent USC regime, while the electron cyclotron resonances just achieved this new regime of physics \cite{LietAl18NP}.

This review presents a general overview of the theoretical and experimental progress in the USC and DSC regimes of light-matter interaction. In the past decade, experimental access to increasingly larger light-matter coupling strengths in different fields has brought forward USC and DSC regimes to the frontiers in quantum optics, both from a theoretical as well as from an experimental point of view. Moreover, beyond the fundamental interest, it is becoming natural to consider the impact of USC regimes in the context of the emerging interdisciplinary aspects of quantum technologies.

The physics of the USC regimes is currently an active research field that is in constant transformation and evolution. In particular, new lines of exploration of USC involving a continuum of modes have already been started~\cite{forn-diaz2017, puertas2018, Magazzu2018}, enabling the exploration of condensed matter models of relevant interest. Additionally, recent work in the two-photon quantum Rabi model~\cite{Felicetti2018} represents a playground for novel physics in nonlinear quantum optics. It is noteworthy to mention that in this review we cover neither open quantum systems nor multiphoton quantum Rabi models, nor the impressive developments in the QRM from a mathematical physics perspective~\cite{Chen2012,Zhong2013,Wakayama2013,Maciejewski2014,Braak2011,Braak2016}. However, we have tried to provide a connection to these growing areas of high theoretical and experimental interest. The USC regimes of light-matter interaction will keep on expanding at the frontier of quantum optics and quantum physics. We envision that all related topics to USC physics will remain a prominent field in the foreseeable future. During the processing of this review, other articles have been published with overviews on the field of USC~\cite{Gu2017, Kockum2019}, again demonstrating the impact this field has attained.

The contents of this review can be summarized as follows. Section~\ref{sec:2} presents an overview of the different light-matter interaction models. We follow a historical approach along the lines of cavity QED and the recent progress in theory and experiments related to the USC regimes. Section~\ref{sec:3} reviews the most relevant experiments having unveiled the physics related to the USC and DSC regimes. In Sec.~\ref{sec:4}, the quantum simulations of USC regimes are reviewed from a theoretical perspective. Section~\ref{sec:5} reviews a variety of potential applications of USC regimes from the point of view of quantum optics and quantum computation. Finally, Sec.~\ref{sec:6} presents our conclusions and outlook. 

\section{The quantum Rabi model}\label{sec:2}

The Rabi model~\cite{PhysRev.49.324} was introduced by Isidor Rabi in 1936 to describe the semiclassical coupling of a two-level atom with a classical monochromatic electromagnetic wave. In its fully quantized version, the model is given by the Hamiltonian
\begin{equation} 
\label{QRH}
\hat{\cal H}_R=\hbar (\Omega/2) \hat{\sigma}_{z} + \hbar \omega \hat{a}^{\dagger} \hat{a} + \hbar g \hat{\sigma}_{x} \left( \hat{a} + \hat{a}^{\dagger} \right) ,
\end{equation}
which is nowadays known as the quantum Rabi model. Here $\Omega$ and $\omega$ are the frequencies of the atomic transition and the electromagnetic field, respectively, and $g$ is the light-matter coupling strength. $\hat{\sigma}_{x,z}$ are Pauli matrices describing the atomic spin, while $\hat{a}$ and $\hat{a}^\dag$ are the annihilation and creation operators of the bosonic field mode, respectively. 

Equation~(\ref{QRH}) describes the dipolar coupling between a two-level atom, which could be a natural atom or an effective two-level system engineered from a solid-state device, and a quantized electromagnetic field mode. This Hamiltonian appropriately describes a plethora of quantum systems, several of which are laid out in Sec.~\ref{sec:3}. Alternative, equivalent forms of the quantum Rabi model have been studied in the literature using gauge transformations \cite{Drummond1987, Stokes2012, Stokes2017, Stokes2018a, Stokes2018b}. We have omitted a constant term $\hbar\omega/2$ in Eq.~(\ref{QRH}) as it does not modify the physics being discussed in this review.

In atomic systems, the achievable ratio $g/\omega$ between the coupling strength and the bosonic field mode frequency is orders of magnitude lower than unity [see~\cite{Kimble2008} for an overview of the achievements in cavity QED experiments]. One can easily understand the order of magnitude of the dipole interaction energy $\hbar g = -\vec{d}\cdot\vec{E}$, by expressing it as a function of system parameters (normalized to cavity frequency), $g/\omega = |\vec{d}|(2\hbar\epsilon_0V_{\rm m}\omega)^{-1/2}$, where $\vec{d}$ is the transition dipole moment between the relevant atomic states of transition frequency $\omega_A$, $\omega = \omega_A$ is the resonant frequency of the cavity, $\epsilon_0$ is the electric permittivity of vacuum, and $V_{\rm m}$ is the cavity mode volume. A typical Fabry-Perot optical cavity such as the ones used in experiments with cold atoms has mode volumes on the order of $V_{\rm m}\sim10^{-15}\rm{m}^{3}$~\cite{Rempe92}. The dipole moments of cesium and rubidium, which are heavy alkali atoms typically used in cavity QED experiments, are on the order of $|\vec{d}|\sim10^{-29}\mathrm{C\,m}$. For a cavity in resonance with cesium at 351.7~THz, this yields $g/\omega\sim10^{-7}.$ The only parameter which can be optimized further is the mode volume $V_{\rm m}$. The efforts by several groups engineering increasingly smaller mode volume cavities \cite{Vahala2003} based on evanescent fields near dielectric photonic microstructures \cite{Aoki2006} and nanostructures~\cite{Tiecke2014}, where $V_{\rm m}$ scales as $\sim\lambda^3$, have brought $g/\omega$ down to $10^{-6}$, which is a very large number for atomic systems but is still far from what has been achieved with solid-state devices (cf. Fig.~\ref{fig:g-w_r}).

Therefore, the QRM has been historically considered for cavity QED systems \cite{Raimond2001} in the so-called JC regime~\cite{jaynes1963}, where one performs the rotating-wave approximation and neglects the terms $\hat{a}^{\dag}\hat{\sigma}_+$ and $\hat{a}\hat{\sigma}_-$, which contribute weakly to the dynamics when $g/\omega\ll 1$. These terms are also known as counterrotating terms, since the other two interacting terms $\hat{a}^{\dag}\hat{\sigma}_-$ and $\hat{a}\hat{\sigma}_+$ are stationary in the interaction picture, therefore corotating with the uncoupled system Hamiltonian $\mathcal{H}_0 \equiv\hbar (\Omega/2) \hat{\sigma}_{z} + \hbar \omega \hat{a}^{\dagger} \hat{a}$. Here $\hat{\sigma}_{+}$ and $\hat{\sigma}_{-}$ are the raising and lowering atomic operators, respectively. The JC Hamiltonian therefore is given by
\begin{equation} \label{eq:JC}
\hat{\cal H}_{\rm JC}=\hbar (\Omega/2) \hat{\sigma}_{z} + \hbar \omega \hat{a}^{\dagger} \hat{a} + \hbar g \left( \hat{\sigma}_{+} \hat{a} + \hat{\sigma}_{-}\hat{a}^{\dagger} \right).
\end{equation}
The interaction term in the Hamiltonian $\hat{\cal H}_{\rm JC}$ is of an exchange type, leading to a conservation of the number of excitations in the system. This implies that only states with the same number of excitations interact, leading to a full diagonalization of $\hat{\cal H}_{\rm JC}$ in subspaces of $n$ number of excitations with JC doublets $|\pm\rangle_n$ as its eigenstates. By contrast, Eq.~(\ref{QRH}) contains only a parity symmetry and its exact diagonalization presents important difficulties see discussion leading to equations (\ref{eq:2-6}) and (\ref{eq:2-7}). \cite{Braak2011}. The JC model has been a cornerstone of quantum optics in the past 50 years. This model has had widespread use in a variety of physical platforms, ranging from neutral atoms in optical and microwave cavities, trapped ions with quantized motion, to superconducting qubits coupled to electromagnetic cavities, transmission line resonators and nanomechanical resonators. Recent implementations of small-scale quantum processors use the physics from Eq.~(\ref{eq:JC}) as the basis for the coherent quantum control of coupled quantum systems \cite{Corcoles2015}. 

In the regime where a detuning $\delta\equiv\Omega-\omega$ exists between the frequencies of the atom and the field mode, a Schrieffer-Wolf transformation can be applied to Eq.~(\ref{eq:JC}) if the dispersive condition is satisfied $g/\delta\ll1$, to become, up to second order in $g$ \cite{blais2004},
\begin{equation}\label{eq:AC}
\mathcal{\hat{H}}_{\rm ac}/\hbar = \frac{1}{2}\left[\Omega+\frac{g^2}{\delta}\right]\hat{\sigma}_z + \left[\omega + \frac{g^2}{\delta}\hat{\sigma}_z\right]\hat{a}^{\dag}\hat{a}.
\end{equation}
Equation~(\ref{eq:AC}) is known as the ac Stark Hamiltonian as well as the dispersive Hamiltonian. The atom-photon interaction is manifested in the nonradiative energy shifts that atom and field mode exert on each other. A detection of the field frequency yields information about the qubit state. This property is being widely exploited in quantum computing approaches, particularly with superconducting qubits \cite{Schuster2005}.

\begin{figure}[!hbt]
\centering
\includegraphics[width = \columnwidth]{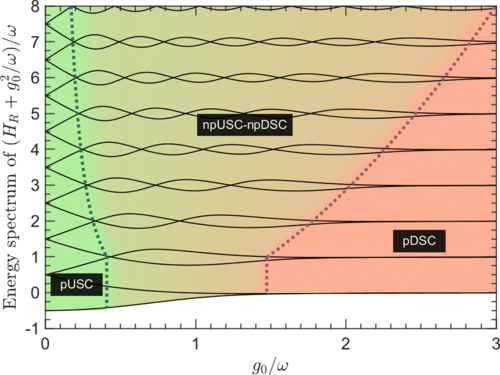}
\caption{\label{fig:class} Classification of the different coupling regimes of the quantum Rabi model (QRM). $g_0$ in the figure corresponds to $g$ as defined in the main text. The leftmost region at lowest couplings stands for the perturbative ultrastrong coupling (pUSC), which includes the Bloch-Siegert Hamiltonian regime. For the lowest-energy eigenstates it extends up to $g/\omega\sim0.4$. The intermediate region symbolizes the nonperturbative ultrastrong and deep strong coupling (npUSC and npDSC) regime. The color gradient around the boundaries symbolizes the lack of an abrupt transition in the physical properties of the QRM. The rightmost area is the perturbative deep strong coupling regime pDSC, where the qubit becomes a perturbation to the system. From~\cite{Rossatto2017}.}
\end{figure}

However, in the past decade, two novel regimes of light-matter interaction have emerged, namely, the USC regime, where $0.1\leq g/\omega< 1$, and the DSC regime, where $g/\omega> 1$. The lower limit $g/\omega=0.1$ has been by now well established as the regime where effects related to the counterrotating terms become sizable and, hence, observable. These new regimes exhibit a variety of physics which are not easily detectable with lower light-matter coupling strengths. In addition, one may take advantage of such new phenomena for quantum information applications as will be shown in Sec.~\ref{sec:5}. Figure~\ref{fig:class} displays the classification of different coupling regimes of the QRM~\cite{Rossatto2017} as a function of $g/\omega$ and for increasing energy eigenstates of Eq.~(\ref{QRH}).

The USC regime $0.1\leq g/\omega< 1$ can be divided into a perturbative region $0.1 \lesssim g/\omega \lesssim 0.3$ and a nonperturbative region $0.3 \lesssim g/\omega \lesssim 1$ \cite{Rossatto2017}. The perturbative region consists of a deviation from the JC model that accepts an analytical treatment by considering the counterrotating terms $\hat{a}^{\dag}\hat{\sigma}_+$ and $\hat{a}\hat{\sigma}_-$ as an off-resonant driving field. Applying perturbation theory to the quantum Rabi Hamiltonian [Eq.~(\ref{QRH})] up to second order on the perturbative parameter $\lambda\equiv g/(\Omega+\omega)$ yields the following Hamiltonian \cite{klimov2009}:
\begin{multline}\label{eq:BS}
\hat{\mathcal{H}}_{\rm BS}/\hbar = \frac{1}{2}\left(\Omega+\omega_{\rm BS}\right) \hat{\sigma}_{z} + \left(\omega + \omega_{\rm BS}\hat{\sigma}_z\right) \hat{a}^{\dagger} \hat{a} \\ -\frac{\omega_{\rm BS}}{2} + f(\hat{a}^{\dag}\hat{a})\hat{\sigma}_{-}\hat{a}^{\dagger} + \hat{\sigma}_{+} \hat{a}f(\hat{a}^{\dag}\hat{a}),
\end{multline}
where $\omega_{\rm BS}\equiv g^2/(\omega + \Omega)$ is the Bloch-Siegert shift. The coupling constant $g$ is renormalized to $f(\hat{a}^{\dag}\hat{a}) \equiv -g[1 - \hat{a}^{\dag}\hat{a}\omega_{\rm BS}/(\omega + \Omega)]$. The additional terms appearing in Eq.~(\ref{eq:BS}) compared to Eq.~(\ref{eq:JC}) are analogous to the ac Stark Hamiltonian [cf. Eq.~(\ref{eq:AC})], arising from having treated the counterrotating terms as an off-resonant driving field. Equation~(\ref{eq:BS}) is known as the Bloch-Siegert Hamiltonian, in analogy to the case of a strongly driven single spin~\cite{bloch1940}. 

The nonperturbative region $0.3 \lesssim g/\omega \lesssim 1$ departs from the standard quantum optical treatment of light-matter interaction. In this region, one has to resort to the exact solution for arbitrary coupling~\cite{Braak2011}. The JC model contains a conserved quantity which corresponds to the total number of excitations, $\mathcal{\hat{C}} = \hat{a}^{\dag}\hat{a} + (1/2)(\hat{\sigma}_z+1)$, leading to the solvability of the model. In contrast to the approximations in Eqs.~(\ref{eq:JC}) and (\ref{eq:BS}), the energy eigenvalues in the nonperturbative region are no longer given in closed form. The conservation of $\mathcal{\hat{C}}$ generates a continuous $U(1)$ symmetry of the JC model which in the nonperturbative region is broken down to a discrete $\mathbb{Z}_{2}$ symmetry, usually called parity, due to the presence of the counterrotating terms $\hat{a}\hat{\sigma}^-+\hat{a}^{\dag}\hat{\sigma}^+$ in Eq.~(\ref{QRH}). This is further evidenced by noting that the quantum Rabi Hamiltonian commutes with the parity operator $\hat{P}=\hat{\sigma}_{z} e^{i \pi \hat{a}^{\dagger}\hat{a}}$. This symmetry leads to a decomposition of the state space into two subspaces and is still sufficient to solve the model exactly \cite{Braak2011}, albeit in a nonanalytical form. However, the spectrum can be analyzed qualitatively, leading to the unification of quasiexact crossing points~\cite{Judd1979,Kus1986} and avoided crossings (see Fig.~\ref{fig:class}). 

In the first-ever work coining the USC regime~\cite{CiutietAl05PRB}, it was found that the ground state of an ultrastrongly coupled system in the nonperturbative region consists of a squeezed vacuum. Later works \cite{Ashhab2010} further explored the ground-state properties of the USC regime. In the ordinary vacuum $|g0\rangle$, in the zero- or weak-coupling regime, it is required that $\hat{\sigma}_- |g0\rangle = \hat{a}|g0\rangle = 0$. However, in the USC regime, the ground state $\widetilde{|g0\rangle}$ is a squeezed state, which contains a finite number of cavity photons and atomic population. Approximate solutions have been found to $\widetilde{|g0\rangle}$ (valid in the perturbative USC regime) \cite{Beaudoin2011}
\begin{equation}
\widetilde{|g0\rangle}\simeq\left(1-\frac{\Lambda^2}{2}\right)|g0\rangle - \Lambda|e1\rangle + \xi\sqrt{2}|g2\rangle,
\end{equation}
where $\Lambda\equiv\omega_{\rm BS}/g$, $\xi = g\Lambda/2\omega$, explicitly showing qubit-resonator excitations and a small degree of squeezing. At larger interaction strengths, the degree of squeezing is enhanced \cite{Ashhab2010}. %Therefore, $\hat{p} |G\rangle = 0$, where $\hat{p}$ is a linear combination of $\hat{\sigma}_+$, $\hat{a}^{\dagger}$, $\hat{\sigma}_-$, and $\hat{a}$. 
Further studies have looked into the possibility to release such a squeezed photon field by modulating different system parameters~\cite{CiutiCarusotto06PRA, LiberatoetAl07PRL, DeLiberato2009}.

As shown in Fig.~\ref{fig:class}, the nonperturbative USC regime merges in a continuous manner with the nonperturbative DSC regime. On the other hand, the perturbative DSC regime represents the extreme coupling condition $g/\omega\gg1$. Here the effective QRM Hamiltonian, in the spirit of spin-dependent forces, can be solved analytically while unitarily creating Schr\"odinger cat states.

In an important step to unveil the physics of the DSC regime~\cite{casanova2010}, new light was shed on the structure of the QRM following an analysis based on the symmetries of Eq.~(\ref{QRH}). As already mentioned, the quantum Rabi Hamiltonian contains a discrete $\mathbb{Z}_{2}$ symmetry. This symmetry is characterized by the parity operator $\hat{P}=\hat{\sigma}_{z} e^{i \pi \hat{a}^{\dagger}\hat{a}}$, which can take values $\pm 1$~\cite{casanova2010,wolf2013}. Therefore, the total Hilbert space splits into two infinite-dimensional invariant chains labeled by the parity eigenvalues
\begin{equation}
\label{eq:2-6}
\begin{split}
&|g0\rangle \leftrightarrow |e1\rangle \leftrightarrow |g2\rangle \leftrightarrow |e3\rangle \leftrightarrow \cdots \left( p = -1 \right)  , \\
&|e0\rangle \leftrightarrow |g1\rangle \leftrightarrow |e2\rangle \leftrightarrow |g3\rangle \leftrightarrow \cdots \left( p = +1 \right)  .
\end{split}
\end{equation}
The quantum Rabi Hamiltonian can be rewritten using the parity operator $\hat{P}$ and a composite bosonic mode $\hat{b} \equiv \hat{\sigma}_{x}\hat{a}$ as 
\begin{equation}
\label{eq:2-7}
\hat{\cal H}_{R}=\hbar \omega \hat{b}^{\dagger} \hat{b} + \hbar g \left( \hat{b} + \hat{b}^{\dagger} \right) -  \hbar(\Omega/2) \left( -1 \right)^{\hat{b}^{\dagger}\hat{b}} \hat{P}. 
\end{equation}
In the slow qubit limit $\Omega \to 0$, $\hat{\cal H}_{R} \to \left[ \hbar \omega \left( \hat{b}^{\dagger} + g/\omega \right)  \left( \hat{b} + g/\omega \right)  - \hbar g^{2}/\omega \right]$, which corresponds to a simple harmonic oscillator displaced by the ratio of the coupling with the frequency of the cavity~$g/\omega$.
\begin{figure}[!hbt]
\centering
\includegraphics[width = 8.5cm]{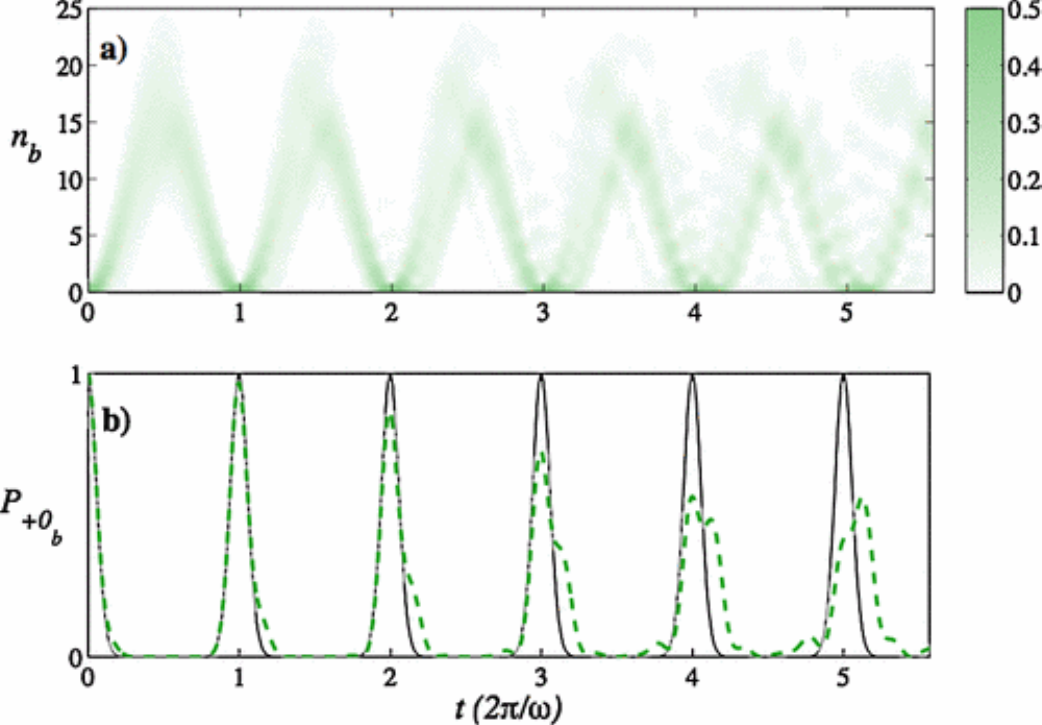}
\caption{\label{fig:jorge_Dsc} (Color online) Dynamics of the deep strong coupling (DSC) regime. (a) Photon statistics at different times of the evolution for $\Omega = 0.5 \omega$. When the qubit frequency $\Omega \neq 0$, the photon number wave packet suffers self-interference and is distorted. (b) Comparison of revival probability of the initial state $P_{+0_{b}}\left( t \right) = |\langle g , 0_{a} |\psi \left( t \right) \rangle | $ calculated for $\Omega = 0$ (solid line) and $\Omega = 0.5 \omega$ (dashed line). In the case $\Omega \neq 0$, full collapses and partial revivals are observed where the initial probability is not completely restored, with a maximum value that deteriorates as time evolves. In all simulations the initial state is $|g,0_{a}\rangle$ and $g/\omega = 2$. From~\cite{casanova2010}.}
\end{figure}

Figure~\ref{fig:jorge_Dsc} shows the time evolution of a state initially prepared in the uncoupled vacuum $|0,g\rangle$. Since this state is not an eigenstate of the quantum Rabi Hamiltonian, the system evolves as a wave packet climbing up and down the parity chains, displaying photon number wave packet oscillations. When the qubit frequency is finite, it effectively dephases the photon number oscillations which decay in amplitude over time. Also, the temporal development of qubit operators depends crucially on the presence of parity chain mixing~\cite{Wolf2012}.

The DSC regime requires a specific theoretical treatment due to its distinctive character when compared to USC physics, both in the discrete \cite{yoshihara2017a,BayeretAl17NL} and in the continuous mode approaches~\cite{forn-diaz2017}. In the latter, the description of a two-level system coupled to a continuum of modes has been traditionally the domain of study of the spin-boson model \cite{leggett1987, weiss}. Recent experiments have reached the nonperturbative interaction regime \cite{forn-diaz2017, Magazzu2018}, where the qubit becomes dressed by the photonic modes, resulting in a polariton with renormalized frequency \cite{Tao2018}. 

Within the QRM, in the regime where the coupling strength dominates over any other term, the limit of spin-dependent forces is expected. Such a limit was previously studied in trapped ion systems in order to achieve faster quantum computing operations, among other applications~\cite{Solano2003, haljan2005}. Finally, it is noteworthy to mention another surprising limit of the QRM when the mode frequency is negligible, giving rise to the emergence of the (1+1)-dimensional Dirac equation~\cite{LamataDirac2007,Gerritsma2010}. This connection was further explored in the literature~\cite{Gerritsma2011} and may still produce important analogies for quantum simulations of relativistic quantum models encoded in nonrelativistic quantum systems~\cite{Pedernales2018}.

Note that in the USC regime, the complete cavity QED Hamiltonian contains an additional term, the so-called $A^2$ term which represents the self-interaction energy of the field. This term usually contains a part that looks like $(g^2/\omega)\hat{a}^{\dag}\hat{a}$, so it is usually neglected due to the smallness of $g/\omega$. In the USC regime, however, it has an important role in most physical systems. An historical dispute in the context of cavity QED has surrounded the discussions about the $A^2$ term due to an initial prediction of a superradiant phase transition \cite{Dicke54PR, HeppLieb73AP, WangHioe73PRA} followed by a no-go theorem \cite{RzazewskietAl75PRL}. More recently, the dispute has surged back in discussing different quantum systems such as superconducting qubits \cite{nataf2010, ViehmannetAl11PRL, jaako2016} and polaritons \cite{HagenmullerCiuti12PRL, ChirollietAl12PRL}. Therefore, the study of the USC regime unavoidably leads to the exploration of the influence of the $A^2$ term in different physical systems as was highlighted in a recent theoretical work which also included direct dipole interactions between the two-level systems \cite{Bernardis2018}. Other theoretical works considered superradiance in a system with a single  \cite{Ashhab2013} and many \cite{Ashhab2017, Bamba2016} superconducting qubits in a cavity. Learning information about this term would lead to profound insight in the ultimate nature of light-matter interaction. Extensions of the QRM considering the anisotropic Rabi model \cite{Xie2014} including discussions of the $A^2$ term \cite{Maoxin2017} have also been investigated. In this modified QRM, the counterrotating terms are assumed with a different coupling strength $g_{cr}$ than the corotating terms $g$.

In systems based on a dense electron gas, such as polaritons in semiconductor quantum wells (see Sec.~\ref{sec:3B}), many identical electronic transitions are resonant with a single cavity mode. In that limit, the material excitation behaves as a bosonic quasiparticle, and a more adequate description is provided by the Hopfield Hamiltonian (boson-boson coupling) \cite{Hopfield58PR}, rather than the QRM (spin-boson coupling). It has been theoretically demonstrated \cite{Todorov2014} how an electronic system can evolve from the quantum Rabi Hamiltonian toward the Hopfield model, by changing the number of electrons. In comparing the two models, the multiple polariton branches of the dressed states in the QRM are progressively washed out, leaving only two polariton branches as observed in experiments with polaritons in semiconductor quantum wells. In describing such dense electron gas systems, alternative Hamiltonians were used in the literature in a different gauge rather than the usual minimal coupling Hamiltonian where the $A^2$ term previously mentioned appears. In the Coulomb gauge and the dipole representation, the $A^2$ term is replaced by a $P^2$ term. The resulting Hamiltonian was used to study nonperturbative superradiant emission of collective excitations in a two-dimensional electron gas \cite{Huppert2016}. These modified Hamiltonians better capture the effects in condensed matter systems, such as those described in Sec.~\ref{sec:3B}.

\section{Experiments in the USC and DSC regimes}\label{sec:3}

Ultrastrong coupling regimes have been the focus of theoretical studies for many decades \cite{Shirley1965, Cohen1973, Zela1997, irish2007}. It was not until the late 2000s that the first truly experimental sightings of light-matter interactions in the USC regime were realized~\cite{niemczyk2010, forn-diaz2010, AnapparaetAl07SSC, DupontetAl07PRB}. This first round of experimental results triggered a period of intense theoretical exploration. Therefore, the experimental progress has marked the pace at which the field has evolved. Coincidentally, the exploration of the USC regime in several physical systems has taken place at about the same period of time. In this section, we overview the most relevant of these fields, namely, superconducting quantum circuits (Sec.~\ref{sec:3A}), semiconductor quantum wells (Sec.~\ref{sec:3B}), and other hybrid quantum systems (Sec.~\ref{sec:3C}).

\subsection{Superconducting quantum circuits}\label{sec:3A}

Superconducting circuits in the quantum regime were shown to be an excellent platform to study light-matter interactions in the microwave regime of frequencies. Early studies of qubit-resonator systems~ \cite{blais2004, wallraff2004} found that a superconducting qubit interacting with the mode of a microwave resonator follows the exact same physics as that of cavity QED, with the qubit playing the role of an artificially engineered atom and the resonator mode emulating the cavity. By analogy, this platform of light-matter interactions on a superconducting circuit was defined as circuit QED. 

The experimental exploration of ultrastrong interactions in superconducting quantum circuits was initiated in 2010, following several years of development of circuit QED \cite{Gu2017}. Early experiments in the strong coupling regime used capacitive \cite{schuster2007, bishop2009}, mutual geometric \cite{johansson2006}, and galvanic inductive couplings \cite{chiorescu2004}. The first two experiments reaching USC regimes used galvanic couplings instead~\cite{niemczyk2010, forn-diaz2010}. Both experiments reported clear evidence of deviations from the conventional model used in quantum optics, the JC model introduced in Sec.~\ref{sec:2}~\cite{jaynes1963}. The couplings achieved are nowadays cast in the perturbative USC regime~\cite{Rossatto2017}. The experiments in 2010 were followed by several studies addressing distinct features related to counterrotating wave physics inherent to the perturbative USC regime~\cite{chen2017, forn-diaz2016, baust2016}. In 2016, two independent experiments attained a qualitative jump in the light-matter interaction strength, pushing the boundaries into the nonperturbative USC domain by using Josephson junctions as coupling elements. These experiments spanned both closed~\cite{yoshihara2017a} and open system settings~\cite{forn-diaz2017} and entered the DSC regime~\cite{casanova2010, Rossatto2017}. In parallel to the engineering of circuits showing USC and DSC physics, novel techniques of digital and analog quantum simulation using superconducting circuits and trapped ions studied the QRM in these extreme coupling regimes~\cite{langford2016, braumuller2016, Lv2018}. Altogether, the year 2016 consolidated the field of research on USC regimes in superconducting circuits from both a fundamental and an applied point of view \cite{Braak2016}.

A summary of the milestones in coupling strength achieved in experiments with superconducting quantum circuits is reported in Table~\ref{table:sqc}.

\begin{table*}[t]
  \centering
  \begin{tabular}{|c|c|c|c|c|c|c|c|c|c|c|}
    \hline 
     & Qubit & Cavity & Interaction & $\gamma/2\pi$ & $\kappa/2\pi$ & $g/2\pi$ & $\omega_r/2\pi$ & $g/\omega_r$ & $U$ & \\ Reference
     & type & type & type & (MHz) & (MHz) & (MHz) & (GHz) & (\%) & & Notes\\
    \hline
    \hline
	\cite{wallraff2004}  & CPB & TL  & Capacitive & 0.7 & 0.8 & 5.8 & 6.044 &  0.1 & 0.24 & First strong coupling \\
	\hline
	\cite{chiorescu2004}  & FQ & LE  & Galvanic, external & 27 & 1.6 & 200 & 2.91 & 6.9 & 7.97 & Resonator SQUID \\
	\hline
	\cite{johansson2006}  & FQ & LE  & Galvanic, external & 0.2 & 0.2 & 216 & 4.35 & 5 & 241 & First vacuum oscillations \\
	\hline
	\cite{schuster2007}  & TR & TL  & Capacitive & 0.25 & 1.6 & 105 & 5.7 & 2 & 22.9 & First transmon work \\
	\hline
	\cite{bishop2009} & TR & TL & Capacitive & 0.3 & 0.09 & 173.5 & 6.92 & 2.5 & 167 & \\
	\hline
        \cite{fedorov2010}  & FQ & LE & Galvanic, external & 2.9 & 0.1 & 119.5 & 2.723 & 4.4 & 46.5 & \\
	\hline
	\cite{niemczyk2010}  & FQ & TL & Galvanic, external & 2.5 & $<2$ & 636 & 5.357 & 12 & 98 & First USC work\\
	\hline
	\cite{forn-diaz2010}  & FQ & LE & Galvanic, external & $<10$ & 10 & 810 & 8.13 & 10 & 25.6 & Bloch-Siegert in USC\\
	\hline
	\cite{baust2016}  & FQ & TL & Galvanic, external & $\sim10$ & $\cdots$ & 775 & 13.3 & 17.2 & $\cdots$ & Dressed mode coupling\\
	\hline
	\cite{chen2017} & FQ & TL & Galvanic, external & $\sim1$ & $\cdots$ & 306 & 3.143 & 9.7 & $\cdots$ & \\
	\hline
	\cite{yoshihara2017a}  & FQ & LE & Galvanic, internal & $\sim1$ & $\sim1$ & 7630 & 5.711 & 134 & 8819 & First DSC work\\
	\hline
	\cite{yoshihara2017b}  & FQ & LE & Galvanic, internal & $\sim1$ & $\sim1$ & 5310 & 6.203 & 86 & 4913 & \\
	\hline
	\cite{bosman2017a} & TR & TL & Capacitive & 29.3 & 38 & 455 & 6.23 & 7.1 & 3.7 & \\
	\hline
	\cite{bosman2017b} & TR & TL & Capacitive & 3.1 & $<0.1$ & 897 & 4.268 & 19 & 739 & First USC transmon\\
	\hline
	\cite{Yoshihara2017c} & FQ & LE & Galvanic, internal & $\sim1$ & $\sim1$ & 7480 & 6335 & 118 & 16256& \\
	\hline
  \end{tabular}
    \caption{\label{table:sqc}Experimental observations of ultrastrong light-matter coupling in superconducting quantum circuits. CPB: Cooper pair box. FQ: flux qubit. TR: transmon qubit. TL: transmission line resonator. LE: lumped-element resonator. $\gamma$: qubit decay rate.  $\kappa$: photon decay rate.  $g$: coupling strength.  $\omega_r$: resonator frequency. $U\equiv\sqrt{(g/\omega_r)4g^2/\kappa\gamma}$: geometric mean between cooperativity and normalized coupling strength. SQUID: superconducting quantum interference device.}
\end{table*}

\subsubsection{Circuit considerations: Qubit-resonator systems}
\label{sec:3A1}
\begin{figure}[!hbt]
\centering
\includegraphics{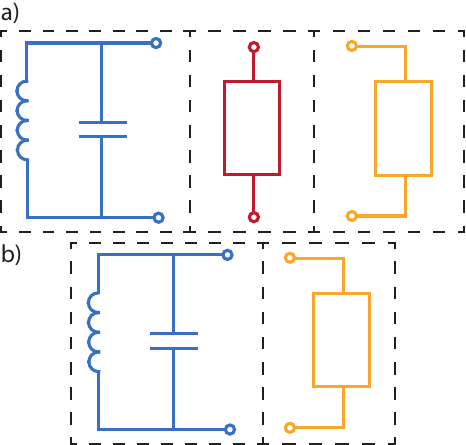}
\caption{\label{fig:types} (Color online) a) Circuit schematic of external coupling, with a circuit element (center, red) which couples resonator (left, blue) and qubit (right, yellow). Capacitors or inductors are examples of possible coupling elements. %\emph{External couplings} are included in the regime of strong coupling. 
(b) Internal coupling where the qubit (right, yellow) and resonator (blue, left) shunt each other and share internal degrees of freedom.}
\end{figure}

The interaction between light and matter is fundamentally manifested as a modification of a property of one of the interacting subsystems due to the presence of the other one. Consider a single atom placed in a dielectric. The presence of the atom represents a sudden modification of the medium through which light propagates. This pointlike discontinuity in the dielectric causes a modification of the electromagnetic field distribution of photons, resulting in a net light-matter interaction. In the case of circuits, superconducting qubits play the role of effective artificial atoms. In analogy to natural atoms, the presence of a qubit induces a strong change in the impedance of the circuit through which microwave photons propagate, enabling qubit-photon interactions. The interaction in this case may be capacitive or inductive, depending on the circuit design, and generally will be determined by the geometry of a coupling circuit element, a capacitor or an inductor, respectively [Fig.~\ref{fig:types}(a)]. We define this type of coupling as \emph{external}. Within the \emph{strong coupling regime} where the interaction strength $g$ dominates over qubit loss $\gamma$ and cavity loss $\kappa$, the qubit-photon interaction is perturbative with respect to the cavity mode frequency $\omega$, $\kappa,\gamma\ll g\ll\omega$, leaving the bare eigenstates of the interacting subsystems unmodified. The eigenstates of the total system will still consist of superpositions of qubit and photon in a dressed-state basis \cite{jaynes1963}. So far, it has been possible to attain the perturbative USC regime with external couplings. %For this reason we refer to strong coupling regimes as~\emph{external} in this review. 

There exists an important difference between atomic systems and superconducting circuits: superconducting qubits are circuits themselves, allowing the possibility to directly embed the artificial atom in the medium of propagation of photons [Fig.~\ref{fig:types}(b)]. In this way, the two coupled systems share more than just mutual geometric elements of the circuit (capacitive and/or inductive) which store the interaction energy, as is the case for external couplings [Fig.~\ref{fig:types}(a)]. As described later in this section, circuit engineering permits sharing an actual \emph{internal} degree of freedom between the artificial atom and the resonator, which becomes the actual source of coupling. We refer to this type of coupling as internal. In such a scheme, the qubit degrees of freedom become renormalized by the elements of the coupling resonator circuit \cite{Manucharyan2017}, such that it is difficult to talk about separate qubit or resonator degrees of freedom. With such a strong interaction, the natural basis of eigenstates of the qubit circuit is modified, for both charge-type [Cooper pair box (CPB), and transmon qubit] and flux-type qubits (flux qubit and fluxonium qubit). This is the fundamental key point that permitted attaining coupling strengths well above the excitation frequencies of the interacting subsystems, i.e., the nonperturbative USC and DSC regimes \cite{yoshihara2017a, forn-diaz2017}. 

Superconducting qubits are generally classified into two types: flux type and charge type. The qubit-resonator interaction can be of inductive (which includes galvanic coupling) or capacitive nature. All types of superconducting qubits developed so far have been shown to couple with either type of interaction. Generally speaking, the capacitive interaction is determined by the mutual capacitance between the two coupled circuits.   Similarly, geometric inductive couplings are given by the mutual qubit-resonator inductance. Galvanic couplings are given by the superconducting phase drop that is developed across the shared mutual inductance between the two circuits (see Sec.~\ref{sec:3.1.b}). It is possible to reach ultrastrong couplings with both capacitive and galvanic interactions, with quite different fundamental limits imposed for each type, as detailed in the next sections.

We emphasize that all formulas shown in this section are specific to a lumped-element resonator for which there is no spatial dependence on the amplitude of the electromagnetic field fluctuations, and only a single resonant mode exists. This is in contrast to distributed resonators made of a section of a transmission line. In the latter, the presence of the qubit modifies the amplitude of the resonator field at that location, leading to a decrease of the interaction strength. This is due to the appearance of additional coupling mechanisms. For example, a flux qubit inductively coupled to a transmission line resonator develops a capacitive coupling at the expense of the inductive interaction \cite{bourassa2012}.

Each superconducting qubit is defined within a subset of a larger Hilbert space of eigenstates of the whole quantum circuit. A recent theoretical study considered the complete circuit Hamiltonian of both flux-type and charge-type superconducting qubits embedded in a resonator \cite{Manucharyan2017}. Deviations from the QRM were evidenced but found to not alter the main qualitative properties of the model, particularly for the ground state. The conclusions of this study will be presented in Sec.~\ref{sec:3.1.b}. 

In the following sections we explore the limits to capacitive and galvanic interactions. Mutual geometric inductive couplings are less interesting as one requires very large qubits, hundreds of $\mu$m long, to attain sufficiently large mutual inductance. This in turn modifies the qubit eigenstates and eventually reduces the qubit persistent current so the coupling starts to decrease. Therefore, in practice the largest attainable qubit-resonator interaction strength is lower than using galvanic interactions.

\subsubsection{Capacitive couplings}\label{sec:3.1.a}

Capacitive couplings have been widely used with all types of superconducting qubits engineered so far \cite{wallraff2004, inomata2012, hofheinz2009, manucharyan2009}. This type of coupling is proportional to the root mean square (rms)~voltage $\hat{V}$ in the ground state of the resonator mode with frequency $\omega_r$ and capacitance $C_r$:
\be
V_{\rm rms}\equiv\langle0| \hat{V}^2|0\rangle^{1/2} = \sqrt{\frac{\hbar\omega_r}{2C_r}}=\omega_r\sqrt{\frac{\hbar Z}{2}},
\ee
which scales as $\sqrt{Z}$, where $Z$ is the impedance of the resonator mode coupled to the qubit \cite{devoret2007, andersen2016, jaako2016}. This scaling already points to high-impedance resonators to reach the USC regime.

\begin{figure}[!hbt]
\centering
\includegraphics[width = 7cm]{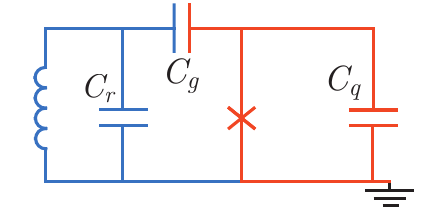}
\caption{\label{fig:cqb} Circuit model of a charge qubit shunted with capacitance $C_q$ coupled with a capacitor $C_g$ to a lumped resonator of capacitance $C_r$. The cross corresponds to the circuit element of a Josephson junction. Lumped resonator (left) is depicted in blue, charge qubit (right) in red. This model is valid both for Cooper pair boxes as well as transmon qubits.}
\end{figure}

The most common type of charge qubit is known as the Cooper pair box. This qubit consists of a superconducting island connected to a large reservoir by a Josephson junction. The island may be connected to another circuit by additional capacitors, as shown in the circuit in Fig.~\ref{fig:cqb}. The qubit junction capacitance $C_q$ may consist of the self-capacitance of the junction or a shunt capacitor externally defined. The CPB Hamiltonian is given by \cite{Bouchiat98}
\begin{multline}
\label{eq:CPBH}
\hat{\mathcal{H}}_{\rm CPB} = 4E_C\sum_{N\in \mathbb{Z}}(\hat{N} - N_{\rm ext})^2|N\rangle\langle N|  \\ + E_J\sum_{N\in \mathbb{Z}}(|N\rangle\langle N+1| + \rm{h.c.}).
\end{multline}
Here $\hat{N}$ is the Cooper pair number operator, and $E_C = e^2/2C_{\Sigma}$ is the charging energy of the Cooper pair island of total capacitance $C_{\Sigma}$, which is equal to $C_q + C_g$ in the circuit in Fig.~\ref{fig:cqb}. $E_J$ is the Josephson energy of the junction connecting the box to the reservoir. $N_{\rm ext} = C_gV_{\rm ext}/2e$ is the charge externally induced on the island via the capacitor $C_g$. ``h.c." stands for Hermitian conjugate. When the qubit is connected to a resonator, as in Fig.~\ref{fig:cqb}, the external voltage corresponds to the quantized voltage from the resonator $V_{\rm ext} = \hat{V}_r = V_{\rm rms}(\hat{a} + \hat{a}^{\dag})$ \cite{blais2004}. When writing out explicitly all terms in Eq.~(\ref{eq:CPBH}), the cross term results in the interaction energy between the charge qubit and the resonator, 
\begin{equation}
\label{eq:cpb_int}
\hat{\mathcal{H}}_{\rm int} = -2e\hat{N}\frac{C_g}{C_{\Sigma}}V_{\rm rms}(\hat{a} + \hat{a}^{\dag}).
\end{equation}
Equation (\ref{eq:cpb_int}) is general and applies to all types of charge-based qubits, such as the CPB and the transmon. In Eq.~(\ref{eq:cpb_int}), the factor $2e\hat{N}$ plays the role of the qubit dipole moment. One can picture this dipole moment as a charge $2e$ moving between the two plates of the capacitor where an external voltage $\hat{V}_{\rm ext}$ has been induced by the external circuit \cite{devoret2007}. 

For a CPB in the charging regime $4E_C\gg E_J$ and for low enough temperatures $E_C\gg k_BT$ that the system lies in its ground state, the Cooper pair number operator may be represented in the basis defined by the two states $|0\rangle$ and $|1\rangle$, representing excess Cooper pairs on the island. Using the Pauli matrix representation $\hat{\sigma}_x = |N\rangle\langle N+1| + \rm{h.c.}$, the Cooper pair number operator is now represented as $\hat{N}\simeq\hat{\sigma}_z$. Equation~(\ref{eq:CPBH}) can be rewritten as $\hat{H}_{\rm CPB} = -(E_{\rm el}/2)\hat{\sigma}_z -(E_J/2)\hat{\sigma}_x$, with $E_{\rm el} \equiv 4E_C(1-2N_g)$. In this charging regime, Eq.~(\ref{eq:cpb_int}) has a modified form 
\be
\hat{\mathcal{H}}_C^{\rm CPB} = 2e\frac{C_g}{C_g+C_q}V_{\rm rms} \hat{\sigma}_x(\hat{a}+\hat{a}^{\dag}).
\ee
The equivalent of the qubit dipole moment here takes the simple form $|\langle 0|2e\hat{N}_{\rm CPB}|1\rangle| = 2e$.

If we now consider the limit $E_J\gg E_C$, we enter the transmon regime \cite{Koch2007}. In this regime, the CPB Hamiltonian can be approximated by a harmonic oscillator with some nonlinearity which introduces anharmonicity in the spectrum. Now, the analog of the dipole moment of the qubit, calculated in the transmon basis, takes a different form $|\langle 0|2e\hat{N}_{\rm tr}|1\rangle| = e(E_J/2E_C)^{1/4}$, leading to a modified interaction Hamiltonian
\be\label{eq:tr_cap}
\hat{\mathcal{H}}_C^{\rm tr} = e\frac{C_g}{C_g+C_q}\left(\frac{E_J}{2E_C}\right)^{1/4}V_{\rm rms}\hat{\sigma}_x(\hat{a}+\hat{a}^{\dag}).
\ee
The coupling strength $g$ in the last expression can be rewritten in a reduced form \cite{devoret2007}
\be\label{eq:gc}
\frac{g_C^{\rm{tr}}}{\omega_r} = \frac{1}{\sqrt{2\pi^3}}\left(\frac{E_J}{2E_C}\right)^{1/4}\sqrt{\frac{Z}{Z_{\rm{vac}}}}\frac{C_g}{C_g+C_q}\alpha^{1/2}.
\ee
$Z_{\rm{vac}}=\sqrt{\mu_0/\epsilon_0}\simeq377\,\Omega$ is the vacuum impedance while $\alpha\simeq1/137$ is the fine structure constant. Note that in conventional cavity QED experiments where a Rydberg atom interacts with a photon, $g/\omega$ is proportional to $\alpha^{3/2}$  \cite{devoret2007}. The different scaling obtained in circuit QED $\alpha^{1/2}$ is related to the different dimensionality of the dipole moment, being 3D for Rydberg atoms and 1D for circuit QED. Equation~(\ref{eq:gc}) shows the fundamental limitations for transmon qubits and capacitive couplings. It has been shown \cite{jaako2016} that this type of coupling cannot reach the DSC regime $g/\omega_r>1$ as the coupling is bound by 
\be\label{eq:jaako}
\frac{g^{\rm{tr}}_C}{\omega_r} = \frac{C_g}{\sqrt{C_r(C_q+C_g) + C_g(C_g+C_q)}}<1,
\ee
for exact qubit-photon resonance. The capacitances refer to the circuit in Fig.~\ref{fig:cqb}. Typical circuit parameters limit this quantity to $g_C^{\rm{tr}}/\omega_r\approx0.01$ for $Z = 50\Omega$.

\begin{figure*}[!hbt]
\begin{centering}
\includegraphics[width = 11cm]{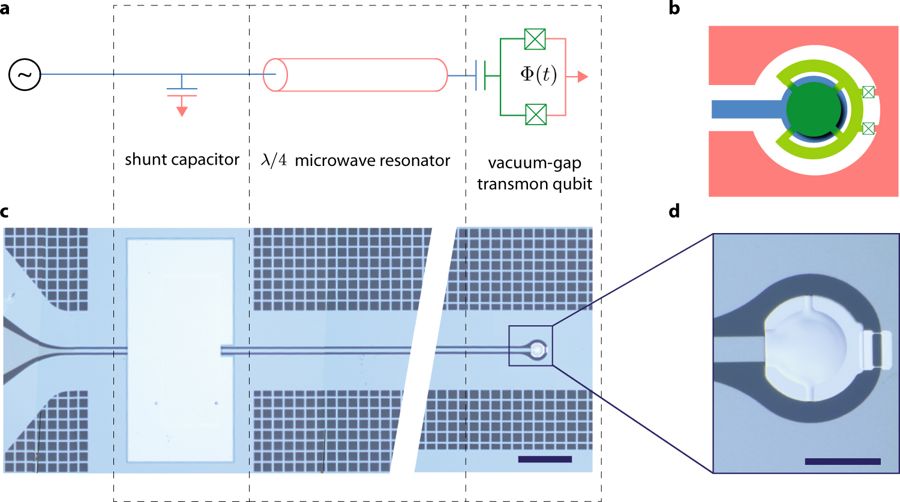}
\caption{(Color online) USC with capacitive coupling. (a) Device schematic of a transmission line resonator capacitively coupled to a transmon qubit. (b) Schematic of the vacuum gap capacitor shunting the qubit junctions. (c) Scanning electron micrograph (SEM) of the device, showing the shunt capacitor that defines the resonator port. (d) SEM zoom-in of the qubit, displaying the vacuum gap capacitor and the Josephson junctions.}
\label{fig:sal_circ}
\end{centering}
\end{figure*}
The same analysis for pure charge qubits (CPB) gives a reduced coupling of 
\be\label{eq:g_c_cpb}
\frac{g_C^{\rm CPB}}{\omega_r} = \sqrt{\frac{2}{\pi^3}}\sqrt{\frac{Z}{Z_{\rm vac}}}\frac{C_g}{C_g+C_J}\alpha^{1/2}.
\ee
Using a lumped-element resonator model, the reduced coupling can be recast using circuit parameters in analogy to the transmon case \cite{jaako2016}
\be\label{eq:g_cpb}
\frac{g^{\rm{CPB}}_C}{\omega_r} = \frac{2C_g}{\sqrt{C_r(C_q+C_g) + C_g(C_g+C_q)}}\sqrt{\frac{E_C}{E_J}}.
\ee
Note that the frequency of a CPB is assumed here to be $\hbar\omega_q= E_J$. Equation~(\ref{eq:g_cpb}) shows that it is in principle possible to reach the DSC regime with a CPB with $E_C\gg E_J$. In practice, the limitation on charge qubit lifetime makes this circuit implementation challenging. The circuit parameters used so far in experiments involving CPBs and resonators \cite{wallraff2004} achieved values of $g_C^{\rm{CPB}}/\omega_r\approx0.01$ with a resonator impedance $Z = 50\Omega$. 

We point out that the limits imposed by Eqs.~(\ref{eq:gc})-(\ref{eq:g_cpb}) are specific to the circuit\footnote{Equations (\ref{eq:jaako}) and (\ref{eq:g_cpb}) are obtained from a modified but similar circuit to that shown in Fig.~{\ref{fig:cqb}} \cite{jaako2016}.} shown in Fig.~\ref{fig:cqb}. However, as will be shown in Sec.~\ref{sec:3.1.b}, a charge qubit, either transmon or CPB, shunted by an $LC$ circuit presents a chargelike interaction with a coupling strength which can reach well into the $g/\omega>1$ regime \cite{Manucharyan2017}.

The $\sqrt{Z}$ scaling of the coupling in Eqs.~(\ref{eq:gc}) and (\ref{eq:g_c_cpb}) is originated from the resonator voltage fluctuations $V_{\rm rms}$, favoring high-$Z$ resonators. Employing high kinetic inductance films or Josephson junction arrays \cite{masluk2012, andersen2016}, impedances of several k$\Omega$ would allow one to reach the regime $g_C^{\rm CPB, tr}\approx\omega_r$. 

\begin{figure*}[!hbt]
\begin{centering}
\includegraphics[width = 13 cm]{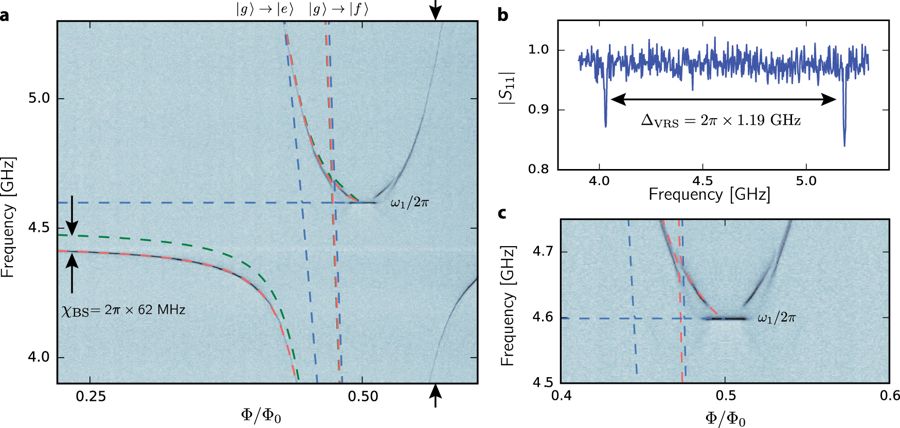}
\caption{(Color online) Spectrum of the capacitively coupled transmon-resonator device in the USC regime. (a) Spectrum displaying an avoided-level crossing. The green dashed line shows the JC model. The blue dashed lines show the uncoupled qubit and resonator transitions. The red dashed line is the QRM for a multimode system. A Bloch-Siegert shift of 62 MHz is clearly displayed as a deviation from the JC model. (b) Vacuum Rabi splitting. (c) Zoom-in of the anticrossing area showing additional avoided-level crossings of the qubit.}
\label{fig:sal}
\end{centering}
\end{figure*}

The first experiment reporting USC with a capacitive coupling consisted of a superconducting transmon qubit coupled to a transmission line resonator \cite{bosman2017b}. The strength of the coupling was attained by implementing a vacuum gap parallel-plate geometry (see Fig.~\ref{fig:sal_circ}) in which the qubit shunt capacitor was suspended over the ground plane, enhancing in this way the ratio of coupling capacitance $C_g$ to total capacitance $C_g+C_q$ in Eq.~(\ref{eq:tr_cap}). Being an effective drum 30~$\mu\rm{m}$ in diameter suspended less than 1~$\mu\rm{m}$ over the resonator ground led to a coupling capacitance nearly an order of magnitude larger than planar capacitance designs. Combined with a high-impedance superconducting transmission line resonator, an USC of up to $g/\omega_r\sim0.19$ was observed with the fundamental resonator mode [Figs.~\ref{fig:sal}(a)-(c)]. The high resonator impedance was achieved by narrowing the center line of the resonator and in this way reducing the capacitance per unit length between the ground planes and the center line. The spectrum of the transmon qubit shown in Fig.~\ref{fig:sal}(a) displayed dispersive effects from the multiple modes of the resonator coupling to the qubit, including qubit-mediated mode-mode interactions. Clear deviations from the JC model were observed, reporting a single-photon Bloch-Siegert shift of $\omega_{\rm{BS}}/2\pi = 62$~MHz. 

Despite Eq.~(\ref{eq:jaako}) limiting the ratio $g_C^{\rm{tr}}/\omega_r$ to lie below~1, transmon-based devices approaching the DSC regime may be demonstrated in the near future. One possible avenue to reach that goal is to engineer the impedance of the transmission line resonator to even higher values. In a separate work \cite{puertas2018}, Puertas-Mart\'inez \emph{et al.} demonstrated a USC coupling strength between a qubit and multiple modes of a superconducting quatum interference device (SQUID) array acting as a high-impedance transmission line. The impedance of the array was measured to lie in the k$\Omega$ range. Even though the experiment was designed as an open system, and therefore modeling the spin-boson model rather than the QRM, the scaling of the qubit-line coupling followed closely that from Eq.~(\ref{eq:gc}). As already mentioned, another straightforward way to enter the DSC regime with charge qubits is to directly shunt the qubit by an $LC$ resonator.

\subsubsection{Galvanic couplings}\label{sec:3.1.b}

\begin{figure}[!hbt]
\centering
\includegraphics[width = 7cm]{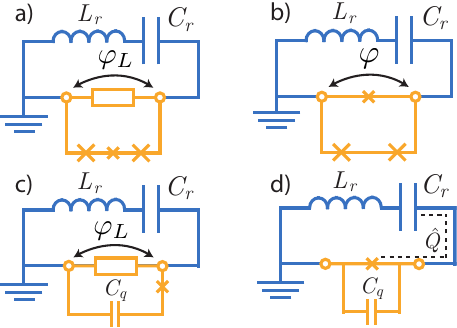}
\caption{\label{fig:galvanic} (Color online) Circuit model for galvanic couplings. (a) Flux qubit sharing a section of its loop with a resonator. The coupling element consists of a linear inductance. (b) Flux qubit embedded into the resonator loop. The coupling is given by the phase across the shared junction. (c) Charge qubit sharing an inductance with a resonator. The coupling element is given by the shared inductance. (d) Charge qubit embedded in the resonator loop. The coupling operator is related to the charge $\hat{Q}$ stored in the superconducting island shared between qubit and resonator, highlighted by the dashed line. (a), (c) External coupling elements, while (b) and (d) are internal couplings.}
\end{figure}

Two systems are galvanically coupled when they share a portion of their respective circuits. Here we distinguish two types of galvanic couplings based on the amount of circuit shared: (a)~sharing a linear inductance and (b)~embedding the qubit directly into the resonator circuit. The general picture is that the qubit and resonator share a circuit element, the latter case being the entire qubit itself. In both situations, the qubit-resonator coupling is then given by the superconducting phase drop across the shared circuit element $\hat{\varphi}$, which itself is a new degree of freedom of the circuit; see Fig.~\ref{fig:galvanic}. For flux-type qubits [Figs.~\ref{fig:galvanic}(a), \ref{fig:galvanic}(b)], $\hat{\varphi}$ can be represented in the basis of eigenstates of the qubit $\langle i|\hat{\varphi}|j\rangle$, which relates to the current running across the inductive element [see Eq.~(\ref{eq:liqir})]. For charge-type qubits, an inductor in series with the qubit junction may be shared with a resonator, as shown in Fig.~\ref{fig:galvanic}(c). Increasing the coupling strength in this configuration will be at the expense of the qubit anharmonicity~\footnote{See related literature for a more detailed calculation of the effects of linear inductors in transmon qubits~\cite{bourassa2012}.}, since the linear inductance dilutes the effect of the Josephson junction and brings the qubit closer to a linear oscillator. Therefore, it is not very favorable for reaching ultrastrong interaction strengths, and we will not discuss this configuration further. In practice, this type of interaction has been implemented only in coupled-qubit circuits~\cite{Chen2014}. The other possibility~\footnote{Here we are discussing only transverse-type couplings. For both flux-type and charge-type qubits, a longitudinal coupling can be instead engineered by replacing one of the qubit junctions by a SQUID loop and galvanically attaching a fraction of this loop to a resonator circuit. We will not discuss longitudinal couplings in this review.} is to embed the qubit in the resonator circuit [Fig.~\ref{fig:galvanic}(d)], where the coupling is to the charge degree of freedom $\hat{Q}$ on the island formed on one side of the qubit junction. 

Coupling to the phase $\hat{\varphi}$ involves the rms~current $\hat{I}$ in the ground state of the resonator mode with frequency $\omega$ and inductance $L_r$:
\be\label{eq:irms}
I_{\rm rms} \equiv\langle0|\hat{I}^2|0\rangle^{1/2} = \sqrt{\frac{\hbar\omega_r}{2L_r}} = \omega_r\sqrt{\frac{\hbar}{2Z}}.
\ee
Clearly, in order to maximize the coupling strength, low resonator impedance $Z$ is desirable.

In what follows, we will use the three-junction flux qubit \cite{Mooij1999} to analyze the different types of galvanic couplings. The description can be easily extended to the fluxonium \cite{manucharyan2009} and other flux-type qubit circuits. 

\paragraph{Linear inductance.-} Here we focus only on flux-type qubits, but the discussion can be extended to charge qubits in the configuration shown in Fig.~\ref{fig:galvanic}(c). The circuit topology of a flux-type qubit consists of one or more junctions interrupting a superconducting loop, a section of which can be shared with a resonator circuit; see Fig.~\ref{fig:galvanic}(a). The coupling element is then the shared linear inductor~$L$, which adds a degree of freedom to the circuit, the phase drop across it $\hat{\varphi}_L$. In the perturbative USC regime, which corresponds to the experiments described in this section, the value of the coupling inductance is typically small compared to the resonator inductance $L_r$ and the qubit loop inductance. Therefore, $\hat{\varphi}_L$ is frozen in its ground state and is treated as a constant which becomes a perturbation to the qubit-resonator system.\footnote{The linear coupling inductance in typical flux qubit loops a few micrometers in size does not significantly contribute to the energy spectrum and is usually neglected.} Therefore, in this regime of small coupling inductance, the coupling element does not modify the bare qubit or resonator spectra and is therefore an external coupling as defined in Sec.~\ref{sec:3A1}.

The inductance of a superconducting wire has a geometric as well as a kinetic origin. The inductance from a Josephson junction may also be used as a linear inductor, provided that its critical current is much larger than the current flowing through it. The geometric inductance is typically calculated from $L_G = (\mu_0l/2\pi)\left[\ln\left(2l/w+t\right)+1/2\right]$. Here $l$, $w$, and $t$ are the wire length, width, and thickness, respectively. The kinetic inductance has the origin in the inertia of Cooper pairs. In the dirty superconductor limit, it takes the form \cite{tinkham2004} $L_K = \mu_0\lambda_L^2 l/wt$, where $\lambda_L$ is the London penetration depth, which for thin films can reach values several times the bulk value. The kinetic inductance can also be expressed as a function of the normal state resistance of the wire $R_n$, $L_K = 0.14\hbar R_n/k_{\rm B}T_c$, with $T_c$ being the superconductor critical temperature. For a large, unbiased Josephson junction, the inductance is given by $L_J = \Phi_0/2\pi I_C$ \cite{orlando_book}, with $I_C$ being the junction critical current, and $\Phi_0=h/2e$ the flux quantum. Irrespective of the type of coupling inductor, the phase across it can be treated as a constant operator with off-diagonal matrix elements which are directly calculated in the qubit eigenbasis $\langle 0|\hat{\varphi}_L|1\rangle\simeq LI_p(\Phi_0/2\pi)$. Here $I_p \equiv\langle 0 |\hat{I}|1\rangle$ is the persistent current in the qubit loop. The interaction strength in this case is given by the magnetic dipolar energy $\mathcal{H}_{\rm int} = -\vec{m}\cdot\vec{B}$, which for a superconducting quantum circuit is rewritten as 
\be\label{eq:liqir}
\hat{\mathcal{H}}_{\rm int} = LI_pI_{\rm rms}\hat{\sigma}_x(\hat{a}+\hat{a}^{\dag}),
\ee 
leading to the definition of the coupling strength $g \equiv LI_pI_{\rm rms}/\hbar$. Here $L$ represents the sum of all linear inductance contributions shared between qubit and resonator, including galvanic and mutual geometric inductance. 

An important remark needs to be made at this point regarding flux qubits and their type of interactions to resonators. The qubit Hamiltonian in the persistent current basis is given by $\hat{\mathcal{H}}_{\rm{FQ}}/\hbar = -(\Delta/2)\sigma_x - (\epsilon/2)\sigma_z$, where $\Delta$ is the tunnel coupling between the persistent current states, and $\hbar\epsilon=2I_p(\Phi_{\rm ext}-\Phi_0/2)$ corresponds to the magnetic energy proportional to the external magnetic flux $\Phi_{\rm ext}$. The effective magnetic dipole interaction [Eq.~(\ref{eq:liqir})] written in the persistent current basis is given by $\hat{\mathcal{H}}_{\rm{int}} = \hbar g\sigma_z(a + a^{\dag})$. In the diagonal basis of the qubits, the interaction Hamiltonian is rotated in such a way that both transverse $\sim\sigma_x$ as well as longitudinal $\sim\sigma_z$ interactions exist
\begin{equation}
\label{eq:Hfq}
\hat{\mathcal{H}}_{\rm int} = \hbar g\left(\frac{\epsilon}{\omega_q}\sigma_z - \frac{\Delta}{\omega_q}\sigma_x\right)(a + a^{\dag}),
\end{equation}
where $\omega_q\equiv\sqrt{\Delta^2 + \epsilon^2}$ is the qubit transition frequency. However, as the flux qubit is normally operated in the neighborhood of the symmetry point $\Phi_{\rm ext}  = \Phi_0/2$ where $\epsilon = 0$, the longitudinal contribution is normally neglected. 

\begin{figure}[!hbt]
\includegraphics[width = 8.5cm]{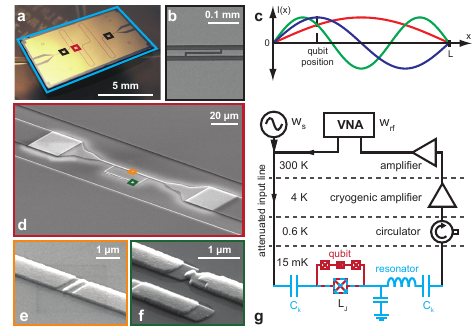}
\caption{\label{fig:NM1} (Color online) First experiment that reported breakdown of the rotating-wave approximation in a superconducting qubit circuit. (a) Optical image of the circuit, (b) scanning electron micrograph (SEM) from coupling capacitor, (c) resonator mode profiles coupling to the qubit, (d)--(f) SEM images showing qubit circuit and qubit junctions, (g) circuit schematic. From~\cite{niemczyk2010}.}
\end{figure}

\begin{figure*}[!hbt]
\includegraphics[width = 16cm]{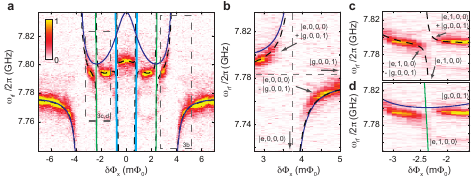}
\caption{\label{fig:NM3} (Color online) Observation of transitions which do not conserve the number of excitations in a flux qubit-resonator spectrum. Plots display transmission through the circuit, with $\omega_{\rm rf}$ being the probe frequency. $\delta\Phi_x$ corresponds to the flux applied to the qubit using an external coil. (a) Full circuit spectrum near the second resonator mode frequency. Dashed lines fitting the data correspond to the full Hamiltonian, the green vertical lines represent the case of no qubit-resonator coupling, while the solid magenta line is the prediction of the Jaynes-Cummings (JC) model; (b) zoom-in near the avoided qubit-resonator level crossing; (c) avoided-level crossings not included in the JC model. The presence of the counterrotating wave terms introduce hybridization between the indicated eigenstates that otherwise would not couple. From~\cite{niemczyk2010}.}
\end{figure*}

The first two experiments demonstrating USC in superconducting circuits used linear inductors as coupling elements. In the first experiment \cite{niemczyk2010}, a flux qubit was coupled to a transmission line resonator by means of the large inductance of a shared Josephson junction operated in the linear regime; see Figs.~\ref{fig:NM1}(a), \ref{fig:NM1}(b), and \ref{fig:NM1}(d)-\ref{fig:NM1}(f). Figure~\ref{fig:NM1}(c) shows the spatial profile of the lowest three resonator modes coupling to the qubit. The measurement setup is shown in Fig.~\ref{fig:NM1}(g), where a vector network analyzer (VNA) used to perform spectroscopy of the system is directly connected to the input capacitor of the resonator (shown in light blue), while the output capacitor couples to an amplifier chain before entering back into the second port of the VNA. A signal generator is combined with the VNA at the input line to perform two-tone spectroscopy and extract in this way the whole qubit spectrum. This experimental setup has become rather ubiquitous nowadays in circuit QED experiments. The spectrum of the system showed clear signatures of qubit-photon interactions in different modes of the resonator. The extracted qubit-resonator coupling rates to the first three resonator modes were $g_0/2\pi = 314~$MHz, $g_1/2\pi = 636~$MHz, and $g_2/2\pi = 568~$MHz, respectively. The maximum normalized coupling strength was achieved by the second mode, i.e., $g_1/\omega_1 = 0.12$. Deviations from the JC model were clearly observed with the appearance of avoided-level crossings corresponding to a breakdown of the conservation of the number of excitations. Because of the presence of the counterrotating terms, the states $|e,1,0,0\rangle$ and $|g,0,0,1\rangle$, which are degenerate under the RWA, hybridize and result in visible avoided-level crossings, as seen in Fig.~\ref{fig:NM3}. 

\begin{figure}[!hbt]
\includegraphics[width = 8cm]{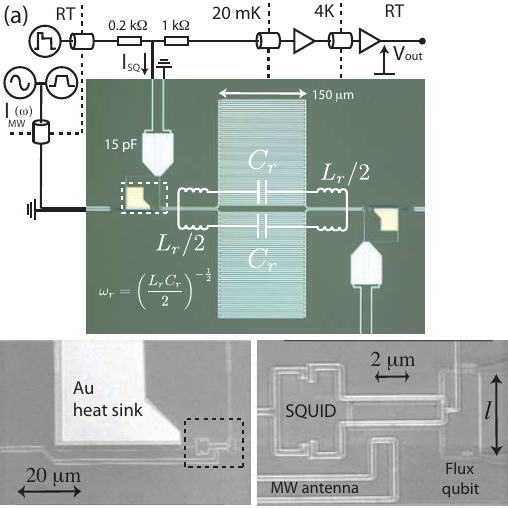}
\includegraphics{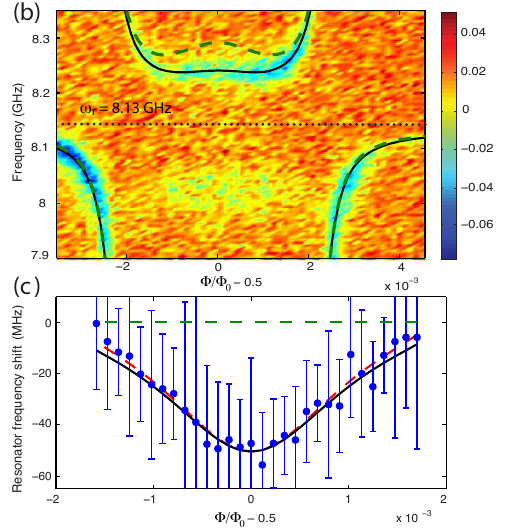}
\caption{\label{fig:BS} (Color online) Observation of physics beyond the rotating-wave approximation: the Bloch-Siegert shift. (a) Circuit schematic and scanning electron micrograph images. (b) Spectrum near resonator frequency $\omega_r/2\pi = 8.13~$GHz as a function of the magnetic flux in the qubit. The acquired signal represents the magnetic flux sensed by the SQUID coupled to the qubit. (c) Resonator frequency shift with respect to the prediction of the Jaynes-Cummings model, identified here as the Bloch-Siegert shift. The horizontal dashed line is the prediction from the Jaynes-Cummings model, the solid line is the full Hamiltonian without approximations, and the dashed line fitting the data is the approximated Hamiltonian in the perturbative USC regime. From~\cite{forn-diaz2010}.}
\end{figure}
In the second experiment \cite{forn-diaz2010}, a flux qubit was galvanically attached to a lumped-element $LC$ resonator, such that both systems were coupled by the inductance of the shared wire; see Fig.~\ref{fig:BS}(a). The qubit spectrum showed a large avoided-level crossing at the resonance point, yielding a coupling strength of $g/2\pi = 810$~MHz for a resonator frequency of $\omega_r/2\pi = 8.13$~GHz. This resulted in a normalized coupling of $g/\omega_r = 0.1$. Deviations from the RWA were identified as a frequency shift in the resonator when the qubit was flux biased near its symmetry point $\Phi=\Phi_0/2$; see Figs.~\ref{fig:BS}(b), and \ref{fig:BS}(c). At this bias point, the effective qubit-resonator coupling is maximal. The frequency shift of the resonator compared to the JC model, also known as the Bloch-Siegert shift \cite{bloch1940}, was attributed to the dispersive effect of the counterrotating terms, as explained in Sec.~\ref{sec:2}. Its existence had long been predicted \cite{Cohen1973, zakrzewski1991} and this experiment represented its first observation. The maximum Bloch-Siegert shift attained in this experiment was $\omega_{\rm{BS}}\equiv g^2/(\omega_r+\omega_q) = 2\pi\times52~$MHz.

The two experiments previously described above were performed in the perturbative USC regime, defined when the normalized coupling constant is $0.1 \lesssim g/\omega \lesssim 0.3$ \cite{Rossatto2017}. The experiments achieved \cite{niemczyk2010} $g/\omega = 0.12$ and \cite{forn-diaz2010} $g/\omega = 0.10$, respectively, satisfying the condition of perturbative USC.

In later experiments, a two-resonator circuit was coupled to a single flux qubit by sharing a section of the qubit loop, several $\mu\rm{m}$ long \cite{baust2016}. The coupling strength observed was of $g/\omega_r = 0.17$, attained using a collective mode between the two resonators. 

Follow-up work on the Bloch-Siegert shift observation experiment studied the energy-level transitions between excited states as a function of coupling strength \cite{forn-diaz2016}. In the RWA regime, the excited states of the JC model appear in doublets $|n, \pm\rangle$ for each photon number $n$. In circuit QED, the qubit is sometimes driven via the resonator. With this indirect driving, a selection rule exists under the RWA between eigenstates of different manifolds $|n,\pm\rangle$ and $|n\pm1,\pm\rangle$. The observation of a transition between dressed states $|1,-\rangle$ and $|2,+\rangle$ belonging to different manifolds was identified in this work as another distinct feature of the USC regime. 

In another experiment in the perturbative USC regime, \cite{chen2017} explored multiphoton red sidebands in an experiment consisting of a flux qubit coupled to a transmission line resonator. These higher-order sidebands could be unambiguously detected only in the USC regime, where the counterrotating terms modify the selection rules. The largest coupling in this experiment was attained between the flux qubit and the fundamental mode of the resonator, reaching a value of $g/\omega_0 = 0.097$. 
\paragraph{Embedded qubit circuit.-}
Up to this point, the description of galvanic couplings as a perturbation of the qubit-resonator system has been valid in the range $0 < g/\omega \lesssim 0.1$. Increasing the coupling strength toward the nonperturbative regime would be analogous to considering the phase drop of the inductive element $\hat{\varphi}_L$ as a degree of freedom shared between the qubit and resonator with dynamics of its own. While in principle it should be possible to increase the shared inductance and enter the nonperturbative USC regime \cite{Rossatto2017}, in practice this would result in a very large qubit geometry, hence susceptible to flux noise, and a decrease of the persistent current in the qubit loop that would eventually decrease the coupling strength. 

The natural way to further enhance the interaction strength is to share a junction of the qubit circuit with the resonator; see Fig.~\ref{fig:galvanic}(b). In other words, the qubit needs to be embedded ``in parallel" to the resonator. This circuit will require full quantization in order to be properly described. In that case, the interaction term becomes of a dipoletype \cite{peropadre2013} 
\be
\hat{\mathcal{H}}_{\rm{int}} = \sum_{\alpha=x,y,z}\hbar g_G^{\alpha}(\hat{a}^{\dag}+\hat{a})\hat{\sigma}_{\alpha}.
\ee
The coupling operators are here defined as 
\begin{align}\label{eq:gx}
\hbar g_G^x &= \sqrt{\frac{\hbar\omega_r}{2L_r}}\frac{\Phi_0}{2\pi}\langle 0|\hat{\varphi}|1\rangle,\\
\label{eq:gz}
\hbar g_G^z &= \sqrt{\frac{\hbar\omega_r}{2L_r}}\frac{1}{2}\left(\frac{\Phi_0}{2\pi}\right)\left(\langle 1|\hat{\varphi}|1\rangle - \langle0|\hat{\varphi}|0\rangle\right).
\end{align}
The prefactor $\sqrt{\hbar\omega_r/2L_r}$ corresponds to the rms~of the resonator current in its ground state, Eq.~(\ref{eq:irms}). The last factors in Eqs.~(\ref{eq:gx}) and (\ref{eq:gz}) correspond to the magnetic dipole moment and the net magnetic flux generated by the qubit, respectively. Near the qubit symmetry point, where the qubit is usually operated to maximize quantum coherence, the net flux generated is null. Therefore, we may neglect the coupling term $g_G^z$. Equation~(\ref{eq:gx}) includes the case of a shared linear inductor, since in that case we can write the dipole moment as $(\Phi_0/2\pi)\langle 0|\hat{\varphi}|1\rangle\simeq LI_p$ so that the coupling becomes the mutual inductive energy $LI_pI_{\rm{rms}}$, as in Eq.~(\ref{eq:liqir}). Equation~(\ref{eq:gx}) can be recast as a function of the resonator impedance $Z$:
\be\label{eq:gw}
\frac{g_G^x}{\omega_r} = \frac{1}{8}\sqrt{\frac{Z_{\rm{vac}}}{\pi Z}}\alpha^{-1/2}\langle 0|\hat{\varphi}|1\rangle.
\ee
Notice the different scaling compared to Eqs.~(\ref{eq:gc}) and (\ref{eq:g_c_cpb}). In Eq.~(\ref{eq:gw}), the fine structure constant appears with a negative power, which is a consequence of coupling the flux qubit to the fluctuations of the magnetic field generated by the resonator (in fact, here the coupling is directly to the current in the resonator). Comparing to Rydberg atoms, atomic magnetic dipole couplings are typically an order of magnitude smaller than electric dipole couplings and are therefore usually not considered. 

\cite{Manucharyan2017} showed that for a fluxonium qubit $g_G^x/\omega_r$ yields an identical result. Using a linear inductance as a coupler, the matrix element of the phase operator is of the order of $\langle 0|\hat{\varphi}|1\rangle\approx10^{-2}$ \cite{forn-diaz2010, chen2017, baust2016} so that Eq.~(\ref{eq:gw}) leads to $g_G^x/\omega_r\approx0.1$, just entering the perturbative USC regime. Maximizing Eq.~(\ref{eq:gx}) may be accomplished by sharing a qubit junction, as shown in Fig.~\ref{fig:galvanic}(b). In that case, $\langle1|\hat{\varphi}|0\rangle\approx1$, so $g_G^x/\omega_r\simeq2$, which lies well in the DSC regime. Increasing the coupling further is possible by using low-impedance resonators. 

Following the initial experiments in the perturbative USC regime, a new wave of results was reported when two experiments demonstrated DSC regimes between both a flux qubit and a resonator \cite{yoshihara2017a} and a transmission line in an open-space setting~\cite{forn-diaz2017}. In both experiments, the qubit was embedded in the resonator and transmission line circuit, with the coupling element being a Josephson junction of the qubit loop. Contrary to the first experiment reporting USC \cite{niemczyk2010}, the coupling junction was part of the qubit internal dynamics, therefore corresponding to an \emph{internal coupling} as defined in Sec.~\ref{sec:3.1.b}. The effective  inductance stored in the junction enabled coupling strengths all the way into the DSC regime.
\begin{figure}[!hbt]
\includegraphics[width = 8cm]{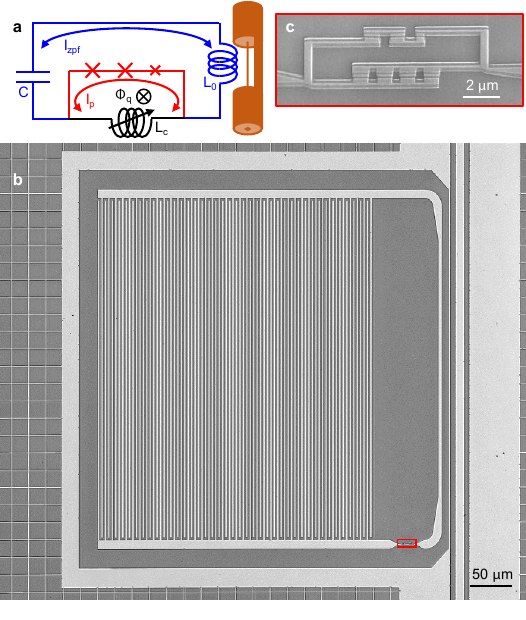}
\caption{\label{fig:NTTcirc} (Color online) DSC regime circuitry of a superconducting flux qubit coupled to an $LC$ resonator. (a) Circuit schematic. (b) Scanning electron micrograph of the device. The large interdigitated-finger capacitor occupies most of the image. The probing transmission line can be seen to the right of the image. (c) Zoom-in of the qubit, with the 4-junction SQUID coupler in the bottom arm. From~\cite{yoshihara2017a}.}
\end{figure}

\begin{figure*}[!hbt]
\includegraphics{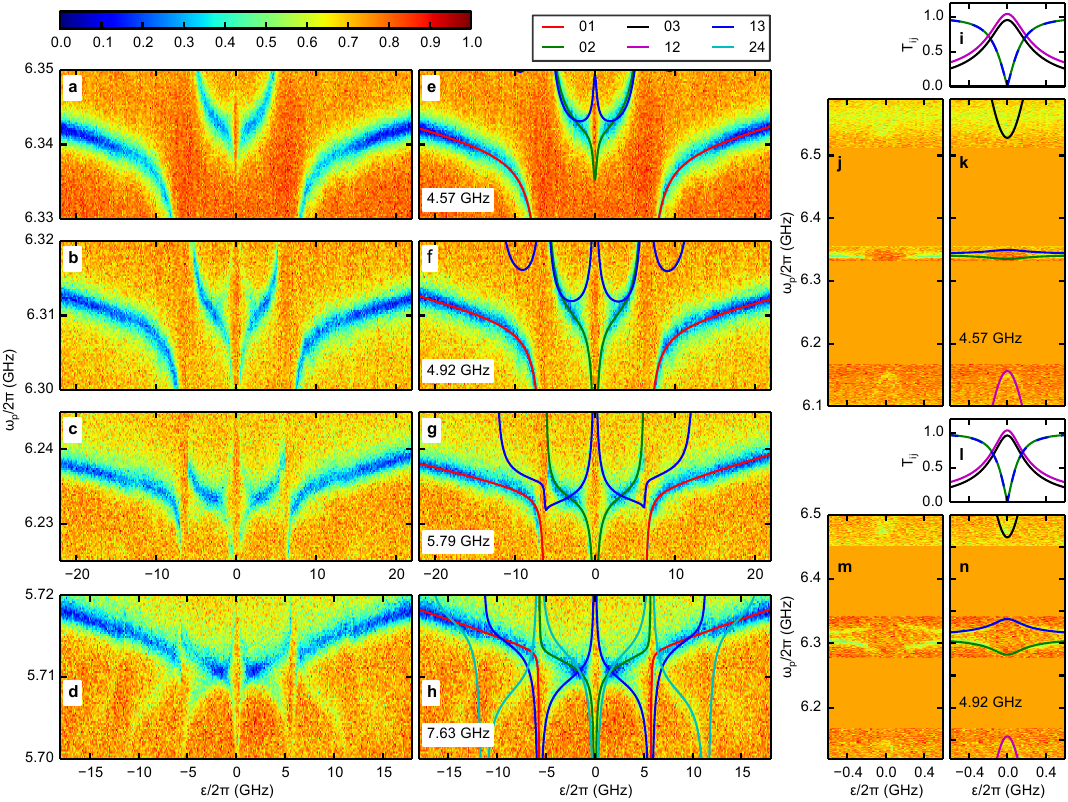}
\caption{\label{fig:NTT} (Color online) DSC regime spectrum at different coupling strengths. (a)-(d) The spectrum near the bare resonator frequency. The signal represents transmission through the resonator. (e)-(h) The same spectra with fitted theory calculations using the full quantum Rabi model, finding an excellent agreement with the experiments. (j)-(n) Broader frequency ranges corresponding to the same coupling strengths in (a)-(d), where additional transitions are identified which confirm the large size of the coupling strength of the system. Certain transitions vanish due to the symmetry of the system Hamiltonian. The inset shows several transition matrix elements coinciding with resonances in the experiment. From~\cite{yoshihara2017a}.}
\end{figure*}

The qubit-resonator experiment consisted of an $LC$ circuit galvanically coupled to a flux qubit by sharing an array of four Josephson junctions in parallel, acting as an effective SQUID, which allowed tuning of the interaction strength \cite{peropadre2010}; see Fig.~\ref{fig:NTTcirc}. The resonator was inductively coupled to a transmission line to allow probing the system in transmission. In order to enhance the coupling strength, a very large resonator capacitor was used to decrease its impedance $Z=\sqrt{L/C}$ and enhance in this way the ground-state current fluctuations $\langle I_{\rm{rms}}^2\rangle^{1/2}=\omega_r\sqrt{\hbar/2Z}$, as explained in Sec.~\ref{sec:3.1.b}. The spectrum of the system showed energy-level transitions that agreed with the full QRM; see Fig.~\ref{fig:NTT}. The coupling strengths reported spanned the region $0.72\leq g/\omega_r \leq 1.34$, with coupling strength values up to $g/2\pi = 7.63~$GHz. These remarkable results exceeded all previous reports of ultrastrong couplings and entered the DSC regime $g/\omega>1$, where the interaction operator starts to dominate the system spectrum and its dynamics \cite{casanova2010}. Given the coupling strength achieved, the system ground state should exhibit a large degree of qubit-resonator entanglement. The results from Yoshihara \emph{et al.} represented the largest normalized atom-photon interaction strength reported in any physical system to date. Within the same work, they found a way to quantify the effect of the so-called $A^2$ term in their particular system. As alluded to in Sec.~\ref{sec:2}, a debate exists whether in circuit QED the $A^2$ term precludes the existence of a superradiant phase transition in the system ground state. Based on the parameters extracted, they were able to demonstrate that the $A^2$ term in their setup did not satisfy the condition of the no-go theorem which led them to claim that a superradiant state may exist.

In follow-up experiments, Yoshihara \emph{et al.} demonstrated insight into the energy spectrum of the QRM to more accurately characterize the relative coupling strength $g/\omega_r$ of the system. By looking at higher-energy level transitions, a method was developed to qualitatively estimate the regime of coupling $g/\omega_r$ in which the system lies without the need for complex fits of the whole spectrum \cite{yoshihara2017b}. Using two-tone spectroscopy, they were able to map out the QRM spectrum up to six levels, finding excellent agreement with Eq.~(\ref{QRH}) \cite{Yoshihara2017c} and demonstrating in this way the validity of circuit QED implementations to faithfully represent the QRM \cite{Manucharyan2017}. The observations were consistent with remarkable Lamb shifts of up to 90\% of the bare qubit energy splitting, together with 1-photon and 2-photon Stark shifts of higher-energy levels, which resulted in the inversion of the qubit states as the interaction strength grows well into the DSC regime, which they were able to demonstrate using devices tunable over a wide range \cite{Yoshihara2017c}. This important work from the National Institute of Information and Communications Technology (Tokyo, Japan) group represents the first steps into the observation of novel DSC physics in upcoming circuit QED experiments.

As discussed in Sec.~II, the following natural step would be to start exploring the dynamics of the QRM in the nonperturbative regime, the coherence time of the system~\cite{nataf2011}, its internal dynamics \cite{casanova2010}, and possibly phase transitions with multiple qubits involved \cite{nataf2010, jaako2016}.

We turn now to galvanic couplings using charge qubits embedded in the resonator circuit. In such a configuration, the qubit couples directly to the charge operator of the resonator. Recently \cite{Manucharyan2017}, a circuit consisting of a charge qubit embedded in an $LC$ resonator circuit [Fig.~\ref{fig:galvanic}(c)] was inspected, and the following normalized coupling strength was obtained:
\be
\frac{g_G^{\rm{ch}}}{\omega_r'} = \frac{C_r}{C_q+C_r}\frac{\langle0|\hat{Q}|1\rangle}{e}\sqrt{2\pi\frac{Z_r'}{Z_{\rm{vac}}}}\alpha^{1/2}.
\ee
Here, the resonator frequency is renormalized due to the qubit capacitor $C_q$, $\omega_r'=1/\sqrt{L_rC_p}$, with $C_p^{-1}= C_r^{-1} + C_q^{-1}$. The resonator impedance is also renormalized as $Z_r'=\sqrt{L_r/C_p}$. $\langle1|\hat{Q}|0\rangle$ is the qubit electric dipole in units of the electron charge. For a Cooper pair box, $\langle1|\hat{Q}|0\rangle\sim1$. With sufficiently large resonator capacitance, it is possible to reach the DSC regime $g_G^{\rm{ch}}/\omega_r'>1$ by employing very high-impedance resonators \cite{masluk2012}. 

A different circuit configuration was analyzed by Bourassa \emph{et al.} \cite{bourassa2012}. The circuit consisted of galvanically attaching a charge qubit to a transmission line resonator. For charge qubits in the transmon regime $E_J/E_C\gg1$, the coupling to such a resonator was calculated to be
\be\label{eq:g_tr_g}
\frac{g_G^{\rm{tr}}}{\omega_r} = \frac{1}{\sqrt{8\pi}}\left(\frac{E_C}{8(E_J+E_L)}\right)^{1/4}\sqrt{\frac{Z_{\rm{vac}}}{Z}}\alpha^{-1/2}.
\ee
In Eq.~(\ref{eq:g_tr_g}), $E_L = (\Phi_0/2\pi)^2/L_r$ corresponds to the inductive energy of the resonator which dilutes the anharmonicity of the transmon qubit and reduces the effective maximum coupling. This inductive term was omitted in the first analysis of this circuit \cite{devoret2007}. Given that the inductive energy of resonators is usually much larger than the Josephson energy, achieving the DSC regime $g_G^{\rm{tr}}/\omega_r>1$ compromises the transmon condition $E_J\gg E_C$ that is required to derive Eq.~(\ref{eq:g_tr_g}). In addition, the presence of the qubit junction was shown to reduce the resonator current, leading to a maximum coupling of $g_G^{\rm{tr}}/\omega_r\sim0.2$ \cite{bourassa2012}, which is far from the DSC regime.

It is worth at this point referring to the analysis carried out by \cite{Manucharyan2017}. They considered the full quantum circuit of both a fluxonium and a CPB qubit and compared them to the QRM. It turns out that both fluxlike and chargelike qubits display a spectrum that resembles very closely with that of the QRM. In particular, the two lowest-energy levels become nearly degenerate in the DSC regime $g/\omega_r>1$. Although a large number of bare qubit states are involved in the qubit-resonator ground state, the entanglement spectrum is dominated by the lowest two eigenvalues even though the qubits are multilevel systems. The analysis for fluxlike qubits using many of the circuit levels shows similar features to the QRM even though the calculated low energy-level splittings differ quantitatively. By contrast, the CPB ultrastrongly coupled to a resonator results in a much more faithful reproduction of the energy-level spectrum of the QRM. Manucharyan \emph{et al.} interpreted the vacuum level degeneracy as an environmental suppression of flux and charge tunneling due to dressing of the qubit with low-- or high--impedance photons in the resonator. In fluxlike qubits, the flux tunneling suppression was understood as the qubit circuit being shunted by the large resonator capacitor, which increases the effective qubit mass and suppresses quantum tunneling. In other words, the system localizes itself in one of the two minima of the qubit potential, suppressing in this way the qubit transition frequency. The CPB ultrastrongly coupled to a resonator has a less obvious circuit model interpretation since no simple circuit elements represent the system at high coupling values. The charge tunneling suppression was related to the manifestation of the dynamical Coulomb effect of transport in tunnel junctions connected to resistive leads. In conclusion, Manucharyan \emph{et al.} found the description of the QRM by superconducting qubits to be quite faithful, despite the presence of the multilevel spectrum. The CPB is the most suitable qubit despite the fact that charge noise has so far hindered the exploration of ultrastrong couplings, even though the USC features may be robust against dissipation \cite{DeLiberato2017}.

\subsection{Semiconductor quantum wells}
\label{sec:3B}

Semiconductor quantum wells (QWs) provide one of the cleanest and most tunable solid-state environments with quantum-engineered electronic and optical properties. In the context of cavity QED, microcavity exciton polaritons in QWs have served as a model system for highlighting and understanding the striking differences between light-atom coupling and light-condensed-matter coupling~\cite{WeisbuchetAl92PRL,KhitrovaetAl99RMP,DengetAl10RMP,GibbsetAl11NP}.  However, the large values of resonance frequency (typically in the near-infrared or visible) and relatively small dipole moments for interband transitions make it impractical to achieve USC using exciton polaritons [see, however, the cases of microcavity exciton polaritons (MEPs) in organic semiconductors, carbon nanotubes, and two-dimensional materials described in Sec.~\ref{sec:3c2}].

\begin{figure}
\begin{center}
\includegraphics[scale=0.62]{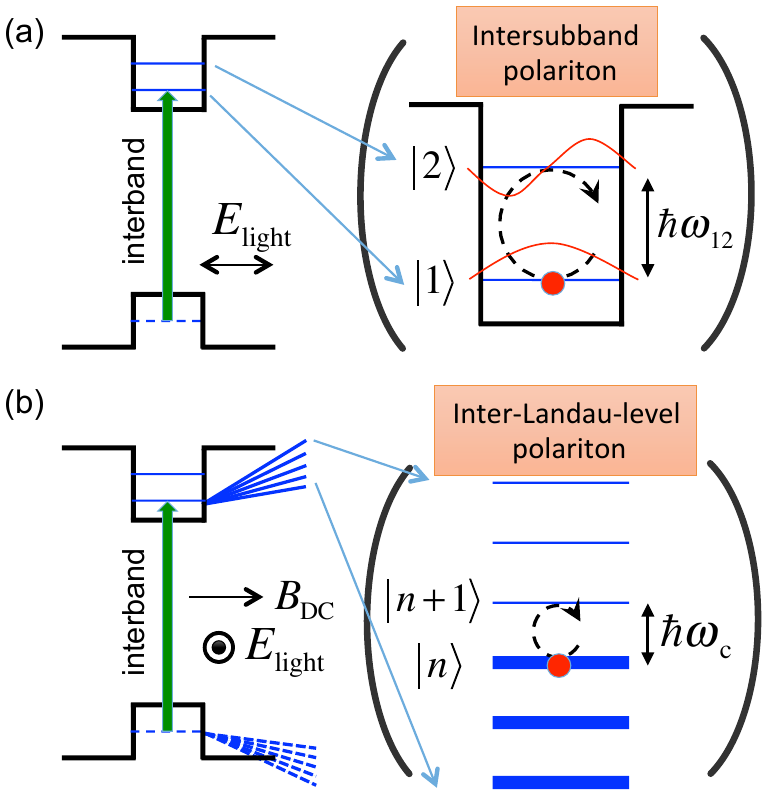}
\caption{\small (Color online) Semiconductor quantum well transitions. Two types of \emph{intraband} transitions in semiconductor quantum wells are shown that have been demonstrated to exhibit USC: (a)~intersubband polaritons and (b)~inter-Landau-level (or cyclotron) polaritons. In contrast to interband transitions, which typically occur in the near-infrared or visible range, these intraband transitions occur in the midinfrared or THz range, with enormous dipole moments. (a) The lowest two subbands of opposite parity, with an energy separation of $\hbar\omega_{12}$, within the conduction or valence band are resonantly coupled with a light field ($E_\mathrm{light}$) polarized in the growth direction (TM polarization), to form \emph{intersubband polaritons}. (b) A magnetic field ($B_\mathrm{dc}$) applied in the growth direction quantizes each subband into Landau levels with an enery separation of $\hbar\omega_\mathrm{c}$, where $\omega_\mathrm{c} = eB_\mathrm{dc}/m^*$ is the cyclotron frequency, $e$ is the electronic charge, and $m^*$ is the effective mass; the highest occupied Landau level and the lowest unoccupied Landau level are resonantly coupled with a light field ($E_\mathrm{light}$) polarized in the quantum well plane (TE-polarization) to form \emph{inter-Landau-level polaritons}.
}
\label{intraband}
\end{center}
\end{figure}

\emph{Intraband transitions}, such as intersubband transitions (ISBTs)~\cite{Helm00Chapter,Paielle06Book} or inter-Landau-level transitions (ILLTs) (colloquially known as a cyclotron resonance, CR)~\cite{Kono01MMR,HiltonetAl12Book}, are much better candidates for realizing USC regimes in QWs.  Shown schematically in Fig.~\ref{intraband}, they have small resonance frequencies, typically in the midinfared (MIR) and terahertz (THz) range, and enormous dipole moments (tens of $e$\AA).  

\begin{figure}
\begin{center}
\includegraphics[scale=1.39]{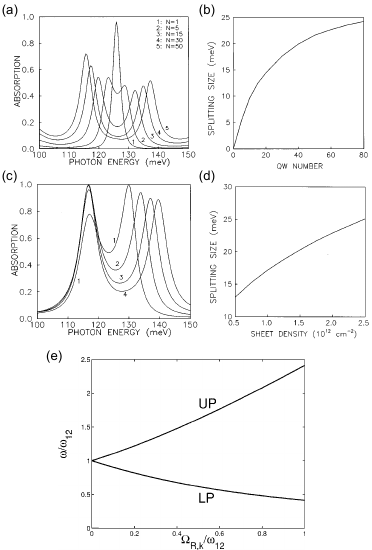}
\caption{\small Theoretically predicted intersubband polaritons. (a)~Absorption spectra showing intersubband polaritons for different numbers of QWs (1--50). (b)~QW number dependence of the vacuum Rabi splitting. (c)~Absorption spectra for intersubband polaritons for different electron densities: 0.5 $\times$ 10$^{12}$\,cm$^2$ (curve 1), 1.0 $\times$ 10$^{12}$\,cm$^2$ (curve 2), 1.5 $\times$ 10$^{12}$\,cm$^2$ (curve 3), and 2.0 $\times$ 10$^{12}$\,cm$^2$ (curve 4). (d)~Electron density dependence of the vacuum Rabi splitting. (e)~Calculated upper polariton (UP) and lower polariton (LP) frequencies as a function of coupling strength, where $\omega_{12}$ is the transition frequency. (a)--(d) Adapted from \cite{Liu97PRB}. (e) Adapted from \cite{CiutietAl05PRB}.
}
\label{LiuCiuti}
\end{center}
\end{figure}

Theoretically, \cite{Liu96JAP,Liu97PRB} was the first to propose and analyze \emph{intersubband (ISB) polaritons} in QWs. He demonstrated that the vacuum Rabi splitting (VRS) increases with the electron density as well as the number of QWs. Figure~\ref{LiuCiuti}(a) shows calculated absorption spectra, displaying ISB polaritons for QWs for different numbers of QWs, while in Fig.~\ref{LiuCiuti}(b) the QW number dependence of the vacuum Rabi splitting is calculated; Fig.~\ref{LiuCiuti}(c) shows absorption spectra for different electron densities, while in Fig.~\ref{LiuCiuti}(d) the electron density dependence of the vacuum Rabi splitting is displayed \cite{Liu96JAP,Liu97PRB}.
Unique electrically driven MIR emission devices based on quantum cascade structures incorporating ISB polaritons have also been proposed \cite{ColombellietAl05SST}. In particular, it was predicted that in InP-based multiple-QW structures a polariton splitting $2\hbar g$ of 40~meV can be obtained for an ISBT at $\hbar\omega_{12} \approx$ 130~meV, resonant with a cavity frequency $\omega$, i.e., $g/\omega\approx 0.15$.
\cite{CiutietAl05PRB} used a Bogoliubov transformation to diagonalize the full Hamiltonian and obtained the energies of the upper polariton (UP) and lower polariton (LP) branches. Figure~\ref{LiuCiuti}(e) shows the calculated UP and LP energies as a function of normalized coupling strength, where $\omega_{12}$ is the ISBT frequency, for zero detuning $\omega = \omega_{12}$, demonstrating that USC is possible.
Similarly, for (ILL) polaritons, \cite{HagenmulleretAl10PRB} derived and diagonalized an effective Hamiltonian describing the resonant excitation of a two-dimensional electron gas (2DEG) by cavity photons in the integer quantum Hall regime. The dimensionless vacuum Rabi frequency in a 2DEG resonant with a cavity of frequency $\omega$, $g/\omega_\mathrm{c}$, was shown to scale as $\sqrt{\alpha N_\mathrm{QW}\nu}$. Here $\omega_\mathrm{c} = eB_\mathrm{dc}/m^*$ is the cyclotron frequency, $B_\mathrm{dc}$ is the dc magnetic field applied perpendicular to the 2DEG, $e$ is the electronic charge, $m^*$ is the effective mass, $\alpha$ is the fine structure constant, $N_\mathrm{QW}$ is the number of QWs, and $\nu$ is the Landau-level filling factor in each well. It was shown that $g/\omega_\mathrm{c} > 1$ could be achieved when $\nu \gg 1$ with realistic parameters of a high-mobility 2DEG.

Furthermore, as mentioned in Sec.~\ref{sec:2}, \cite{CiutietAl05PRB} provided much physical insight into the ground-state properties of ISB polaritons.  They found that the ground state consists of a \emph{two-mode squeezed vacuum}. 

Various experimental schemes have been proposed to experimentally probe these special properties of the ground state of ISB polaritons in the USC regime. \cite{CiutietAl05PRB} specifically considered a system in which a cavity photon mode was strongly coupled to an ISBT. They showed that the system could be brought into the USC regime, where correlated photon pairs can be generated, by tuning the quantum properties of the ground state. The tuning could be achieved by changing the Rabi frequency via an electrostatic gate.

Similarly, \cite{LiberatoetAl07PRL} proposed to modulate the vacuum Rabi frequency in time and calculated the spectra expected for the emitted radiation.  More recently,~\cite{StassietAl13PRL} described a three-level system ($|0\rangle$, $|1\rangle$, $|2\rangle$) in which a spontaneous $|1\rangle \rightarrow |0\rangle$ transition was accompanied by the creation of real cavity photons out of virtual photons resonant with the $|1\rangle \rightarrow |2\rangle$ transition. Finally, \cite{Hagenmuller16PRB} has recently proposed an all-optical scheme for observing the dynamical Casimir effect in a THz photonic band gap using ILL polaritons.

These theoretical studies have stimulated much interest in experimentally probing ultrastrong light-matter coupling phenomena in semiconductor QWs.

The design and nature of photonic cavities used in the context of semiconductor USC physics depend on, with respect to the QW plane, whether the in-plane or out-of-plane electric field component needs to be enhanced to couple with the electronic excitations. ISBTs and ILLTs couple with the out-of-plane and in-plane cavity electric field component. Examples of typical cavities and their working principles are described next.

\smallskip

{\emph{Intersubband polariton cavities}:
\begin{itemize}	
\item[i)] A planar waveguide microcavity [Fig.~\ref{fig:cavities}(a)] consists of, from bottom to top, an undoped GaAs layer, an AlAs+$n$-doped GaAs cladding layer, a QW layer, and a metal layer. Light is obliquely incident onto the side of the waveguide, and is confined through multiple reflections between the top metal layer and the AlAs cladding layer. The photonic resonance leads to enhancement of the out-of-plane electric field component around the metal layer [$E_z$ plotted with blue lines in Fig.~\ref{fig:cavities}(a)]. The metal layer also serves as an electrical gate to tune the electron density in the QW.
\item[ii)] A metal-dielectric-metal microcavity is shown in the left panel in Fig.~\ref{fig:cavities}(b). It contains a QW sandwiched between a planar metallic mirror and a metallic rectangular strip grating. The grating defines a lateral photonic confinement while at the same time ensures efficient coupling of incident light into the double-metal regions. Both obliquely incident ($\theta \ne 0$) and normal incident ($\theta = 0$) light are able to excite the ISBT in the QW due to enhancement of $E_z$ [right panel of Fig.~\ref{fig:cavities}(b)] in the cavities.
\item[iii)] An inductor-capacitor resonator substitutes the top metallic strip gating in a metal-dielectric-metal cavity with a microstructure where a wire with finite inductance connects two circular capacitor elements. The electric and magnetic field distributions at resonance are plotted in Fig.~\ref{fig:cavities}(c).
\item[iv)] A surface plasmon photonic crystal replaces the bottom planar metallic mirror of a metal-dielectric-metal cavity with a cladding semiconductor layer [left panel of Fig.~\ref{fig:cavities}(d)]. The device can be considered as a 1D metallic photonic crystal, which folds the modes guided by the cladding layer and the QW into the first Brillouin zone. The full dispersion can be mapped out by recording light transmittance at various incident angles [right panel of Fig.~\ref{fig:cavities}(d)].
\end{itemize}
 
{\emph{Landau polariton cavities}:
\begin{itemize}
\item[i)] Depending on the applied magnetic field strength and electron effective mass, ILLTs of typical semiconductor QWs occur in the microwave or terahertz frequency range. Resonators that are standard in the microwave technology, such as coplanar microresonators [Fig.~\ref{fig:cavities}(e)], and metallic patch resonators [Fig.~\ref{fig:cavities}(f)] can be easily integrated with QWs to study the microwave dynamics of Landau polaritons.
\item[ii)] Metamaterial cavities are an array of metallic resonance microstructures, typically split-ring resonators (SRRs), that are patterned and evaporated on top of the semiconductor capping layer of the QW [left panel of Fig.~\ref{fig:cavities}(g)]. The resonance frequencies and quality factors can be adjusted by properly designing the structure within a unit cell. In-plane electric fields [right panel of Fig.~\ref{fig:cavities}(g)] are enhanced around the gaps of the SRRs.
\item[iii)] A photonic-crystal cavity [left panel of Fig.~\ref{fig:cavities}(h)] consists of  a QW that is sandwiched by silicon Bragg mirrors; each Bragg mirror consists of several silicon wafers aligned parallel and at controllable distances from each other. The in-plane electric field at cavity resonance reaches maximum at the position of the QW to ensure maximum coupling strength. 
\end{itemize}
\begin{figure*}
\begin{center}
\includegraphics[scale = 0.3]{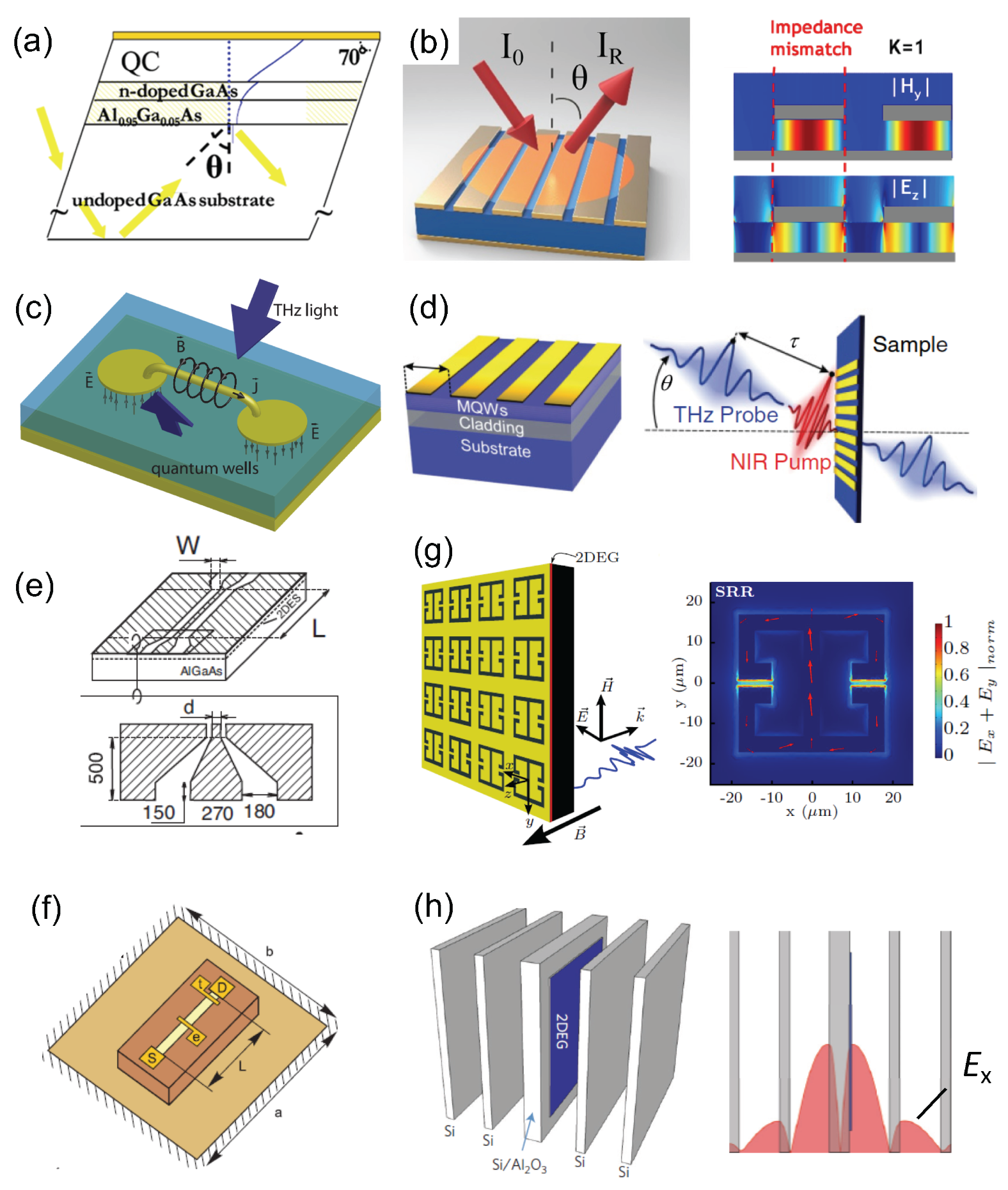}
\caption{\small (Color online) Assorted cavities used in semiconductor-based light-matter ultrastrong coupling experiments. (a) A planar waveguide cavity. Form~\cite{SapienzaetAl08PRL}. (b) A metal-dielectric-metal cavity. From~\cite{LaurentetAL17APL}. (c) An inductor-capacitor resonator. From~\cite{GeiseretAl12PRL}. (d) A surface plasmon photonic crystal. From~\cite{PoreretAl12PRB}. (e) A coplanar microresonator. From~\cite{MuravevetAl11PRB}. (f) A metallic patch resonator. From~\cite{MuravevetAl13PRB}. (g) A metamaterial cavity. From~\cite{MaissenetAl14PRB}. (h) A photonic-crystal cavity. From~\cite{ZhangetAl16NP}.
} 
\label{fig:cavities}
\end{center}
\end{figure*}

\subsubsection{Intersubband transitions}
\label{sec:3B1}

\begin{figure}[b]
\begin{center}
\includegraphics[scale=0.91]{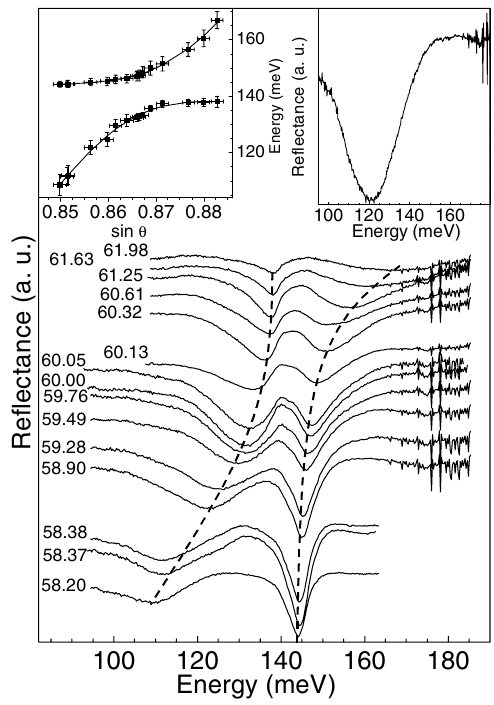}
\caption{\small First experimental observation of intersubband polaritons. Reflectance spectra are shown for a GaAs quantum well sample at 10~K for different angles of incidence for TM-polarized light. The spectra are offset from each other for clarity. Top-left inset: the dip position versus the angle shows a level anticrossing. Top-right inset: a spectrum recorded for TE-polarized light, showing only a dip due to the cavity mode. From~\cite{DinietAl03PRL}.
}
\label{Dini}
\end{center}
\end{figure}

Experimentally, the first observation of polariton splitting of an ISBT was reported by Dini \textit{et al}.\ in 2003~\cite{DinietAl03PRL}. The dispersion of the ISB polaritons in GaAs QWs was measured through angle-dependent reflectance measurements using a prismlike geometry, as shown in Fig.~\ref{fig:cavities}(a). Figure~\ref{Dini} shows measured reflectance spectra at 10\,K for TM-polarized waves for different incidence angles. Two dips are clearly displayed, exhibiting anticrossing behavior with a splitting ($2\hbar g$) of 14~meV as a function of incident angle. With an ISBT resonance energy of $\hbar\omega_{12}=142~{\rm meV}$, $g/\omega_{12} \sim 0.05$ at zero detuning $\omega = \omega_{12}$ was achieved even in this early work.  As a comparison, in the top-right inset of Fig.~\ref{Dini}, a TE reflectance spectrum is shown; only a single dip corresponding to the cavity mode is observed, as the ISBT is dipole forbidden for this polarization. In the top-left inset, the energies of the UP and LP dips are plotted as a function of the incidence angle, highlighting the anticrossing behavior.

\begin{table*}
  \centering
  \begin{tabular}{|c|c|c|c|c|c|c|c|c|c|c|c|}
    \hline 
     & Transition & Cavity & $d_\mathrm{QW}$ & $N_\mathrm{QW}$ & $\hbar\gamma$ & $\hbar\kappa$ & $\hbar g$ & $\hbar\omega$ & $g/\omega$ & $U$ & \\
   Reference & type & type & (nm) &  & (meV) & (meV) & (meV) & (meV) & (\%) & &Notes\\
    \hline
    \hline
	\cite{DinietAl03PRL} & ISBT & PWM & 7.2 & 18 & 5 & 15 & 7 & 142 & 5 & 0.62 &\\
	\hline
	\cite{DupontetAl03PRB} & ISBT & PWM & 6.0 & 140 & 2.2 & 11 & 6 & 115 & 5 & 0.54 & Bound to quasibound\\
	\hline
	\cite{AnapparaetAl05APL} & ISBT & PWM & 7.5 & 10 & $\cdots$ & $\cdots$ & 7 & 135 & 5 & $\cdots$ & Electrical control\\
	\hline
	\cite{AnapparaetAl06APL} & ISBT & PWM & 7.2+14 & 9 & $\cdots$ & $\cdots$ & 10.5 & 150 & 7 & $\cdots$ & Coupled double QWs\\
	\hline
	\cite{AnapparaetAl07SSC} & ISBT & PWM & 13.7 & 10 & $\cdots$ & $\cdots$ & 16.5 & 123 & 14 & $\cdots$ & InAs/AlSb QWs\\
	\hline
        \cite{DupontetAl07PRB} & ISBT & PWM & 7.5 & 160 & 6.9 & 12 & 21 & 123 & 17 & 1.9 & 
        \\
	\hline
	\cite{SapienzaetAl07APL} & ISBT & PWM & QC & 30 & $\sim$10 & $\cdots$ & 8 & 163 & 5 & $\cdots$ & QC photovoltaic\\
	\hline
	\cite{SapienzaetAl08PRL} & ISBT & PWM & QC & 30 & 8 & 15 & 11 & 150 & 7 & 0.54 & QC LED\\
	\hline
	\cite{TodorovetAl09PRL} & ISBT & MDM & 32 & 15 & 2 & 3 & 1.6 & 14.4 & 11 & 0.44 & First THz ISB polariton\\
	\hline
	\cite{AnapparaetAl09PRB} & ISBT & PWM & 6.5 & 70 & 12 & $\sim$15 & 16.5 & 152 & 11 & $\sim$0.82 &\\
	\hline
	\cite{GunteretAl09Nature} & ISBT & PWM & 9 & 50 & $\cdots$ & $\cdots$ & 10 & 113 & 9 & $\cdots$ & Ultrafast buildup\\
	\hline
	\cite{GeiseretAl10APL} & ISBT & ICR & 95 & 8 & 3.3 & 0.8 & 1.9 & 13 & 14 & 0.88 & parabolic QWs\\
	\hline
	\cite{TodorovetAl10PRL} & ISBT & MDM & 32 & 25 & $\cdots$ & $\cdots$ & 2.8 & 12 & 24 & $\cdots$ & 0D polaritons\\
	\hline
	\cite{ZanottoetAl10APL} & ISBT & SPPC & 8.3 & 50 & 5 & 5 & 5.5 & 119 & 5 & 0.47 &\\
	\hline
	\cite{JouyetAl11APL} & ISBT & MDM & 9 & 10 & $\cdots$ & $\cdots$ & 11 & 107 & 10 & $\cdots$ &\\
	\hline
	\cite{GeiseretAl12PRL} & ISBT & ICR & 72 & 8 & $\cdots$ & $\cdots$ & 4.7 & 18 & 27 & $\cdots$ & Parabolic QWs\\
	\hline
	\cite{PoreretAl12PRB} & ISBT & SPPC & 8.3 & 50 & $\cdots$ & $\cdots$ & 6.8 & 113 & 6 & $\cdots$ & Ultrafast buildup\\
	\hline
	\cite{ZanottoetAl12PRB} & ISBT & SPPC & 8.3 & 50 & 5.36 & $\cdots$ & 5.5 & 125 & 4 & $\cdots$ & Ultrafast bleaching\\
	\hline
	\cite{DelteiletAl12PRL} & ISBT & MDM & 18.5 & 5 & $\cdots$ & $\cdots$ & 57 & 166 & 17 & $\cdots$ & Multisubband plasmon\\
	\hline
	\cite{DietzeetAl13APL} & ISBT & MMC & 32 & 25 & $\cdots$ & 2.5 & 1.4 & 13 & 11 & $\cdots$ &\\
	\hline
	\cite{AskenazietAl14NJP} & ISBT & MMC & 148 & 1 & 7.5 & $\cdots$ & 43 & 118 & 37 & $\cdots$ & The Berreman mode\\
	\hline
	\cite{AskenazietAL17ACS} & ISBT & MDM & 148 & 18 & $\cdots$ & $\cdots$ & 45 & 100 & 45 & $\cdots$ & Thermal emission  \\
	\hline
	\cite{LaurentetAL17APL} & ISBT & MDM & 5 & 18 & 77 & 17 & 53 & 403 & 13.1 & 1.06 &   \\
	\hline
	\cite{MuravevetAl11PRB} & ILLT & CMR & 30 & 1 & 0.02 & 0.02 & 0.025 & 0.058 &  46 & 1.64 & \\
	\hline
	\cite{ScalarietAl12Science} & ILLT & MMC &  $\cdots$ & 4 & $>$0.5 & $>$0.5 & 1.2 & 2.1 &  58 & $<$3.66 &\\
	\hline
	\cite{MuravevetAl13PRB} & ILLT & MPR & 20 & 1 & $\cdots$ & 0.002 & 0.01 & 0.05 & 25 & $\cdots$ & \\
	\hline
	\cite{MaissenetAl14PRB} & ILLT & MMC & 20 & 4 & $\sim$0.8 & $\sim$0.2 & 1.11 & 1.28 & 87 & $\sim$5.16 & InAs/AlSb QWs\\
	\hline
	\cite{ZhangetAl16NP} & ILLT & PCC & 30 & 1&$<$0.04 & $<$0.04 & 0.18 & 1.5 & 12 & $>$3.2 & $C = 4g^2/\kappa\gamma > 300$\\
	\hline
	\cite{MaissenetAl17NJP} & ILLT & MMC & $\cdots$ & 1 &  $>$0.5 & $>$0.5 & 0.46 & 1.98 & 23 & $<$0.88 &  \\
	\hline
	\cite{KelleretAl17arXiv} & ILLT & MMC & 20 & 1 & $\cdots$ & $\cdots$ & 0.49 & 0.86 & 57 & $\cdots$ & Strained Ge QWs \\
	\hline
	\cite{BayeretAl17NL} & ILLT & MMC & 25 & 6 & $\cdots$ & $\cdots$ & 2.85 & 1.99 & 143 & $\cdots$ & $g/\omega>1$  \\
	\hline
	\cite{LietAl18NP} & ILLT & PCC & 30 & 10 & 0.024 & 0.019 & 0.62 & 1.7 & 36 & 35.8 &   $C = 4g^2/\kappa\gamma = 3513$ \\
	\hline
	\cite{Paravicini-BaglianietAl18NP} & ILLT & MMC & 20 & 1 & $\cdots$ & $\sim$0.1 & 0.17 & 0.58 & 30 & $\cdots$ &  Magnetotransport \\
	\hline
	
  \end{tabular}
  \caption{Experimental observations of ultrastrong light-matter coupling in semiconductor quantum wells.  $d_\mathrm{QW}$: QW width. $N_\mathrm{QW}$: number of QWs or periods. $\hbar\gamma$: matter decay rate.  $\hbar\kappa$: photon decay rate; cavity $Q: \omega/\kappa$.  $\hbar g$: coupling strength.  $\omega = \omega_{12}$: ISBT; and $\omega = \omega_\mathrm{c}$: ILLT. ISBT = intersubband transition. ILLT = inter-Landau-level transition (i.e., cyclotron resonance).  PWM = planar waveguide microcavity.  MDM = metal-dielectric-metal microcavity. ICR = inductor-capacitor (LC) resonator. SPPC = surface plasmon photonic crystal. CMR = coplanar microresonator. MMC = metamaterial cavity.  MPR = metallic patch resonator. PCC = photonic-crystal cavity. FPC = Fabry-P\'erot cavity. QC = quantum cascade. $U\equiv\sqrt{(4g^2/\kappa\gamma)g/\omega}$ = geometric mean between cooperativity and normalized coupling.}
  \label{Summary}
\end{table*}

This initial ISB polariton work \cite{DinietAl03PRL} was immediately followed by similar observations by~\cite{DupontetAl03PRB}, who measured a bound-to-quasibound transition in a QW-IR-photodetector structure through both reflection and photocurrent spectroscopy. Rabi splittings were demonstrated with $g/\omega_{12}$ values similar to those reported by Dini \textit{et al}. Furthermore, by increasing the doping density, \cite{DupontetAl07PRB} were able to observe a square-root dependence of the VRS on the total electron density ($N_\mathrm{QW} n_\mathrm{e}$). Here $N_\mathrm{QW}$ corresponds to the number of QWs and $n_\mathrm{e}$ is the density per well, i.e., $2g \propto \sqrt{N_\mathrm{QW}n_\mathrm{e}}$, indicating that electrons in QWs interact cooperatively as a single giant atom with cavity photons  \cite{Dicke54PR,KaluznyetAl83PRL,Agarwal84PRL,AmsussetAl11PRL,tabuchi2014,zhang2014}. A coupling of $g/\omega_{12} = 0.17$ at zero detuning $\omega = \omega_{12}$ was achieved at the highest electron density \cite{DupontetAl07PRB}. 

During the past decade, progressively higher values of $g/\omega$ have been reported, as seen in Table~\ref{Summary}, due to the diverse approaches used by different experimental groups.

In a simple approximation, for a parabolic band of mass $m^*$, the $g/\omega_{12}$ ratio can be written as
\begin{align}
\frac{g}{\omega_{12}} \propto \frac{1}{\sqrt{m^* \omega_{12}}}.
\label{g-over-omega}
\end{align}
Therefore, one can immediately see that a lighter-mass material can generally provide larger $g/\omega_{12}$ ratios for a given $\omega_{12}$. \cite{AnapparaetAl07SSC} used QWs composed of InAs (which has a bulk band-edge electron mass of 0.023$m_0$, as compared to 0.069$m_0$ for electrons in GaAs) to achieve $g/\omega_{12} = 0.14$ at zero detuning $\omega = \omega_{12}$. Another guideline for increasing the $g/\omega_{12}$ ratio, hinted at by Eq.~(\ref{g-over-omega}), is to increase the QW width, which naturally decreases $\omega_{12}$. Todorov \textit{et al.}\ used 32-nm-wide GaAs QWs embedded inside a subwavelength metal-dielectric-metal microcavity \cite{TodorovetAl10OE} to demonstrate USC ($g/\omega = 0.11$) in the THz regime \cite{TodorovetAl09PRL}. By further reducing the cavity volume with respect to the wavelength of
the mode $V_\mathrm{cav}/\lambda^3_\mathrm{res}$ to 10$^{-4}$, \cite{TodorovetAl10PRL} achieved $g/\omega = 0.24$.

As one increases the electron density and QW width, more subbands are occupied, which, within a single-particle picture, leads to multiple ISBT peaks due to band nonparabolicity.  However, \cite{DelteiletAl12PRL} showed that due to many-body interactions a single peak appears. Namely, cooperative Coulombic coupling of dipolar oscillators with different frequencies can induce mutual phase locking, lumping together all individual ISBTs into a single collective bright excitation (multisubband plasmon resonance).  Furthermore, \cite{AskenazietAl14NJP} presented a model to describe the crossover from the ISB plasmon to the multisubband plasmon and then eventually to the so-called  Berreman mode in the classical limit as the QW width was increased. In the Berreman mode limit, a record high $g/\omega$ value of 0.37 was experimentally achieved. For a recent review, see \cite{VasanellietAl16CRP}.   

\begin{figure}
\begin{center}
\includegraphics[scale=1.1]{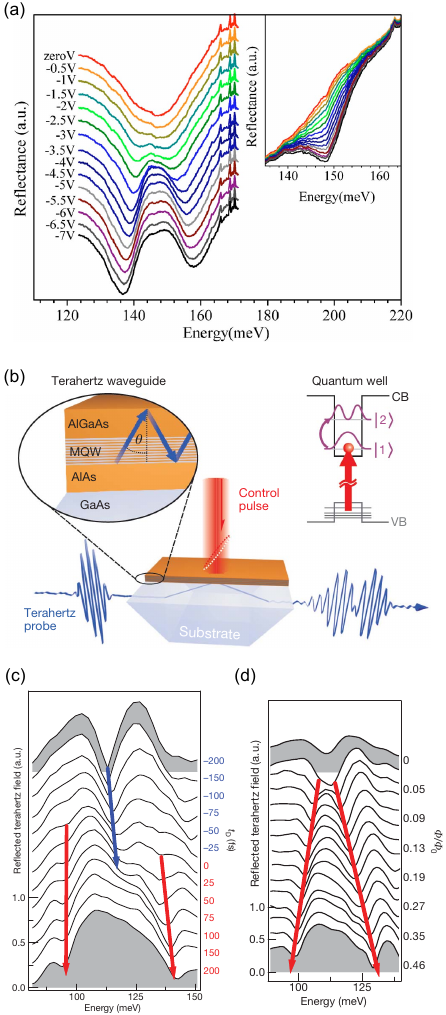}
\caption{\small (Color online) Switchable USC. (a)~Reflectance spectra for GaAs asymmetrically coupled quantum wells at various bias voltages, showing field-tuned vacuum Rabi splitting. The splitting increases with increasing voltage. From reference \cite{AnapparaetAl06APL}. (b)~Setup used for ultrafast control of ultrastrong light-matter coupling. A quantum well structure embedded in a planar waveguide structure is activated by a near-infrared control pulse. Terahertz transients probe the ultrafast build-up of light-matter coupling. (c)~Ultrafast switch-on of ISB polaritons. Spectra of the reflected terahertz field are given for various delay times. (d)~Terahertz reflectance spectra measured at 293~K for various fluences of the control pulse. From~\cite{GunteretAl09Nature}.
}
\label{Control}
\end{center}
\end{figure}

One of the attractive features of ISB polaritons is their controllability via external fields, which can lead to practical devices. Since the vacuum Rabi splitting 2$g$ in a collective system is proportional to $\sqrt{n_\mathrm{e}}$, controlling the electron density $n_\mathrm{e}$ in the QW controls 2$g$. An electric field applied perpendicular to the QW changes the ground state $n_\mathrm{e}$ through gating \cite{AnapparaetAl05APL}, or more quickly through resonant charge transfer via tunneling \cite{AnapparaetAl06APL}. Figure~\ref{Control}(a) shows reflectance spectra for GaAs asymmetrically coupled QWs at a fixed incidence angle at various bias voltages. At zero bias voltage, all electrons are in the wider well, and the spectrum shows a single peak due to the ISBT in the wider well.  As the bias voltage is increased, electrons are increasingly transferred into the ground subband of the narrower quantum well, resulting in the appearance of ISB polaritons. As the bias is further increased, $n_\mathrm{e}$ increases in the narrow well and thus the vacuum Rabi splitting increases  \cite{AnapparaetAl06APL}.  

An ultrafast optical excitation can also be used to control ultrastrong light-matter coupling in ISB polaritons -- an ultrashort laser pulse can either enhance it \cite{GunteretAl09Nature,PoreretAl12PRB} or destroy it \cite{ZanottoetAl12PRB}. For example, ultrafast buildup of ultrastrong light-matter coupling was demonstrated using interband-pump or ISBT-probe measurements in undoped QWs \cite{GunteretAl09Nature}, as shown in Figs.~\ref{Control}(b)--(d). A multiple-QW sample was embedded into a planar waveguide structure based on total internal reflection. The band diagram shows how the $|1\rangle \rightarrow |2\rangle$ ISBT is activated by a near-infrared control pulse, populating level $|1\rangle$. Few-cycle TM-polarized multi-THz transients guided through the prism-shaped substrate are reflected from the waveguide to probe the ultrafast buildup of light-matter coupling, as shown in Fig.~\ref{Control}(c). The blue arrow shows the bare cavity resonance, whereas the red arrows show the ISB LP and UP. Figure~\ref{Control}(d) plots THz reflectance spectra measured for various fluences of the control pulse at a fixed time delay. As the fluence increases, $n_\mathrm{e}$ increases, which in turn increases the VRS.

\subsubsection{Inter-Landau-level transitions (cyclotron resonance)}
\label{sec:3B2}

Strong light-matter coupling has also been actively studied using ILLTs (or cyclotron resonance CR) in 2DEGs formed in GaAs QWs \cite{MuravevetAl11PRB,ScalarietAl12Science,ScalarietAl13JAP,MuravevetAl13PRB,MaissenetAl14PRB,ZhangetAl16NP,MaissenetAl17NJP, BayeretAl17NL, LietAl18NP}, InAs QWs \cite{MaissenetAl14PRB}, and on the surface of liquid helium~\cite{AbdurakhimovetAl16PRL}. \cite{MuravevetAl11PRB} studied the USC of magnetoplasmon (also known as ``cyclotron-plasmon'') excitations with microwave photon modes in a coplanar microresonator and a metallic patch resonator \cite{MuravevetAl13PRB}. An advantage of the straightforward continuous magnetic field tuning of polaritons over ISB polaritons was clearly demonstrated. High values of $g/\omega$ close to 0.5 were achieved \cite{MuravevetAl11PRB} owing to the large dipole moment of ILLTs.

\cite{ScalarietAl12Science} reported experiments showing USC of 2DEG CR with photons in a THz metamaterial cavity consisting of an array of electronic split-ring
resonators shown in Fig.~\ref{ETH-CR}(a) and \ref{ETH-CR}(b). They obtained a $g/\omega$ value of 0.58 and showed potential scalability in frequency to extend to the microwave spectral range, where control of the magnetotransport properties of the 2DEG through light-matter coupling would be possible.  Furthermore, using similar split-ring resonators in the complementary mode, Maissen \textit{et al.}\  obtained $g/\omega=0.87$, shown in Fig.~\ref{ETH-CR}(c)-(d). In addition, a blueshift of both LP and UP was observed due to the diamagnetic term of the interaction Hamiltonian. %Most recently, by carefully tailoring the vacuum mode of a structure using similar metamaterial resonators, Bayer \emph{et al.} were able to push the 2DEG-CR system into the DSC regime, achieving an unprecedentedly high value $g/\omega_0 =$ 1.43 \cite{BayeretAl17NL}.

\begin{figure}
\begin{center}
\includegraphics[scale=0.83]{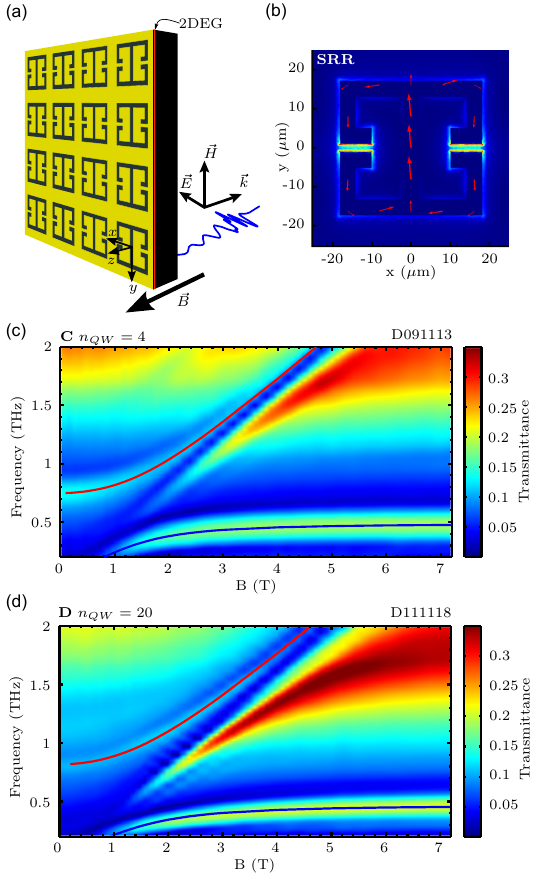}
\caption{\small (Color online) USC of normal-incidence THz radiation with a GaAs 2DEG in a Landau-quantizing magnetic field. (a)~Experimental setup used to observed USC. An array of metamaterial THz cavities is deposited on top of the 2DEG. (b) Scanning electron microscopy picture displays a single cavity unit. Adapted from reference \cite{ScalarietAl12Science}. (c), (d) Transmittance spectra at different magnetic fields showing anticrossing behavior with a $g/\omega$ value of (c) 0.69 and (d) 0.87. From~\cite{MaissenetAl14PRB}.
}
\label{ETH-CR}
\end{center}
\end{figure}

In these CR studies of ultrastrong light-matter coupling using metamaterial split-ring resonators, however, the value of cooperativity $C = 4g^2/\gamma\kappa$ remained small due to ultrafast decoherence (large $\gamma$) and/or lossy cavities (large $\kappa$).  Recently, \cite{ZhangetAl16NP} developed a THz 1D photonic-crystal cavity (PCC), utilizing Si thin slabs and air as the high and low index materials, respectively; see Fig.~\ref{CR-QED}(a). The air-Si combination provided a large index contrast and thus significantly reduced the number of layers needed on each side of the cavity~\cite{YeeSherwin09APL,ChenetAl14APB}. A thin 2DEG film was transferred onto one surface of the central layer, where the electric field maximum was located. Figure~\ref{CR-QED}(b) shows an experimental transmission spectrum measured for one of the empty cavities, demonstrating an ultranarrow photonic mode ($\kappa/2\pi \sim$ 2.6\,GHz).  The highest cavity quality factor $Q$ achieved in this scheme was $\sim$10$^3$. 
\begin{figure}[hb]
\begin{center}
\includegraphics[scale=0.62]{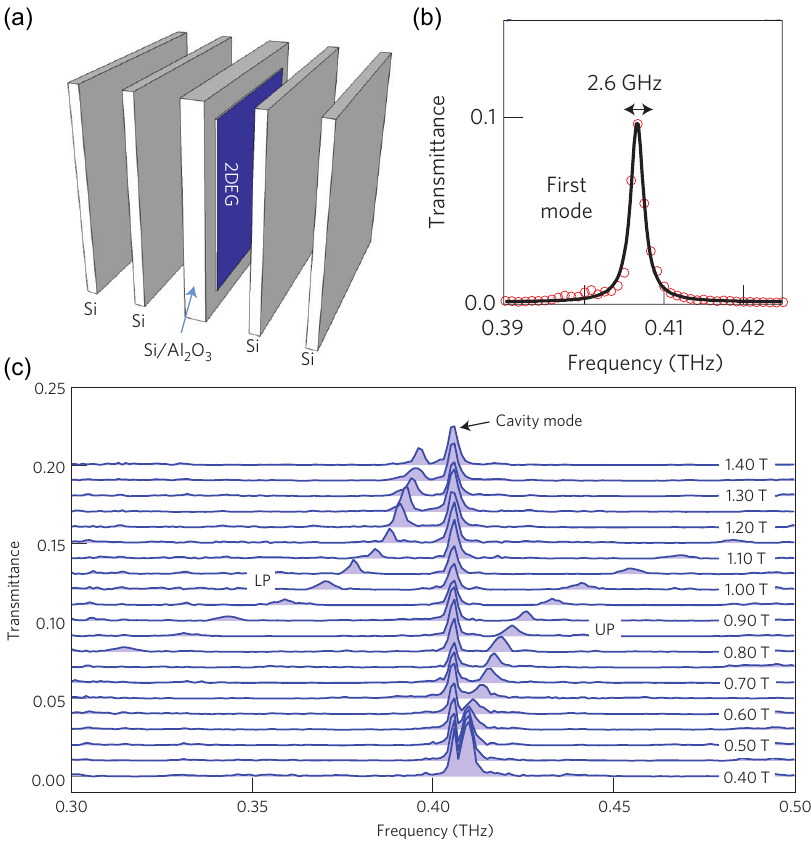}
\caption{\small (Color online) Observation of USC of CR of a 2DEG and high-$Q$ THz cavity photons. (a)~1D terahertz photonic-crystal cavity structure. Two silicon layers are placed on each side of the central defect layer. The blue part is the transferred 2DEG thin film. (b)~Zoom-in spectrum for the first cavity mode, together with a Lorentzian fit with a full-width-at-half-maximum of 2.6\,GHz (c)~Anticrossing of CR and the first cavity mode, exhibiting the LP and UP branches. The central peak due to the cavity mode results from the CR-inactive circularly polarized component of the linearly polarized terahertz beam. Transmission spectra at different magnetic fields are vertically offset for clarity. The magnetic field increases from 0.4\,T (bottom) to 1.4\,T (top). From~\cite{ZhangetAl16NP}.
}
\label{CR-QED}
\end{center}
\end{figure}

Using these high-$Q$ PCCs,~\cite{ZhangetAl16NP} simultaneously achieved small $\gamma$ and small $\kappa$ in ultrahigh-mobility 2DEGs in GaAs QWs in a magnetic field; see Fig.~\ref{CR-QED}(c). High cooperativity values $C > 300$ were achieved, with VRS leading to $g/\omega\sim 0.1$. With these favorable parameters it was possible to observe Rabi oscillations in the time domain.
Zhang \emph{et al.} showed that the influence of such USC extended even to the region with detuning $\delta > \omega$. This effect could occur only when $g^2/\omega\kappa > 1$, which in the experiment was satisfied through a unique combination of strong light-matter coupling, a small resonance frequency, and a high-$Q$ cavity.
Furthermore, the expected $\sqrt{n_\mathrm{e}}$ dependence of 2$g$ on the electron density ($n_\mathrm{e}$) was observed, signifying the collective nature of light-matter coupling \cite{Dicke54PR}. A value of $g/\omega$ $=$ 0.12 was obtained with just a single QW with a moderate $n_\mathrm{e}$ (= 3 $\times$ 10$^{11}$\,cm$^{-2}$). Finally, \cite{ZhangetAl14PRL} observed a significant suppression of a previously identified superradiant decay of CR in high-mobility 2DEGs due to the presence of the high-$Q$ THz cavity. As a result, ultranarrow polariton lines were observed, yielding an intrinsic CR linewidth as small as 5.6\,GHz (or a CR decay time of 57\,ps) at 2\,K.

\begin{figure}%[!hbt]
\centering
\includegraphics[width = 1.0\columnwidth]{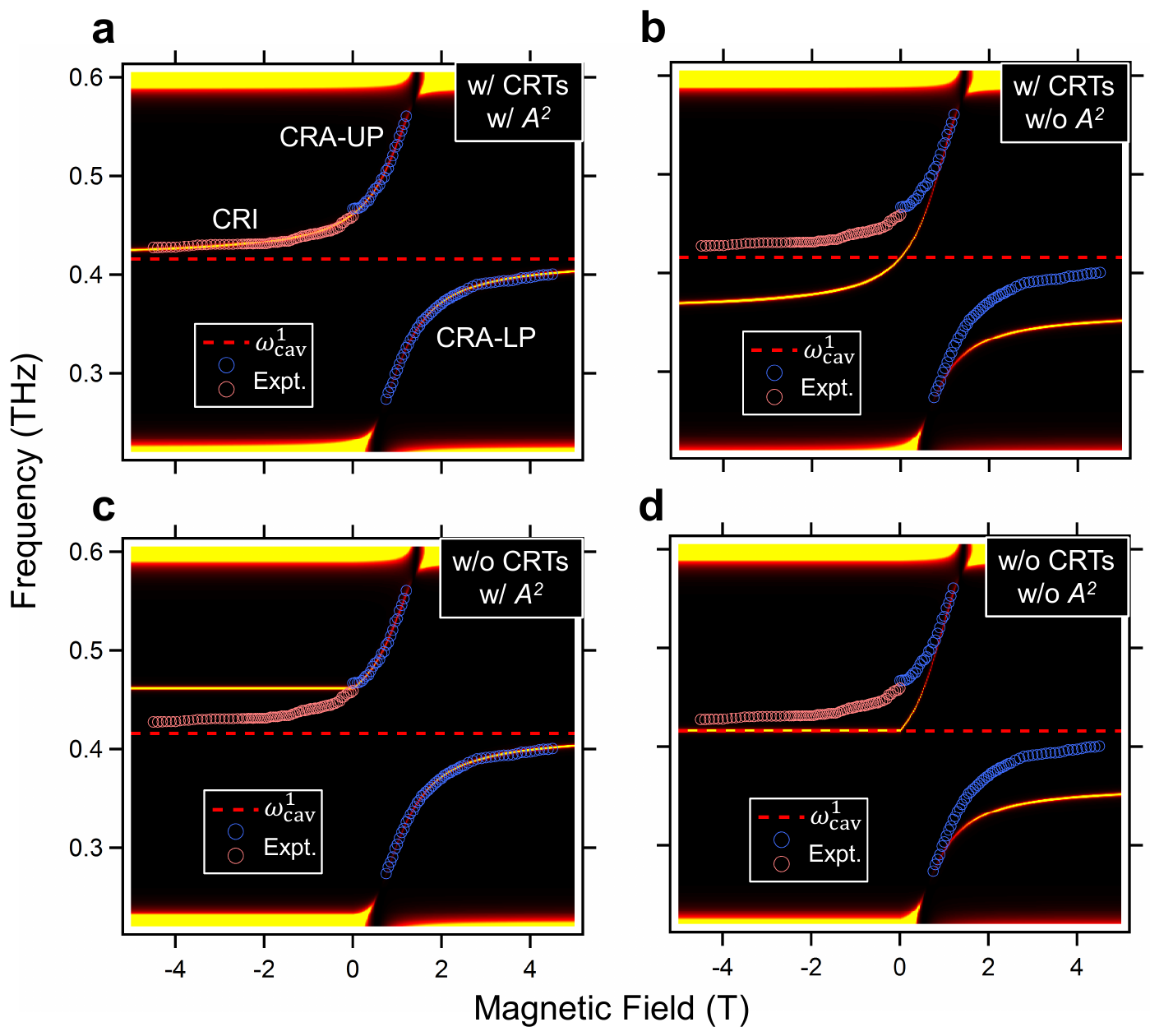}
\caption{\label{fig:BSVR} (Color online) Distinction between the vacuum Bloch-Siegert shift due to the counterrotating terms (CRTs) and the shift due to the $A^2$ terms in the USC regime. Simulated spectra (a)~with both the CRTs and the $A^2$ terms (full Hamiltonian), (b)~with the CRTs but without the $A^2$ terms, (c)~without the CRTs but with the $A^2$ terms, and (d)~without the CRTs and $A^2$ terms.  Each graph includes experimental peak positions as open circles. Adapted from \cite{LietAl18NP}.}
\label{VBS}
\end{figure}

More recently, through optimization of both electronic and photonic components of a 2DEG-metamaterial system, \cite{BayeretAl17NL} have significantly boosted the light-matter coupling strength, entering the DSC regime.  By tailoring the shape of the vacuum mode in the cavity, they achieved a remarkable $g/\omega = 1.43$, the highest result reported to date for semiconductor QWs. This achievement opens up the possibilities of studying vacuum radiation with cutting-edge THz quantum detection techniques~\cite{RieketAl15Science,Benea-ChelmusetAl16PRA,RieketAl17Nature}. \cite{KelleretAl17NL} probed USC at 300 GHz to less than 100 electrons located in the last occupied Landau level of a high-mobility two-dimensional electron gas. By using hybrid dipole antenna-split ring resonator-based cavities with extremely small effective mode volumes and surfaces they achieved a normalized coupling ratio of $g/\omega = 0.36$. Effects of the extremely reduced cavity dimensions were observed as the light-matter coupled system resulted better described by an effective mass heavier than the uncoupled one.

In later work, \cite{KelleretAl17arXiv} studied the USC of the CR of a 2D hole gas in a strained germanium QW with THz metasurface cavity photons. They observed a mode softening of the polariton branches, deviating from the Hopfield model successfully used in studies of GaAs QWs~\cite{HagenmulleretAl10PRB,ScalarietAl12Science}. At the largest coupling strength, the lower polariton branch was observed to move toward zero frequency, raising the exciting perspective of the Dicke superradiant phase transition in equilibrium~\cite{HeppLieb73AP,WangHioe73PRA}. They modeled this behavior by effectively reducing the magnitude of the $A^2$ term in the Hamiltonian. The 2D hole gas exhibits heavy nonparabolicity, strain, and spin-orbit interaction, features differing from the standard GaAs QWs; however, theoretical modeling of the observed deviation remains an open quest.

Most recently,~\cite{LietAl18NP} reported the vacuum Bloch-Siegert shift, which is induced by the coupling of matter with the counterrotating component of the vacuum fluctuation field in a cavity, as explained in Sec.~\ref{sec:2}; see, e.g., Eq.~(\ref{eq:BS}). Using an ultrahigh-mobility 2DEG in a high-$Q$ THz cavity in a magnetic field, they created Landau polaritons with an ultrahigh cooperativity ($C = 3513$), which exhibited a vacuum Bloch-Siegert shift up to 40\,GHz.
They found that the probe polarization plays a critical role in exploring USC physics in this ultrahigh-cooperativity system. The resonant corotating coupling of electrons with CR-active (CRA) circularly polarized radiation leads to the extensively studied VRS. Conversely, the counterrotating coupling of electrons with the CR-inactive (CRI) mode leads to the time-reversed partner of the VRS, i.e., the vacuum Bloch-Siegert shift.

Li \textit{et al}.\  theoretically simulated polariton spectra to explain their data while selectively removing the counterrotating terms (CRTs) and the $A^2$ terms from the full Hamiltonian, as shown in Figs.\,\ref{VBS}(a), \ref{VBS}(d) together with experimental data. From the perfect agreement between experiment and theory shown in Fig.~\ref{VBS}(a), deviations appear when either the CRTs or the $A^2$ terms are removed. By comparing Figs.\,\ref{VBS}(a) and \ref{VBS}(b), one can confirm that the $A^2$ terms produce an overall blueshift for both polariton branches and the CRI mode.  On the other hand, through comparison of Figs.~\ref{VBS}(a) and \ref{VBS}(c), one can confirm that the CRTs only affect the CRI mode, producing the vacuum Bloch-Siegert shift. It is important to note that one of the goals of cavity QED studies using semiconductor QWs, or condensed matter systems in general, is to search for cooperative effects and new ground states. To this aim, adaptation of quantum optical concepts and tools in condensed matter physics is an emerging subject of research \cite{CongetAl16JOSAB, LietAl18Science}, where Hamiltonians traditionally used in atomic quantum optics must be modified through the incorporation of many-body effects and dispersions of collective excitations \cite{Dicke54PR,Hopfield58PR}.

One peculiar aspect of ultrastrong light-matter coupling in a cavity is the conspicuous absence of a strong external light field in the problem.  In other words, no strong field is needed to induce strong-field physics.  Matter placed inside a cavity nonperturbatively couples with the vacuum fluctuation field of the cavity to form polaritons with VRS comparable to the original matter and photon energies.  This is a highly unusual situation for a nonlinear optical process, which would ordinarily increase with increasing strength of an applied light field.
This aspect of USC in a cavity allows one to study USC in unusual ways, sometimes even without using light.  For example, electronic transport properties, such as the electrical conductivity and Hall coefficient, are expected to be affected by the presence of USC in a quantum Hall system~\cite{HagenmulleretAl10PRB,BartoloCiuti18PRB}.  The conductivity of a molecular crystal inside a cavity has indeed been observed to be enhanced by strong coupling with a plasmonic mode~\cite{OrgiuetAl15NM}, and a general theoretical treatment of charge transport in the USC regime has recently been formulated~\cite{HgenmulleretAl17PRL,HagenmulleretAl18PRB}. 
Most recently,~\cite{Paravicini-BaglianietAl18NP} demonstrated the crucial role played by the matter component of polaritons in the USC regime through magnetotransport measurements on a 2DEG embedded in a metamaterial cavity. They showed that the dc resistivity of the 2DEG is substantially modified by the USC to the cavity photons without external irradiation. This observation is consistent with recent theoretical predictions of vacuum-induced modifications of resistivity.~\cite{HagenmulleretAl10PRB,HgenmulleretAl17PRL,BartoloCiuti18PRB,HagenmulleretAl18PRB}. 

\subsection{Hybrid quantum systems}\label{sec:3C}

In Secs.~\ref{sec:3A} and \ref{sec:3B}, we presented the main achievements in experimental USC regimes in the fields of superconducting quantum circuits and semiconductor quantum wells, respectively. 
This section reviews quantum systems of hybrid nature where ultrastrong couplings have also been demonstrated. In these systems, the magnitude of the coupling originates from a collective degree of freedom which is the result of an ensemble of individual systems coupling to the same cavity mode. In such a configuration, a typical scaling of $\sqrt{N}$ is obtained \cite{Dicke54PR, YamamotoImamoglu99Book}, with $N$ being the number of systems participating in the collective degree of freedom. The same scaling is found for intraband transitions in semiconductor QWs (see Sec.~\ref{sec:3B}). 

In particular, the systems described in this section consist of molecular aggregates in optical microcavities, microcavity exciton polaritons in unconventional semiconductors with large binding energies and oscillator strengths, and magnons in magnetic materials coupled to the magnetic field of a microwave cavity. These cases combine quantum systems of a very distinct nature and therefore fall into the category of hybrid systems. Technically speaking, the previous section on conventional III-V semiconducting quantum wells already presented hybrid quantum systems, i.e., intersubband polaritons (Sec.~\ref{sec:3B1}) and inter-Landau-level polaritons (Sec.~\ref{sec:3B2}). This section therefore covers topics of polaritons in ultrastrong coupling regimes in systems other than traditional semiconductor quantum wells.

\subsubsection{Molecules in optical cavities}

The influence of cavity modes on the radiative properties of quantum emitters such as molecules has been the object of study since the early works of \cite{Purcell1946}. In more recent times, the strong coupling regime was reached with ensembles of molecules coupling to a single mode of an optical microcavity \cite{LidzeyetAl98Nature, Holmes2004}. A key element to maximize the coupling strength was the discovery of molecules with a large enough electric dipole coupling to the electric field of the cavity mode.

The electric dipole energy of interaction between an ensemble of molecules and a cavity mode can be calculated from \cite{george2015}
\begin{equation}
\label{eq:mol}
\hbar g = d\sqrt{\frac{\hbar\omega}{2\epsilon_0V_\mathrm{m}}}.
\end{equation}
Here $d$ is the total electric dipole moment of the molecular ensemble and is therefore proportional to $\sqrt{N}$, $d = d_0\sqrt{N}$ with $d_0$ being the electric dipole of a single molecule. $\epsilon_0$ is the vacuum permittivity, and $V_\mathrm{m}$ is the cavity mode volume. The square-root factor in Eq.~(\ref{eq:mol}) corresponds to the rms electric field in the ground state of the cavity mode. 

The first demonstration of a molecular ensemble ultrastrongly coupled to a single mode of a microcavity was carried out by Schwarz \emph{et al.} \cite{schwartz2011}. The experiment consisted of a PMMA (polymethyl methacrylate) matrix sputtered on both sides by a thin Ag layer in a Fabry-Perot configuration, resulting in a low-$Q$ cavity. The PMMA matrix was filled with photochromic spiropyran (SPI) molecules (10, 30-dihydro-10, 30, 30-trimethyl-6-nitrospiro[2H-1-benzopyran-2, 20-(2H)-indole]). These molecules can undergo photoisomerization between a transparent SPI form and a colored merocyanine (MC) form. Schwarz \emph{et al.} observed that molecules in the SPI form were not coupling to the cavity mode. As shown in Fig.~\ref{fig:mol}, upon ultraviolet illumination, a transition between SPI and MC forms was induced, the latter having a strong dipolar coupling to the cavity mode. This was observed as a large mode splitting in the cavity transmission, indicating strong coupling. With longer illumination, more molecules transitioned and the value of $\hbar g$ reached up to 357~meV, being 16.2\% of the cavity resonance and well in the USC regime. In later work \cite{george2015}, other molecules, such as 1,1'-diethyl-3,3'-bis(4-sulfobutyl)-5,5',6,6'-tetrachlorobenzimidazolocarbocyanine (TDBC), 5-(4-(dibutylamino)-benzylidene)-1,3-dimethylpyrimidine-2,4,6
(1H,3H,5H)-trione (BDAB), and fluorescenin, were observed to yield $g/\omega$ values of 13, 24, and 27\% of the cavity resonance, respectively. 
\begin{figure}[!hbt]
\centering
\includegraphics[width = 8.5cm]{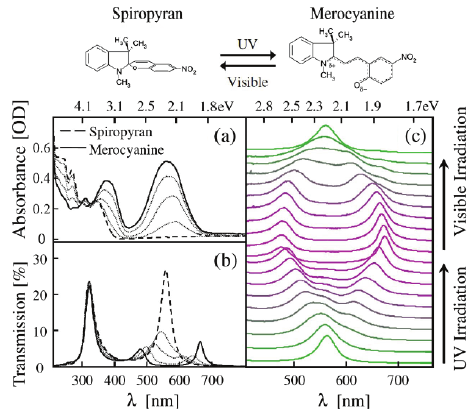}
\caption{\label{fig:mol}(Color online) USC achieved with a molecular ensemble in a Fabry-Perot cavity. By shining ultraviolet (UV) light the molecules change from spiropyran (SP) to merocyanine (MC) form. The latter displays a large dipole moment which couples to the cavity electromagnetic field all the way up to the USC regime. (a) Cavity absorption spectrum. (b) Cavity transmission spectroscopy with no UV illumination. (c) Cavity transmission for varying exposure times. Traces are offset for clarity. Mode splitting increases as the UV light exposes the molecules and closes back with infrared radiation that returns the molecules into the SP state demonstrating the reversibility of the process. From~\cite{schwartz2011}.}
\end{figure}

In a more recent study, the vibrational dipolar strength of a molecular liquid was also shown to simultaneously ultrastrongly couple to several modes of a Fabry-Perot cavity in the infrared \cite{george2016}. The molecules chosen for the study were iron pentacarbonyl [Fe(CO)$_5$] and carbon disulphide (CS$_2$), both showing very strong oscillator strength, which was key to the successful attainment of large coupling strengths to the cavity modes. This work may be important in molecular chemistry as vibrational strong coupling could be used to control chemical reactions given the role played by vibrations in the process. 

Finally, it is also worth mentioning that in a recent study strong coupling ($g/\kappa \sim 0.2$) was achieved in a single molecule level}~\cite{BenzetAl16Science}.  Benz and co-workers demonstrated that individual molecules can be trapped inside the gap of a plasmonic nanoassembly that localizes light to volumes well below 1 nm$^3$ (``picocavities''). Such extreme optical confinement yielded a factor of 10$^6$ enhancement of optomechanical coupling between the picocavity field and vibrations of individual molecular bonds.

\subsubsection{Microcavity exciton polaritons}
\label{sec:3c2}

%%% Table 1 %%%
\begin{table*}
  \centering
  \begin{tabular}{|c|c|c|c|c|c|c|c|c|c|c|c|}
    \hline 
     &  & Exciton &  &  2$\hbar g$ & $\hbar\omega$ & $g/\omega$ &  \\
    Reference&  Material& type & Temperature  &  (meV) & (eV) & (\%) & Notes\\
    \hline
    \hline
	\cite{WeisbuchetAl92PRL} & GaAs & Wannier & 20\,K & 5 & 1.58 & 0.2 &  QWs \\
	\hline
	\cite{BlochetAl98APL} & GaAs & Wannier & 77\,K  & 19 & 1.62 & 1.2 & QWs\\
	\hline
	\cite{DengetAl02Science} & GaAs & Wannier & 4\,K & 15 & 1.61 & 0.46 & QWs\\
	\hline
	\cite{BellessaetAl04PRL} & J aggregates & Frenkel & RT & 180 & 2.1 & 4.3 & Plasmon-exciton coupling\\
	\hline
	\cite{KasprzaketAl06Nature} & CdTe & Wannier & 5\,K & 26 & 1.68 & 0.77 & QWs\\
	\hline
		\cite{vanVugtetAl06PRL} & ZnO & Wannier & RT & 100 & 3.3 & 1.5 & Nanowires\\
	\hline
	\cite{ChristmannetAl08PRB} & GaN & Wannier & RT & 50 & 3.64 & 0.7 & QWs\\
	\hline
		\cite{GuilletetAl11APL} & ZnO & Wannier & 120\,K & 130 & 3.36 & 1.9 & Bulk\\
	\hline
	\cite{WeietAl13OE} & J aggregates & Frenkel & RT & 400 & 2.27 & 8.8 &\\
	\hline
	\cite{Kena-CohenetAl13AOM} & TDAF & Frenkel & RT & 1000 & 3.534 & 14 &\\
	\hline
	\cite{GambinoetAl14ACS} & Squaraine & Frenkel & RT & 1120 & 2.07 & 27 &\\
	\hline
	\cite{LiuetAl15NP} & MoS$_2$ & Wannier & RT & 46 & 1.87 & 1.2 &\\
	\hline
	\cite{FlattenetAl16SR} & WS$_2$ & Wannier & RT & 70 & 2 & 1.75 &\\
	\hline
	\cite{LiuetAl16NL} & MoS$_2$ & Wannier & 77\,K & 116 & 1.87 & 3 & Plasmon-exciton coupling\\
	\hline
	\cite{GrafetAl16NC} & SWCNTs & Wannier & RT & 110 & 1.24 & 4.4 & (6,5)-enriched\\
	\hline
	\cite{BrodbecketAl17PRL} & GaAs & Wannier & 20\,K & 17.4 & 1.61 & 1.1 & QWs, $g/Ry^* = 0.64$\\
	\hline
	\cite{GaoetAl18NP} & SWCNTs & Wannier & RT & 329 & 1.24 & 13.3 & (6,5)-enriched \& aligned\\
	\hline

  \end{tabular}
  \caption{Experimental observations of strong and ultrastrong light-exciton coupling in various microcavity exciton polariton systems. QW: quantum well. $\hbar g$: coupling strength. 2$\hbar g$: vacuum Rabi splitting. $\hbar \omega$:  exciton resonance photon energy. $Ry^*$: exciton binding energy. SWCNTs: single-wall carbon nanotubes. TDAF: 2,7-bis[9,9-di(4-methylphenyl)-fluoren-2-yl]-9,9-di(4-methylphenyl)fluorene. RT: room temperature, 300~K.}
  \label{table:MEP}
\end{table*}

%%%%%%%%%%

As described in Sec.~\ref{sec:3B}, MEPs in semiconductor QWs have long been studied as a model system for investigations of solid-state cavity QED phenomena~\cite{WeisbuchetAl92PRL,SkolnicketAl98SST,KhitrovaetAl99RMP,DengetAl10RMP,GibbsetAl11NP}. However, MEPs based on Wannier excitons in inorganic semiconductors, such as GaAs QWs, have remained in the strong coupling regime, typically with $g/\omega$ $<$ 10$^{-2}$, far from the USC and DSC regimes.  
Wannier excitons in other traditional inorganic semiconductors with larger exciton binding energies (and thus larger band gaps, effective masses, and oscillator strengths) than GaAs, including GaN, CdTe, and ZnO, have been utilized to achieve larger values of $g/\omega$ up to $\sim$0.02; see Table~\ref{table:MEP}.

Frenkel excitons (i.e., excitons with Bohr radii of the same order as the size of the unit cell) in organic semiconductors~\cite{LidzeyetAl98Nature} possess large binding energies and oscillator strengths and have displayed larger VRS than Wannier-exciton-based MEPs, reporting generally larger values of $g/\omega$, as shown in Table~\ref{table:MEP}.  In particular, two groups observed giant VRSs, on the order of 1\,eV, in Fabry-Perot microcavities filled with 2,7-bis[9,9-di(4-methylphenyl)-fluoren-2-yl]-9,9-di(4-methylphenyl)fluorene~\cite{Kena-CohenetAl13AOM} and squaraine~\cite{GambinoetAl14ACS}, respectively.  Representative spectra are shown in Fig.\,\ref{Frenkel}. The corresponding $g/\omega$ values are 0.14 and 0.27, respectively, indicating that these systems are in the USC regime.

\begin{figure}[!hbt]
\centering
\includegraphics[width = 1.0\columnwidth]{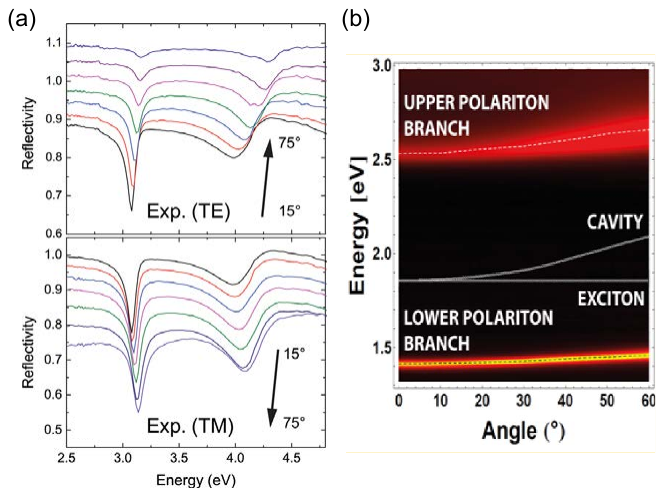}
\caption{\label{fig:GVR} (Color online) Observation of giant vacuum Rabi splitting ($\sim$1\,eV) in microcavity exciton polariton systems based on Frenkel-type excitons. (a)~Angle-resolved reflectivity spectra for a 67-nm-thick cavity containing a thin film of 2,7-bis[9,9-di(4-methylphenyl)-fluoren-2-yl]-9,9-di(4-methylphenyl)fluorene measured using TE (upper panel) and TM (lower panel) polarized light. Adapted from \cite{Kena-CohenetAl13AOM}. (b)~Contour plots of angle-resolved transmission spectra for a 140-nm-thick microcavity entirely filled with squaraine. Adapted from \cite{GambinoetAl14ACS}.}
\label{Frenkel}
\end{figure}

Moreover, nanomaterials with large binding energy Wannier excitons have recently emerged, including atomically thin transition metal dichalcogenide layers~\cite{LiuetAl15NP,LiuetAl16NL,FlattenetAl16SR} and single-wall carbon nanotubes (SWCNTs)~\cite{GrafetAl16NC,GrafetAl17NM}. These novel materials provide a platform for studying strong coupling physics under extreme quantum confinement. 
In particular, one-dimensional (1D) excitons in SWCNTs have enormous oscillator strengths, revealing a very large VRS exceeding 100\,meV in microcavity devices containing a film of single-chirality SWCNTs~\cite{GrafetAl16NC}; the VRS showed a $g \propto \sqrt{N}$ behavior, where $N$ is the number of dipoles (i.e., excitons in the present case), evidencing cooperative enhancement of light-matter coupling~\cite{Dicke54PR,ZhangetAl16NP}, as shown in Fig.\,\ref{SWCNTs}(a). Furthermore,~\cite{GrafetAl17NM} recently demonstrated electrical pumping and tuning of exciton polaritons in SWCNTs, making impressive progress toward creating polaritonic devices~\cite{Sanvitto-KenaCohen16NM}.

\begin{figure*}[!hbt]
\centering
\includegraphics[width = 1.4\columnwidth]{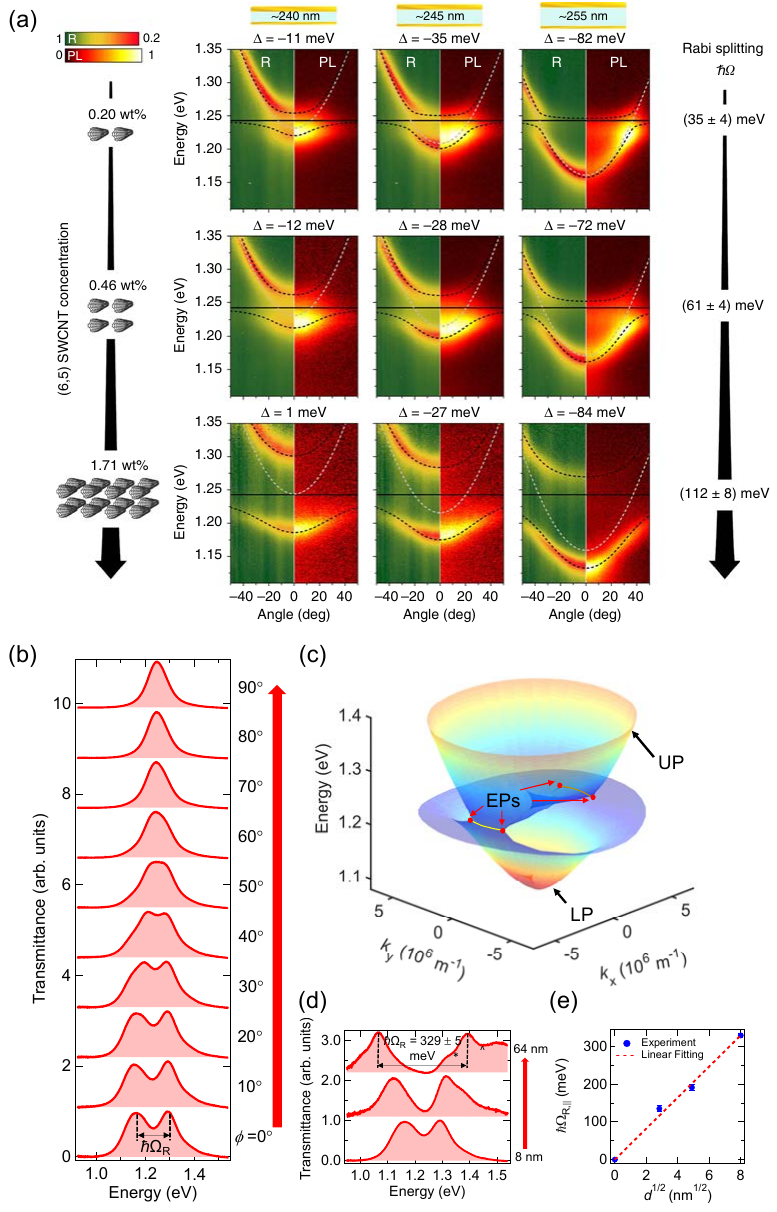}
\caption{\label{fig:SWCNTs} (Color online) Single-wall carbon nanotube microcavity exciton polaritons exhibiting ultrastrong coupling. (a)~Angle-resolved reflectivity and photoluminescence spectra for (6,5) SWCNT microcavity exciton polaritons with increasing nanotube concentrations (from top to bottom) and increasing cavity thickness and detuning from (left to right). Adapted with permission from \cite{GrafetAl16NC}. (b)~Transmittance spectra for a cavity containing aligned (6,5) SWCNTs at zero detuning for various polarization angles from 0$^\circ$ to 90$^\circ$. (c)~Continuous mapping of the dispersion surfaces of the upper polartion (UP) and lower polartion (LP) for the device in (b). EP: exceptional points.  (d)~Transmittance spectra for parallel polarization at zero detuning for devices containing aligned SWCNT films of different thicknesses. The device containing a 64-nm-thick aligned SWCNT film demonstrates the largest VRS of 329~meV. (e)~VRS for parallel polarization at zero detuning vs the square root of the film thickness, demonstrating the $\sqrt{N}$-fold enhancement of collective light-matter coupling. 
Adapted with permission from \cite{GaoetAl18NP}.}
\label{SWCNTs}
\end{figure*}

Most recently,~\cite{GaoetAl18NP} developed a unique architecture in which 1D excitons in an aligned SWCNT film interact with cavity photons in two distinct manners. The system reveals ultrastrong coupling (VRS up to 329\,meV) for probe light with polarization parallel to the nanotube axis, whereas VRS is absent for perpendicular polarization. Between these two extreme situations, the coupling strength is continuously tunable through facile polarization rotation; see Fig.\,\ref{SWCNTs}(b).  Figure~\ref{SWCNTs}(c) shows complete mapping of polariton dispersions, which demonstrates the existence of exceptional points (EPs), spectral singularities that lie at the border of crossing and anticrossing; the points bounded by a pair of EPs formed two equienergy arcs in momentum space, onto which the upper and lower polariton branches coalesced. This unique system with {\em on-demand USC} can be used for exploring exotic topological properties~\cite{Yuen-ZhouetAl14NM,Yuen-ZhouetAl16NC} and exploring applications in quantum technologies. Similar to \cite{GrafetAl16NC}, the VRS exhibited cooperative enhancement, proportional to the square root of the film thickness, as shown in Figs.\,\ref{SWCNTs}(d) and (e).  Figure~\ref{SWCNTs}(d) shows transmittance spectra for the three samples with different thicknesses; the VRS for the thickest sample is 329 $\pm$ 5\,meV, corresponding to $g/\omega$  = 0.13, the highest value for MEPs based on Wannier excitons.

\subsubsection{Magnons in microwave cavities}

In recent years, a new platform of coherent light-matter interaction has been developed by combining magnetic fields from cavity photons and spin waves in magnetic materials \cite{huebl2013, tabuchi2014, zhang2014}. This quantum hybrid system consists of microwave photons residing in a resonant cavity, which interact with a spin wave in a ferromagnetic (ferri)magnetic material, as shown in Fig.~\ref{fig:magnons}(a). At the fundamental level, a microwave photon interacts with a quantum of excitation of such a spin wave, known as a magnon. This emerging platform of quantum magnonics is designed for strong magnon-photon interactions for applications in quantum information such as frequency conversion, quantum memories, and quantum communication \cite{Zhang2016}.

The prototypical system used in these experiments is the ferrimagnetic insulator yttrium iron garnet $\rm Y_3Fe_5O_8$ (YIG). This material exhibits spin waves with the largest quality factors among all magnetic materials explored so far, which explains why it is the most widely used. YIG is often employed in spherical form, with its fundamental mode being the Kittel mode in which all spins oscillate collectively in phase. 
\begin{figure}[!hbt]
\centering
\includegraphics[width = 9cm]{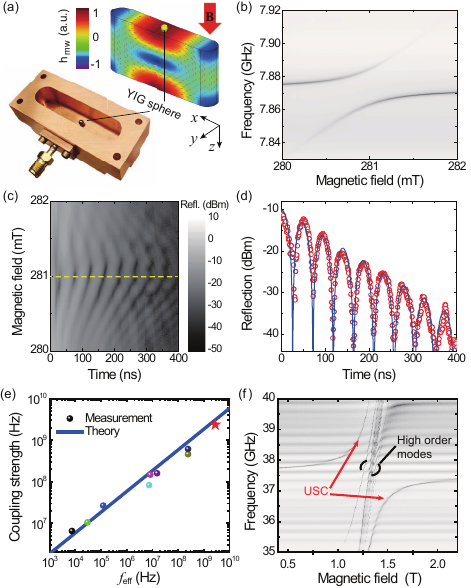}
\caption{\label{fig:magnons}(Color online) Strong coupling between magnons and photons at room temperature. (a) Image of a microwave cavity used in the experiment with a yttrium-iron-garnet (YIG) sphere positioned near a side wall. Simulations show the magnetic field profile of the mode coupling to the magnons in the YIG sphere. The cavity is designed to yield maximum magnetic field amplitude at the position of the sphere. (b) Avoided-level crossing observed at room temperature, indicating strong magnon-photon interactions. The signal displays reflection off the cavity port. (c) Real-time, resonant magnon-photon dynamics being driven by an externally applied microwave field. (d) Cross section of trace indicated in (c). (e) Scaling of coupling strength as a function of cavity mode frequency. The star indicates a device in the USC regime. (f) Spectrum of device exhibiting USC. From~\cite{zhang2014}.}
\end{figure}

The coupling strength $g$ between the Kittel and the cavity modes is proportional to the square root of the number of participating spins $g = g_0\sqrt{N}$, where $g_0$ is the coupling strength of a single Bohr magneton to a cavity photon. The rms magnetic field generated in the cavity in its ground state is given by $\langle \hat{B}^2\rangle^{1/2}=\sqrt{\mu_0\hbar\omega/2V_c}$, with $\omega$ being the cavity frequency, $V_c$ the mode volume occupied by the cavity mode, and $\mu_0$ the vacuum permeability. The single-spin coupling strength is calculated to be \cite{tabuchi2014, zhang2014} 
\begin{equation}\label{g_mag}
g_0/2\pi = \eta\frac{\gamma}{2\pi}\sqrt{\frac{\hbar\omega\mu_0}{2V_c}}.
\end{equation}
Here, $\eta\leq1$ describes the spatial overlap and polarization matching conditions between the
microwave field and the magnon mode \cite{zhang2014}. $\gamma = 2\pi \times 28~$GHz/T is the electron gyromagnetic ratio. 

In the first demonstration of strong coupling between magnons and photons \cite{tabuchi2014}, a collective coupling strength in the range of 100s of MHz was observed using a cavity of 10.7~GHz resonant to a ferromagnetic resonance mode. The $\sqrt{N}$ scaling was further demonstrated by using spheres of different volume (and therefore of a larger number of spins). In a parallel experiment \cite{zhang2014}, real-time magnon-photon oscillations were observed at room temperature; see Figs.~\ref{fig:magnons}(b)--\ref{fig:magnons}(d). The same authors studied the scaling properties of the coupling constant [Eq.~(\ref{g_mag})] to maximize the interaction strength; see Figs.~\ref{fig:magnons}(e)--\ref{fig:magnons}(f). By using a smaller cavity to enhance its frequency and a larger sphere containing more spins, a coupling rate of $g/2\pi = 2.5~$GHz was attained, being $g/\omega = 0.067$ of the magnon resonance frequency resonant with a cavity of $\omega/2\pi=37.5$~GHz. Therefore, the system is approaching the perturbative USC regime, being the only result so far in this field reaching such a high coupling strength.

\section{Quantum simulations}\label{sec:4}

The previous section gave an overview of the most relevant work in all experimental platforms studying ultrastrong light-matter interactions. Besides the remarkable couplings achieved in superconducting quantum circuits (see Sec.~\ref{sec:3A}), these platforms have also been used to explore quantum simulations~\cite{Georgescu2014}. With a quantum simulator, all regimes of coupling between a qubit and a resonator can be implemented in a fully tunable and efficient manner. In this respect, some proposals were put forward in the literature using superconducting circuits, which include the analog quantum simulation of the quantum Rabi model~\cite{daniel_prx,Pedernales15,Felicetti15,Plenio2015,Plenio2017}, Dirac equation physics~\cite{Pedernales13}, the digital-analog quantum simulation of the quantum Rabi model~\cite{Mezzacapo14}, and Dicke physics~\cite{Mezzacapo14,Lamata16}, as well as bosonic modes in the USC regime~\cite{Fedortchenko2017}. In this section, we give an overview of several of these proposals. Experimental realizations of the analog \cite{braumuller2016,Lv2018} and the digital-analog quantum simulation of the quantum Rabi model \cite{langford2016} have recently been carried out, as well as the USC regime of bosonic modes~\cite{Markovic2018}. In addition, an experimental realization of a classical simulation of the quantum Rabi model was performed in photonic chips~\cite{Crespi2012}. Moreover, an analysis of the quantum simulation of the Dicke model with cavity QED was proposed~\cite{PhysRevA.75.013804, PhysRevA.87.033814}, and an early experiment on Dicke physics in this platform was performed~\cite{Baumann2010}. We point out that Secs.~\ref{sec:4A}--\ref{sec:4C} analyze quantum simulations of USC and DSC models, while Sec.~\ref{sec:4D} deals with analog quantum simulations employing devices already in the USC and DSC regimes.

In Fig.~\ref{regionsFigure}, we summarize the different regimes of the QRM that are reproduced by an analog or a digital-analog quantum simulator, following \cite{Pedernales15}.

\begin{figure}[!hbt]
\centering
\includegraphics[width = 9cm]{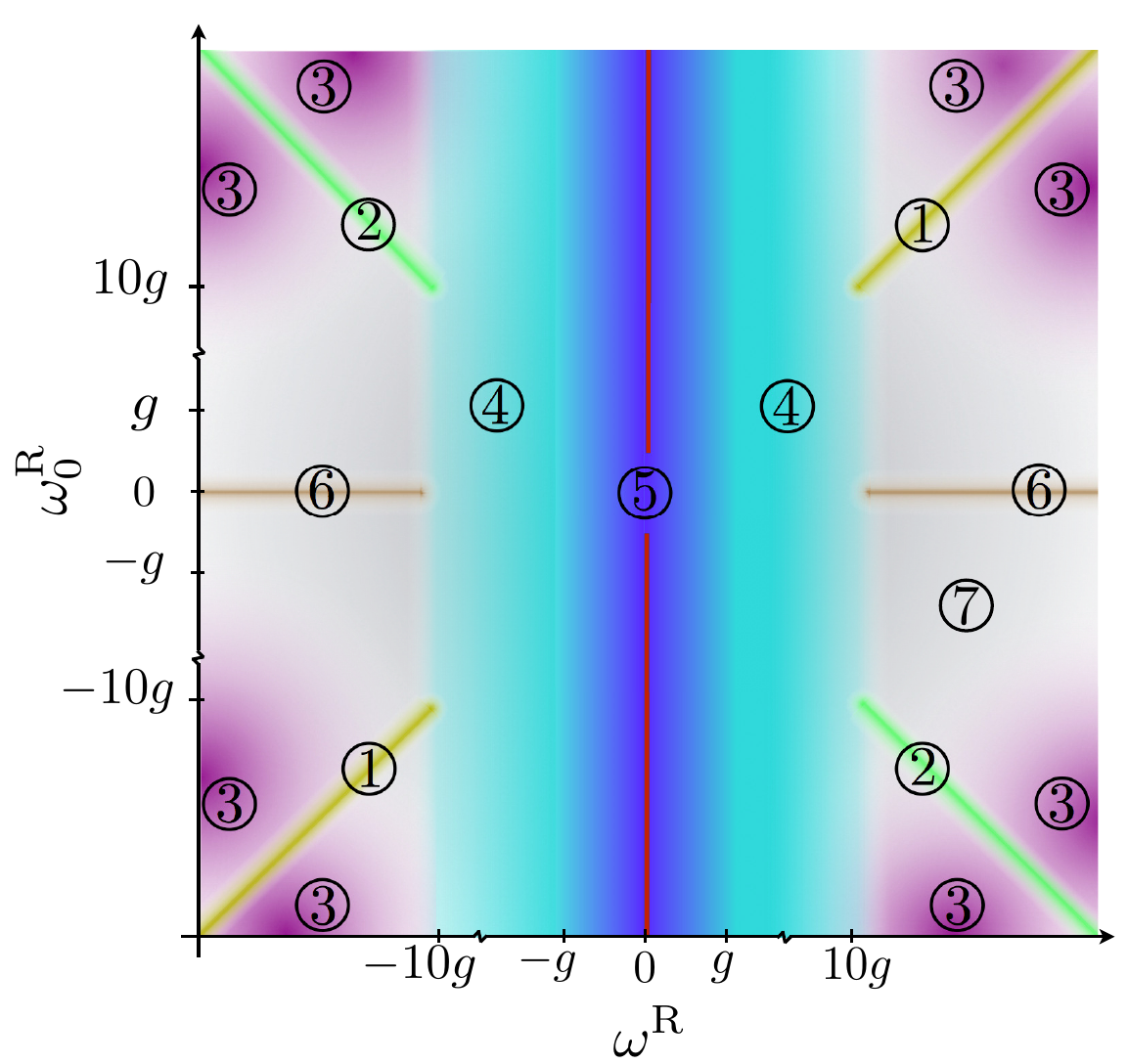}
\caption{(Color online) Different parameter regimes of the quantum Rabi model (QRM). Here $g$ is the light-matter coupling strength, $\omega^R$ represents the resonator frequency, and $\omega_0^R$ the qubit energy splitting, according to the QRM. (1) Jaynes-Cummings (JC) regime: $ g \ll \{|\omega^R|,|\omega_0^R| \}$ and $| \omega^R - \omega_0^R | \ll |\omega^R + \omega_0^R | $.
(2) Anti-JC regime: $ g \ll \{|\omega^R |,|\omega_0^R| \}$ and $|\omega^R - \omega_0^R| \gg | \omega^R + \omega_0^R | $. (3) Twofold dispersive regime: $ g < \{ | \omega^R | , | \omega_0^R | , | \omega^R - \omega_0^R |, |\omega^R + \omega_0^R | \}$. (4) USC regime: $| \omega^R | < 10 g $. (5) DSC regime: $ | \omega^R | < g $. (6) Decoupling regime: $ | \omega_0^R | \ll g \ll |\omega^R| $. (7) The intermediate regime ($ |\omega_0^R| \sim g \ll |\omega^R|$) is still open to analysis.
The (red) vertical central line corresponds to the regime of the Dirac equation. The colors indicate the different regimes of the QRM, color degradation denotes transitions between different regions. From~\cite{Pedernales15}.}\label{regionsFigure}
\end{figure}

\subsection{Analog quantum simulation of the quantum Rabi model}\label{sec:4A}

\subsubsection{Quantum Rabi model with superconducting circuits and the Jaynes-Cummings model}
\label{sec:4A1}

The first analog quantum simulation of the USC and DSC dynamics was proposed by \cite{daniel_prx}. The proposed simulator consists of a superconducting qubit coupled to a cavity mode in the strong coupling regime, with a two-tone orthogonal drive applied to the qubit. It was shown through analytical calculations and numerics that the method can access all regimes of light-matter coupling, including USC ($0.1\lesssim \! g/ \omega \lesssim \! 1$, with $g/ \omega$ the ratio of the coupling strength over the resonator frequency) and DSC~\cite{casanova2010} ($g/ \omega\gtrsim \! 1$). This scheme allows one to realize an analog quantum simulator for a wide range of light-matter coupling regimes~\cite{Braak2011} in platforms where those regimes are unattainable from first principles. This includes, among others, the simulation of Dirac equation physics, the Dicke and spin-boson models, the Kondo model, and the Jahn-Teller instability~\cite{meaney2010}. We use the language of circuit QED \cite{blais2004} to describe the method, although it can also be implemented in microwave cavity QED~\cite{Solano2003}.

Let us consider a physical system consisting of a superconducting qubit strongly coupled to a transmission line microwave resonator. Working at the qubit degeneracy point, the Hamiltonian reads \cite{blais_pra}
\begin{eqnarray}
\hat{\cal H} =  \frac{\hbar \Omega}{2} \hat{\sigma}_z + \hbar  \omega\hat{a}^\dag\hat{a}  -\hbar  g    \hat{\sigma}_x  (\hat{a} + \hat{a}^\dag) \label{HamilDiag}, 
\end{eqnarray}
where $\Omega$ is the qubit frequency, $\omega$ is the photon frequency, and $g$ denotes the coupling strength. Moreover, $\hat{a}$ and $\hat{a}^\dag$ stand for the annihilation and creation operators for the field mode of the photon, while $\hat{\sigma}_x = \hat{\sigma}_+ + \hat{\sigma}_- = \projsm{e}{g}+ \projsm{g}{e}$, $\hat{\sigma}_z = \projsm{e}{e}-\projsm{g}{g}$, where ${\ket{g},\ket{e}}$ denote ground and excited states of the superconducting qubit, respectively. One can apply the RWA in a typical circuit QED implementation to further simplify this Hamiltonian. More specifically~\cite{zueco_dispbs}, if $\{|\omega-\Omega|, g\} \ll\omega+\Omega$, then it can be expressed as
\begin{eqnarray}
\hat{\cal H} &=& \frac{\hbar \Omega}{2} \hat{\sigma}_z +\hbar  \omega\hat{a}^\dag\hat{a} -\hbar  g (\hat{\sigma}_+\hat{a} + \hat{\sigma}_-\hat{a}^\dag) ,\label{HamilRWA}
\end{eqnarray}
which is formally equivalent to the well-known JC model of cavity QED. By performing the RWA, one is neglecting counterrotating terms $\hat{\sigma}_-\hat{a}$ and $\hat{\sigma}_+\hat{a}^\dag$, producing in this way a Hamiltonian [Eq.~(\ref{HamilRWA})] where the number of excitations is conserved.

The Hamiltonian in Eq.~(\ref{HamilRWA}) is the basis for our derivations. Consider now two classical microwave fields driving the superconducting qubit. Adding the drivings to Eq.~(\ref{HamilRWA}) results in the following Hamiltonian:
\begin{multline}
\hat{\cal H} =  \frac{\hbar \Omega}{2} \hat{\sigma}_z +\hbar  \omega\hat{a}^\dag\hat{a} -\hbar  g (\hat{\sigma}_+\hat{a} + \hat{\sigma}_-\hat{a}^\dag)  \\  - \hbar \Omega_1  ( e^{i \omega_1 t} \hat{\sigma}_- + e^{-i \omega_1 t} \hat{\sigma}_+) - \hbar  \Omega_2 ( e^{i \omega_2 t} \hat{\sigma}_- + e^{-i \omega_2 t} \hat{\sigma}_+), \label{HamilDrivRabi}
\end{multline}
where $\omega_j$ and $\Omega_j$ denote the frequency and amplitude of the $j$th driving. We point out that the orthogonal drivings interact with the qubit in a similar manner as the microwave resonator field. To obtain Eq.~(\ref{HamilDrivRabi}), we assumed a RWA not only applied to the qubit-resonator coupling term, but also to the orthogonal drivings.

We then write Eq.~(\ref{HamilDrivRabi}) in a frame rotating with the first driving frequency $\omega_1$, namely,
\begin{multline}
\hat{\cal H}^{L_1} = \hbar  \frac{\Omega-\omega_1}{2} \hat{\sigma}_z +\hbar  (\omega-\omega_1) \hat{a}^\dag\hat{a} \\ - \hbar g  (\hat{\sigma}_+\hat{a} +\hat{\sigma}_-\hat{a}^\dag) - \hbar \Omega_1  ( \hat{\sigma}_- + \hat{\sigma}_+ ) \\- \hbar  \Omega_2   ( e^{i (\omega_2-\omega_1) t} \hat{\sigma}_- + e^{-i (\omega_2-\omega_1) t} \hat{\sigma}_+ ) .
\end{multline}
This transformation permits mapping the original first driving Hamiltonian into a time independent one $\hat{{\cal H}}_0^{L_1} = - \hbar \Omega_1  ( \hat{\sigma}_- + \hat{\sigma}_+) $, while leaving the number of excitations unperturbed. We consider this term to be the most sizable and treat the rest perturbatively by transforming into a rotating frame with respect to $\hat{\cal H}_0^{L_1} $, $\hat{\cal H}^{I} (t) = e^{i \hat{\cal H}_{0}^{L_1} t/\hbar } \pare{\hat{\cal H}^{L_1}   - \hat{\cal H}_0^{L_1} }    e^{-i \hat{\cal H}_{0}^{L_1} t/\hbar } $. By employing the rotated qubit basis $\ket{\pm} = \pare{\ket{g} \pm \ket{e} }/\sqrt2$, we obtain
\begin{multline}
\hat{\cal H}^{I} (t) = -\hbar\frac{\Omega-\omega_1}{2} \pare{ e^{-i 2  \Omega_1  t} \proj{+}{-} + {\rm h.c.}} \\
 + \hbar (\omega-\omega_1) \hat{a}^\dag\hat{a} - \frac{\hbar g }{2} \left( \left\{  \proj{+}{+} - \proj{-}{-} \right.\right.\\
 \left.\left. + e^{-i 2  \Omega_1 t} \proj{+}{-} - e^{i 2  \Omega_1  t}\proj{-}{+}  \right\}\hat{a} + {\rm h.c.} \right) \\
  -  \frac{\hbar \Omega_2 }{2} \left(   \left\{  \proj{+}{+} - \proj{-}{-} - e^{-i 2  \Omega_1  t} \proj{+}{-}  \right. \right.\\   \left. \left. + e^{i 2 \Omega_1  t}\proj{-}{+}  \right\} e^{i (\omega_2-\omega_1) t} + {\rm h.c.} \right). \label{HI1}
\end{multline}
The external driving parameters can be tuned in such a way that $\omega_1-\omega_2=2 \Omega_1$, allowing us to select the resonant terms in the time-dependent Hamiltonian. Therefore, if the first driving $ \Omega_1$ is relatively strong, one can approximate Eq.~(\ref{HI1}) by an effective Hamiltonian which is time independent as
\begin{eqnarray}
\hat{\cal H}_{\rm eff}
=  \hbar (\omega-\omega_1) \hat{a}^\dag\hat{a} + \frac{\hbar\Omega_2}{2} \hat{\sigma}_z   -  \frac{\hbar g}{2} \hat{\sigma}_x \pare{\hat{a}+\hat{a}^\dag} .  \label{HamilEffRabi}
\end{eqnarray}
Note the similarity between the original Hamiltonian (\ref{HamilDiag}) and Eq.~(\ref{HamilEffRabi}). Even though the coupling $g$ is fixed in Eq.~(\ref{HamilEffRabi}), one can still tailor the relative size of the rest of the parameters by tuning frequencies and amplitudes of the drivings. If one can reach $\Omega_2  \sim (\omega-\omega_1) \sim g/2$, the original system dynamics will emulate those of a qubit coupled to a bosonic mode with a relative coupling strength beyond the SC regime, reaching the USC and DSC regimes. The coupling strength attained with the effective Hamiltonian (\ref{HamilEffRabi}) can be estimated by the ratio $g_{\rm eff} / \omega_{\rm eff}$, where $g_{\rm eff} \equiv g/2$ and $ \omega_{\rm eff}\equiv\omega-\omega_1$.

\subsubsection{Quantum Rabi model in the Brillouin zone with ultracold atoms}\label{sec:4A2}

In the following, we present a technique to implement a quantum simulation of the QRM for unprecedented values of the coupling strength using a system of cold atoms freely moving in a periodic lattice. An effective two-level quantum system of frequency $\Omega$ can be simulated by the occupation of lattice Bloch bands, while a single bosonic mode is implemented with the oscillations of the atom in a harmonic optical trap of frequency $\omega$ that confines atoms within the lattice. We see that highly nontrivial dynamics may be feasibly implemented within the validity region of this quantum simulation.

At sufficiently low density, the dynamics of the neutral atoms loaded in an optical lattice can be described by the single-particle Hamiltonian
\[\hat{\cal H} =\frac{\hat{p}^{2}}{2m} + \frac{V}{2} \cos{\left( 4 k_{0} \hat{x} \right)} + \frac{m \omega^{2}}{2} \hat{x}^{2},\] 
where $\hat{p} = - i \hbar \partial/\partial x$, $m$ is the mass of the atom, $\omega$ is the frequency of the harmonic trap, while $V$ and $4k_{0}$ are the depth and wave vector of the periodic potential. Using the Bloch functions, we can identify a discrete quantum number, the band index $n_{b}$, and a continuous variable, the atomic quasimomentum $q$. Fixing our attention to the bands with the two lowest $n_b$, the Hamiltonian can be recast into
\begin{equation}
\begin{split}
\hat{\cal H} =& \frac{1}{2m} \begin{pmatrix} 
q^{2}+4\hbar k_{0} q & 0 \\
0 & q^{2} - 4\hbar k_{0} q
\end{pmatrix}
+ \frac{V}{4} \begin{pmatrix} 
0 & 1 \\
1 & 0
\end{pmatrix} \\
&- \frac{m \omega^{2} \hbar^{2}}{2} \frac{\partial^{2}}{\partial q^{2}} \begin{pmatrix} 
1 & 0 \\
0 & 1
\end{pmatrix}.
\end{split}
\end{equation}
By analogy to the usual QRM, 
\[\hat{\cal H} = \hbar \omega \hat{a}^{\dagger} \hat{a} + \frac{\hbar \Omega}{2} \sigma_{z} + i \hbar g \sigma_{x}\left( \hat{a}^{\dagger} - \hat{a} \right),\] 
we define an effective qubit energy spacing $\Omega \equiv V/2 \hbar$ and an effective light-matter interaction $g \equiv 2k_{0} \sqrt{\hbar \omega/2m}$.

The value of the effective coupling strength is intrinsically linked to the trap frequency $g \sim \sqrt{\omega}$, and since the trap frequency is low (typically kilohertz in actual experiments) the ratio $g/\omega$ is tunable only over a range of extremely high values, $g/\omega \sim 10$. However, the tunability of the ratio $g/\Omega$ allows us to explore a large region of parameters at the transition between resonant and dispersive qubit-oscillator regimes. Indeed, the value of $\Omega$ can be made large enough such that the qubit free Hamiltonian becomes the dominant term or small enough to make its energy contribution negligible.

Given that only very high values of the ratio $g/\omega$ are accessible, the RWA can never be applied and the model cannot be implemented in the JC limit. Interesting dynamics at the crossover between the dispersive and resonant DSC regimes can be observed for values of parameters unattainable so far with available implementations of the QRM. However, the analogy with the QRM breaks down when the value of the simulated momentum exceeds the borders of the first Brillouin zone. When this is the case, the model represents a generalization of the QRM in periodic phase space.

Both the momentum (and correspondingly the state of $\hat{\sigma}_x$) and the atomic cloud position can in principle be measured with absorption imaging techniques. For the former, standard time-of-flight imaging may be used, as performed by simultaneously deactivating both the lattice beams and the dipole trapping potential and then detecting the atoms in the far field after a given free expansion time. While the reconstruction in this way is possible with high precision, achieving the required spatial resolution for an in situ position detection of the oscillation is experimentally challenging. Figure~\ref{fig:coldatom1} shows experimentally accessible quantities like the distribution $\mathcal{P}(p) = | \langle p | \psi(t) \rangle  |^2$ of the atomic physical momentum $\hat{p}$ for different evolution times. The momentum distribution can be experimentally obtained using time-of-flight measurements and gives a clear picture of the system dynamics during the quantum simulation of the QRM. The cloud is initialized in the momentum eigenstate $|q=0\rangle |n_{b}=0\rangle$. When the periodic lattice strength $V$ is large enough, the dynamics are dominated by the coupling between $|n_{b}=0\rangle$ and $|n_{b}=1\rangle$. This case corresponds to the dispersive DSC regime. Otherwise, the dynamics are dominated by the harmonic potential, and the evolution resembles the QRM in the DSC regime.

An alternative implementation of the QRM with cold atoms has been proposed using atomic Zeeman states and vibrational modes of a trapping atomic potential. The coupling is mediated by a suitable fictitious magnetic field pattern and allows accessing a wide parameter regime of the QRM \cite{Schneeweiss2018}.

\begin{figure}[!hbt]
\centering
\includegraphics[width = 7cm]{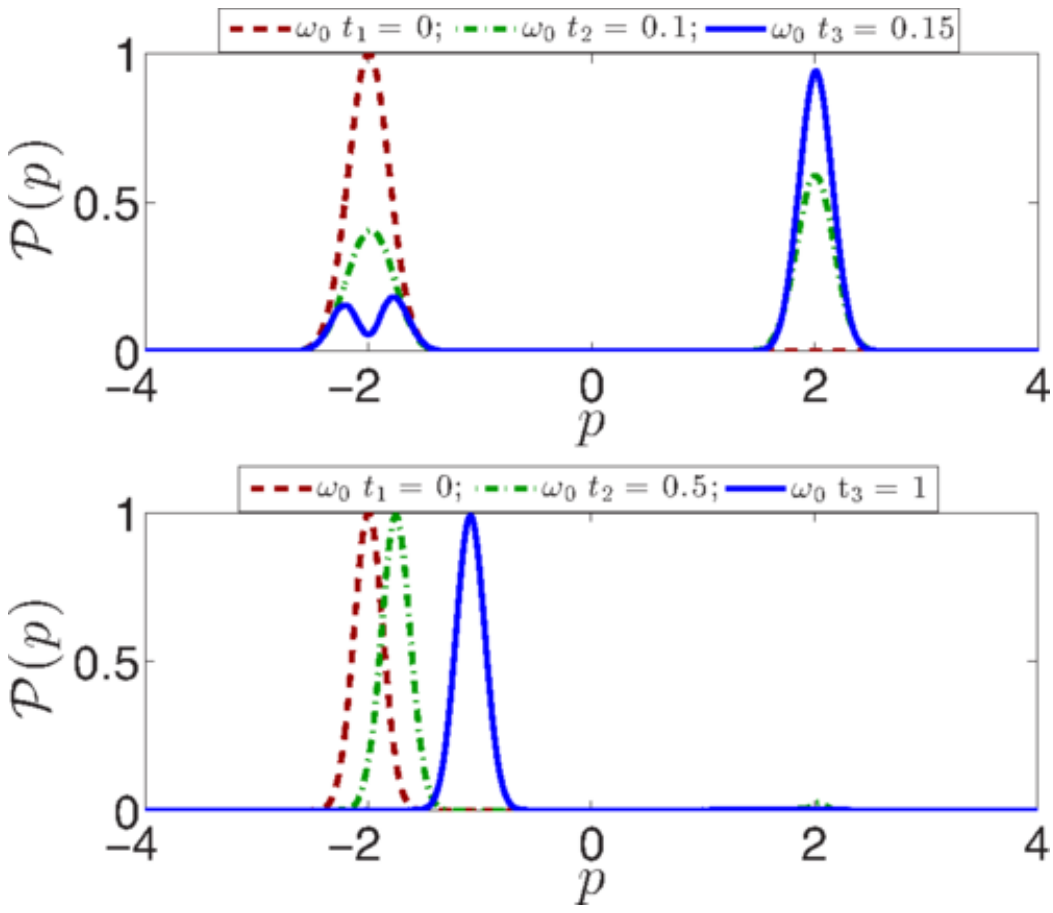}
\caption{\label{fig:coldatom1} (Color online) Quantum Rabi model using ultracold atoms. The distribution $\mathcal{P}(p) = | \langle p | \psi(t) \rangle  |^2$ of the atomic physical momentum $\hat{p}$ for different evolution times is shown. $\omega_0$ corresponds to $\omega$ of the main text. For the dispersive DSC regime (upper panel), the parameters are given by $g/\omega = 7.7$ and $g/\Omega = 0.43$. In this case, the initial wave function is transformed back and forth between two distributions centered around the states $|p=\pm 2 \hbar k_0 \rangle$. For the resonant DSC regime (lower panel), $g/\omega = 10$ and $\omega = \Omega$. In this case, the system is continuously displaced in momentum space up to a maximum value of the momentum. From~\cite{felicetti2017}.}
\end{figure}

\subsection{Analog quantum simulation of Dirac physics}\label{sec:4B}

There exist strong connections between the QRM and the Dirac equation~\cite{Pedernales13, LamataDirac2007}. Therefore, simulating the physics of the Dirac equation is important to connect to physics of the USC and DSC regimes. We review here a particular method employing superconducting quantum circuits. We point out some crucial differences with regards to previous implementations of the Dirac equation Klein paradox in other quantum platforms, particularly ion traps \cite{Gerritsma2010}. Using the method described here, the dynamics of a spin-1/2 relativistic particle are emulated by 2 interacting degrees of freedom from two different subsystems, namely, a standing wave in a transmission line resonator and a superconducting qubit, none of them representing real motion. The position and momentum of the simulated Dirac particle are codified in the field quadratures. Contrary to the ion trap simulator \cite{Gerritsma2010}, this approach paves the way for combining cavity fields with quantum propagating microwaves \cite{edwin_dual-path,Bozyigit,Itinerant} in complex quantum network architectures \cite{Leib12}.

In the protocol described here one requires a superconducting qubit, e.g., a flux qubit~\cite{floor_prl}, working at its degeneracy point strongly coupled to an electromagnetic field mode of a transmission line resonator. The interaction between the two systems can be described by the JC Hamiltonian \cite{jaynes1963,blais_pra,wallraff2004}. Additionally, we consider three classical external microwave drivings, two of them transversal to the resonator \cite{daniel_prx} which will couple only to the qubit, and the third drive coupled longitudinally to the resonator. The Hamiltonian of the system reads
\begin{multline}
\label{HamilDriv} 
\hat{\cal H} =  \frac{\hbar \Omega}{2} \hat{\sigma}_z +\hbar  \omega\hat{a}^\dag\hat{a} -\hbar  g \pare{\hat{\sigma}_+\hat{a} + \hat{\sigma}_-\hat{a}^\dag}  \\ 
-\hbar \Omega_1  \pare{ e^{ i \pare{ \omega t +\varphi} } \hat{\sigma}_- + e^{-i \pare{ \omega t +\varphi } } \hat{\sigma}_+} - \hbar  \lambda   \left( e^{i  \pare{ \nu t +\varphi }} \hat{\sigma}_- \right. \\+ \left. e^{-i  \pare{ \nu t +\varphi } } \hat{\sigma}_+\right) + \hbar \xi \pare{  e^{i \omega t}\hat{a} + e^{-i \omega t}\hat{a}^\dag  },
\end{multline}
where $\hat{\sigma}_y=i (\hat{\sigma}_- -  \hat{\sigma}_+)=i (\proj{g}{e} -  \proj{e}{g})$ and $\hat{\sigma}_z= \proj{e}{e} -  \proj{g}{g}$, with $\ket{g}$, $\ket{e}$ denoting the ground and excited qubit states, respectively. Here $\hbar\omega$ and $\hbar\Omega$ correspond to photon and qubit uncoupled energies, whereas $g$ stands for the qubit-photon coupling strength. The two orthogonal microwave drivings have amplitudes $\Omega_1$, $\lambda$, phase $\varphi$, and frequencies $\omega$ and $\nu$. Additionally, the longitudinal driving has amplitude $\xi$ and frequency $\omega$. Note that two of the drivings are chosen to be resonant with the resonator mode. We also assume that $\Omega=\omega$, i.e., the qubit and the resonator are on resonance as well.

This protocol is based on two transformations. First, the Hamiltonian in Eq.~(\ref{HamilDriv}) can be transformed into the rotating frame with respect to the resonator frequency $\omega$:
\begin{multline}
\label{HL1}
\hat{\cal H}^{L_1} =  - \hbar g  \pare{\hat{\sigma}_+ \hat{a} +\hat{\sigma}_- \hat{a}^\dag} \\ - \hbar \Omega_1 \pare{ e^{ i \varphi } \hat{\sigma}_- + e^{-i \varphi } \hat{\sigma}_+} + \hbar \xi \pare{\hat{a} + \hat{a}^\dag} \\ - \hbar  \lambda   \pare{ e^{i \left[ (\nu-\omega) t + \varphi \right] } \hat{\sigma}_- + e^{-i \left[ (\nu-\omega) t  +  \varphi \right]  } \hat{\sigma}_+ }.
\end{multline}

Second, the Hamiltonian obtained is transformed into another frame rotating with respect to the Hamiltonian $\hat{\cal H}_0^{L_1} = - \hbar \Omega_1   \pare{ e^{ i \varphi } \hat{\sigma}_- + e^{-i \varphi } \hat{\sigma}_+}$,
\begin{multline}
\hat{\cal H}^{I} = -  \frac{\hbar g }{2} \left( \left\{  \proj{+}{+} - \proj{-}{-} + e^{-i 2  \Omega_1 t} \proj{+}{-} \right. \right.\\
-  \left.\left. e^{i 2  \Omega_1  t}\proj{-}{+}  \right\} e^{i \varphi} \hat{a} + {\rm h.c.} \right) \\  
  -  \frac{\hbar \lambda}{2} \Big(   \left\{  \proj{+}{+} - \proj{-}{-} - e^{-i 2  \Omega_1  t} \proj{+}{-}  \right.  \\ \left.  + e^{i 2 \Omega_1  t}\proj{-}{+}  \right\} e^{i (\nu-\omega) t} + {\rm h.c.} \Big)   + \hbar \xi \pare{ \hat{a} + \hat{a}^\dag}   , \label{HI12} 
  \end{multline}
where we considered the rotated qubit basis $\ket{\pm} = \pare{\ket{g} \pm e^{-i \varphi} \ket{e} }/\sqrt2$. We now assume $\omega-\nu=2 \Omega_1$ to simplify the calculation, and also assume the first driving amplitude $\Omega_1$ to be large when compared to the other Rabi frequencies in Eq.~(\ref{HI12}). Therefore, we can apply the RWA, which produces the Hamiltonian
\begin{eqnarray}
\hat{{\cal H}}_{\rm eff}
=  \frac{\hbar\lambda}{2} \hat{\sigma}_z   +  \frac{\hbar g}{\sqrt2} \hat{\sigma}_y \hat{p} + \hbar \xi \sqrt2 \, \hat{x} ,  \label{HamilEff}
\end{eqnarray}
where $\varphi=\pi/2$ and we made use of the electromagnetic field quadratures, i.e. $\hat{x} =(\hat{a}+\hat{a}^\dag)/\sqrt2$, $\hat{p} =-i(\hat{a}-\hat{a}^\dag)/\sqrt2$, obeying the commutation relation $\left[ \hat{x} , \hat{p} \right] = i$. Note that $\Omega_1$ is not present in the effective Hamiltonian equation~(\ref{HamilEff}). This is a consequence of deriving the Hamiltonian in a rotating frame with $\Omega_1$ acting as a large frequency in the strong driving parameter regime.

The Schr\"odinger dynamics of Eq.~(\ref{HamilEff}) are analogous to those of the 1+1 Dirac equation, where the parameters $\hbar g/\sqrt2$ and $\hbar \lambda / 2$ simulate, respectively, the speed of light and the particle mass. Moreover, we also have an external potential $ \Phi = \hbar \xi \sqrt2 \, \hat{x}$ which is linear in the particle position. The simulated dynamics allow one to cover a wide range of physical regimes within this quantum simulation. We point out that, for fixed coupling constant $g$, the simulated mass grows linearly with the amplitude of the weak driving $\lambda$, while the strength of the potential can be adjusted with the longitudinal driving amplitude $\xi$. This is in contrast with respect to the trapped ion implementation, where one needs a second ion to simulate the external potential \cite{Casanova2010b,Gerritsma2011}. In the case of a massless particle, $\lambda = 0$ and $\nu = 0$, such that $\omega = 2 \Omega_1$ in Eq.~(\ref{HI12}). 

In the superconducting quantum circuit implementation, the analysis of relativistic quantum features, such as {\it Zitterbewegung} or Klein paradox, should be carried out by a phase-space description of the electromagnetic field in the transmission line resonator. The initial quantum state of the bosonic degree of freedom of the simulated Dirac particle may be represented by a wave packet with average position $\langle  \hat{x}_0 \rangle$ and average momentum $\langle  \hat{p}_0  \rangle$,
\begin{eqnarray}
\psi (x) =  \pi^{-1/4} \exp\left\{ i \langle  \hat{p}_0  \rangle  x  \right\} \,\,    \exp \left\{ -   \frac{ ( x  -  \langle  \hat{x}_0  \rangle )^2 }{2} \right\}  .    \label{psi}
\end{eqnarray}
The wave packet is analogous to the $x$-quadrature representation of an electromagnetic field coherent state \[\ket{ \frac{ \langle  \hat{x}_0  \rangle + i \langle  \hat{p}_0  \rangle }{\sqrt2} } = \hat{{\cal D}} \pare{ \frac{ \langle  \hat{x}_0  \rangle + i \langle  \hat{p}_0  \rangle }{\sqrt2}  } \ket{0},\]
where $\ket{0}$ is the vacuum state of the bosonic field, and $\hat{{\cal D}} ( \alpha ) = \exp \left\{  \alpha \hat{a}^\dag  -  \alpha^* \hat{a}   \right\} $ is the displacement operator.

\subsection{Digital-analog quantum simulation of the quantum Rabi and Dicke models}\label{sec:4C}

The previous Secs.~\ref{sec:4A} and \ref{sec:4B} described analog simulations of different physical models. We now review the digital-analog quantum simulation of the quantum Rabi and Dicke models implemented in a circuit quantum electrodynamics platform. The simulation employs only JC dynamics and local interactions~\cite{Mezzacapo14,Lamata16}. We describe how the rotating and counterrotating Hamiltonians of the corresponding evolution can be straightforwardly implemented using digital techniques. By interleaving the dynamics of rotating and counterrotating Hamiltonians, the evolution of the quantum Rabi and Dicke models can be implemented in all parameter regimes of light-matter coupling. At the end of this section, we illustrate how a Dirac equation evolution can be achieved in the limit of negligible mode frequency.

We begin by assuming a generic circuit quantum electrodynamics platform composed of a superconducting qubit coupled to a transmission line microwave resonator. This scenario is described by the Hamiltonian~\cite{blais_pra} 
\begin{equation}
\hat{\cal H} = \hbar\omega_r \hat{a}^{\dagger}\hat{a} +\frac{\hbar\omega_q}{2}\hat{\sigma}_z +\hbar g(\hat{a}^{\dagger}\hat{\sigma}_-+\hat{a}\hat{\sigma}_+),\label{QubitResHam}
\end{equation}
where $\omega_r$ and $\omega_q$ are, respectively, the resonator and qubit transition frequencies, $g$ is the qubit-cavity coupling strength, $\hat{a}^{\dagger}$ is the creation bosonic operator for the cavity mode, and $\hat{\sigma}_+,\hat{\sigma}_-$ are raising and lowering spin operators acting on the qubit. 

Let us take a look at the Hamiltonian of the QRM
\begin{equation}
\hat{\cal H}_R=\hbar\omega^R_r \hat{a}^{\dagger}\hat{a} +\frac{\hbar\omega^R_q}{2}\hat{\sigma}_z +\hbar g_R\hat{\sigma}_x(\hat{a}^{\dagger}+\hat{a})\label{RabiHam}.
\end{equation}
It turns out that its evolution can be codified in a superconducting qubit platform with available JC interactions [Eq.~(\ref{QubitResHam})] by a digital decomposition. 
\begin{figure}
\includegraphics[scale=0.26]{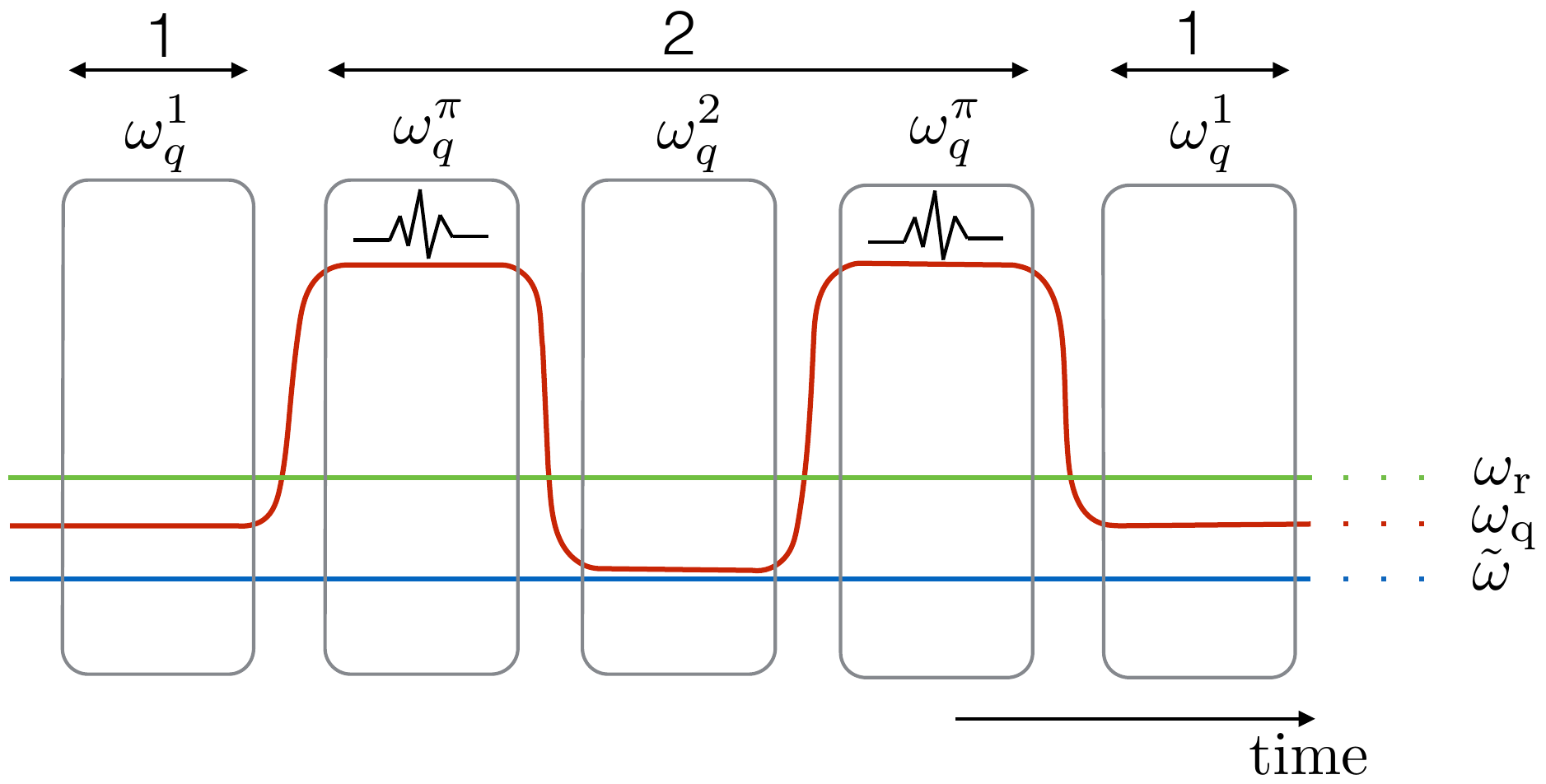}
\caption{(Color online) Frequency diagram of the digital-analog implementation of the quantum Rabi Hamiltonian. A superconducting qubit of frequency $\omega_q$ interacts with a microwave resonator with transition frequency $\omega_r$. The evolution with $\hat{\cal H}_{1,2}$ in Eqs.~(\ref{Ham11}), and (\ref{Ham22}) are implemented, respectively, with a Jaynes-Cummings interaction (step 1), and other Jaynes-Cummings dynamics with a different detuning, interspersed with $\pi$ pulses (step 2), to transform the second Jaynes-Cummings evolution into an anti-Jaynes-Cummings interaction. From~\cite{Mezzacapo14}. \label{FrequencyScheme}} 
\end{figure}
Let us express Eq.~(\ref{RabiHam}) as the sum of two parts, $\hat{\cal H}_R=\hat{\cal H}_1+\hat{\cal H}_2$, with
\begin{align}
\hat{\cal H}_1 &=\frac{\hbar\omega^R_r}{2}\hat{a}^{\dagger}\hat{a} +\frac{\hbar\omega^1_q}{2}\hat{\sigma}_z +\hbar g(\hat{a}^{\dagger}\hat{\sigma}_-+\hat{a}\hat{\sigma}_+) , \label{Ham11}  \\
\hat{\cal H}_2 &=\frac{\hbar\omega^R_r}{2} \hat{a}^{\dagger}\hat{a} -\frac{\hbar\omega^2_q}{2}\hat{\sigma}_z +\hbar g(\hat{a}^{\dagger}\hat{\sigma}_++\hat{a}\hat{\sigma}_-),
\label{Ham22}
\end{align}
where we considered the qubit frequency in the two terms in such a way that $\omega_q^1-\omega_q^2=\omega^R_q$. The dynamics arising from the two Hamiltonians in Eqs.~(\ref{Ham11}), and (\ref{Ham22}) can be implemented in a standard circuit quantum electrodynamics platform that includes the possibility of fast detuning of the qubit frequency; see Fig.~\ref{FrequencyScheme}. Beginning with the qubit-resonator Hamiltonian in Eq.~(\ref{QubitResHam}), we can transform into a frame which rotates at frequency $\tilde{\omega}$, where an effective interaction Hamiltonian results 
\begin{equation}
\tilde{\cal H}=\hbar\tilde{\Delta}_r\hat{a}^{\dagger}\hat{a}+\hbar\tilde{\Delta}_q\hat{\sigma}_z+\hbar g(\hat{a}^{\dagger}\hat{\sigma}_-+\hat{a}\hat{\sigma}_+),\label{IntHam}
\end{equation}  
with $\tilde{\Delta}_r=\omega_r-\tilde{\omega}$ and $\tilde{\Delta}_q=\left(\omega_q-\tilde{\omega}\right)/2$. Accordingly, Eq.~(\ref{IntHam}) coincides with $\hat{{\cal H}}_1$ after redefinition of the coefficients.
The counterrotating Hamiltonian $\hat{{\cal H}}_2$ can be realized by local qubit drivings to $\hat{\tilde{\cal H}}$, employing a different detuning for the qubit frequency,
\begin{equation}
e^{-i \pi\hat{\sigma}_x/2}\tilde{\cal H}e^{i \pi\hat{\sigma}_x/2}=\hbar\tilde{\Delta}_r\hat{a}^{\dagger}\hat{a}-\hbar\tilde{\Delta}_q\hat{\sigma}_z+\hbar g(\hat{a}^{\dagger}\hat{\sigma}_++\hat{a}\hat{\sigma}_-).\label{RotHam}
\end{equation}
By choosing different qubit-resonator detunings in the two steps $\tilde{\Delta}^1_q$ and $\tilde{\Delta}^2_q$, the quantum Rabi Hamiltonian [Eq.~(\ref{RabiHam})] is simulated by a digital expansion~\cite{Lloyd96} by interleaving the different interactions. 

In the protocol described here, customary quasiresonant JC dynamics with different qubit frequencies are combined with single-qubit drivings to perform standard qubit rotations~\cite{blais_pra}. This sequence is repeated following the digital quantum simulation scheme in order to achieve a better fidelity of the quantum Rabi dynamics.

Note the existence of a direct relationship between the effective system parameters and the real circuit variables. The simulated bosonic mode frequency is related to the resonator detuning $\omega_r^R=2\tilde{\Delta}_r$, while the effective two-level system frequency is connected to the qubit frequency considering the two steps $\omega_q^R/2=\tilde{\Delta}_q^1-\tilde{\Delta}_q^2$. Finally, the qubit-resonator coupling strength is the same in both cases $g_R=g$.  

This digital-analog quantum simulation was carried out in a circuit QED experiment~\cite{langford2016}.

\subsection{Quantum simulation with ultrastrong couplings}\label{sec:4D}

In this section, we analyze analog quantum simulator devices in the USC and DSC regimes which are used to study complex phenomena occurring in real systems, such as biologically relevant molecular complexes. This should not be confused with Sec.~\ref{sec:4A1}, dealing with quantum simulations of models in USC and DSC regimes employing superconducting quantum simulators in the strong coupling regime.

\subsubsection{Jahn-Teller transitions in molecules}

Jahn-Teller models describe the interaction of localized electronic states with vibrational modes in crystals or in molecules \cite{bersuker}. Certain molecules contain a degeneracy in their ground state due to their molecular configuration. A spontaneous symmetry breaking of the geometry of the molecule, a process known as a Jahn-Teller transition, results in one favorable stable configuration, becoming the absolute ground state of the system. Interesting molecular systems undergoing a Jahn-Teller transition exist, e.g. fullerene. Therefore, simulating such quantum systems is very attractive.

In a pioneering work \cite{Hines2004}, a connection was made between a class of Jahn-Teller Hamiltonians and a qubit coupled to an oscillator in the USC regime. This initial work was followed by several extensions into other classes of Jahn-Teller models and how to efficiently simulate them using superconducting quantum circuits \cite{meaney2010, dereli2012}. 

Following the original work~\cite{Hines2004, larson2008}, the most general Hamiltonian of a $E\times\epsilon$ Jahn-Teller model implemented in a cavity QED setting using a single two-level system coupled to two degenerate modes of a cavity has the form
\begin{multline}
\hat{\mathcal{H}}_{\epsilon\times E}/\hbar = \omega_c(\hat{a}^{\dag}\hat{a}+\hat{b}^{\dag}\hat{b})+\frac{\Omega_q}{2}\hat{\sigma}_z +\\ \lambda[(\hat{a}^{\dag}+\hat{a})(\hat{\sigma}_+e^{-i\theta}+\hat{\sigma}_-e^{i\theta}) +\\ (\hat{b}^{\dag} + \hat{b})(\hat{\sigma}_+e^{-i\phi} + \hat{\sigma}_-e^{i\phi})].
\end{multline}
Here $\omega_c$ is the frequency of the two cavity modes. $\theta$ and $\phi$ represent different phases of the mode field interacting with the two-level system. $\lambda$ is the interaction strength between the two-level system and each cavity mode. This Hamiltonian has a strong resemblance to the QRM [Eq.~(\ref{QRH})], where the only difference is the presence of the second mode $\hat{b}$. The Jahn-Teller transition occurs for values of the qubit-oscillator coupling strengths which correspond to the DSC regime. Such a regime has recently been attained unambiguously in a superconducting circuit \cite{yoshihara2017a}, as detailed in Sec.~\ref{sec:3A}. The $\epsilon\times E$ Jahn-Teller model is the simplest of its kind. More complex models, and thus more realistic, contain several oscillator modes, with a hopping interaction between those modes. The simplest of such multimode models is the $E\times(\beta_1+\beta_2)$ Jahn-Teller model, also known as the Herzberg-Teller model. \cite{dereli2012} studied the behavior of two coupled modes interacting with the same qubit. Its implementation in a superconducting circuit is presented in Fig.~\ref{fig:JT}. The Hamiltonian of such a system can be expressed as
\begin{multline}
\hat{\mathcal{H}}_{(\beta_1+\beta_2)\times E}/\hbar = \frac{\Omega_q}{2}\hat{\sigma}_z + \Omega_1\hat{a}_1^{\dag}\hat{a}_1+\Omega_2 \hat{a}_2^{\dag}\hat{a}_2 +\\ \left[g_1(\hat{a}_1+\hat{a}_1^{\dag})+ g_2(\hat{a}_2+\hat{a}_2^{\dag})\right]\hat{\sigma}_x  + J(\hat{a}_1^{\dag}\hat{a}_2 + \hat{a}_2^{\dag}\hat{a}_1),
\end{multline}
where $J$ is the mode-mode coupling energy representing the hopping rate of phonons in the simulated system. $g_i$ are the qubit-mode interaction strength coefficients, representing the coupling of a molecular transition to each of the two vibrational modes of the simulated molecule. This type of Hamiltonian can be realized using the technology of superconducting quantum circuits. This simple Hamiltonian already contains the physics of real systems of interest such as the two phonon modes in C$_6$H$_6$$^{\pm}$ and the two phonon modes of Fe$^{2+}$ in ZnS. 

More complex Jahn-Teller models involve the interaction of a qubit to several bosonic modes. A possible candidate to perform an analog simulation would correspond to a qubit ultrastrongly coupled to a coplanar waveguide resonator supporting a collection of modes. By reducing the fundamental mode frequency of the resonator the qubit can simultaneously interact to many modes. Experiments have already been performed using superconducting qubit circuits where such a configuration has been engineered \cite{sundaresan2015, puertas2018}. 

Irrespective of the simulated type of Jahn-Teller model, it is crucial to attain ultrastrong couplings between the two-level system, or qubit, and the bosonic modes involved in order to perform a faithful analog simulation of the actual molecular system.

\begin{figure}
\includegraphics[width = \columnwidth]{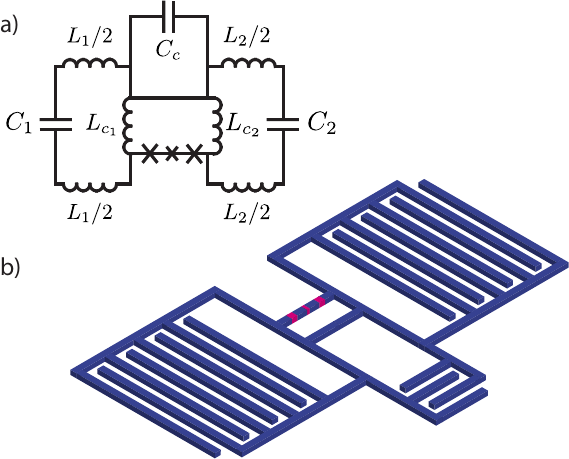}
\caption{\label{fig:JT} Circuit schematic to produce the Jahn-Teller $E\times(\beta_1 + \beta_2)$ model. (a) Circuit diagram of a flux qubit galvanically coupled to two lumped-element resonators, which are capacitively coupled to each other. (b) Circuit representation with the capacitors being interdigitated-finger style. The magenta sections represent the qubit Josephson junctions. From~\cite{dereli2012}.}
\end{figure}

\subsubsection{Energy transfer in photosynthetic complexes}

The transfer of energy in light harvesting systems has been a subject of intense study in the last decade. The observation of excitonic quantum oscillations in molecular complexes as a result of light absorption triggered the birth of a field known now as quantum biology \cite{Engel2007, Lambert2013}. 

Biological systems are inherently complex and particularly hard to describe quantitatively, especially considering the fact that key biological processes, in this case the transfer of energy within the molecular complex, are heavily influenced by the environmental fluctuations and the finite temperature. Therefore, a quantum simulator that aims at simulating such relevant processes needs to include the environmental degrees of freedom. As measured in spectroscopic experiments \cite{Wendling2000}, molecular complexes consist of several nodes which are coupled to each other in a particular network configuration. The most popular of light-harvesting complexes, the Fenna-Matthews-Olson (FMO) complex, contains seven nodes, and the interaction between nodes is in fact ultrastrong. In addition, the correlation time of the bath is found to be of comparable order as the internal dynamics of the molecule. In other words, the system is heavily non-Markovian. The strong effect of the environment is due to an USC of the nodes within the molecular complex to its environmental degrees of freedom, most likely phonons in the case of FMO. 

An analog quantum simulator must then consist of qubits playing the role of the FMO nodes which couple to each other ultrastrongly, with some of the qubits ultrastrongly coupled to the environment. Ultrastrong qubit-qubit interactions are relatively easy to obtain using superconducting circuits \cite{Majer2005}, while ultrastrong qubit-bath interactions have just recently been achieved in experiments using superconducting qubits in transmission lines \cite{forn-diaz2017}. A theoretical proposal of such a quantum simulator was already put forward \cite{Mostame2012} using superconducting flux qubits. Figure~\ref{fig:FMO} shows a schematic of the qubit network proposed to mimic that of the actual FMO complex, along with a circuit representation of the qubit-environment coupling. The interaction to the flux qubit is longitudinal to simulate ultrastrong dephasing. 
\begin{figure}[!hbt]
\centering
\includegraphics{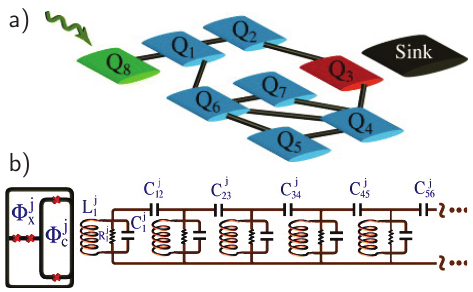}
\caption{\label{fig:FMO} a) Experimental layout for simulating the exciton dynamics and environment-assisted quantum transport in the FMO complex. The different nodes $Q_i$ represent eight qubits emulating the FMO nodes and their connections. The green node $Q_8$ is the receiver of the photon excitation, blue nodes are intermediate paths of the exciton, and node $Q_3$ is the final one where the energy is delivered and couples to the rest of the molecule, labeled as ``Sink". b) Circuit schematic representation of a single qubit coupled to an Ohmic environment. In this circuit, the coupling is longitudinal with respect to the qubit, simulating in this way ultrastrong dephasing. The environment can be simulated by a linear chain of $LCR$ resonators, as in this figure, or by using a transmission line, as demonstrated experimentally \cite{forn-diaz2017}. From~\cite{Mostame2012}.}
\end{figure}

Two recent experiments \cite{Gorman2018, Potocnik2018} have reproduced certain aspects of the basic physics believed to occur in light-harvesting complexes. Poto\v{c}nik \emph{et al.} studied the interplay of quantum interference and environmental fluctuations to lead to a maximal energy transfer in a system of three superconducting qubits. The qubits were directly coupled to each other and subject to different types of environmental noise. They found a maximal efficiency of energy transfer when the qubits were experiencing coherent excitation and Lorentzian noise, conditions which mimic the phononic environment found in molecular complexes such as FMO. Gorman \emph{et al.} used a system of two ions in a linear trap, one of which was coupled to one of its vibrational modes, playing the role of the phononic environment in an actual light-harvesting complex. By application of external laser beams, they simulated the regime where the relative energy splitting between the ions, their interaction strength and the interaction to the environment were of the same order. This regime mimics a realistic environment in a molecular complex such as the FMO. They observed clear signatures of environment-assisted energy transfer between ions, supporting the idea that this process does play an important role in the actual energy transfer of real photosynthetic molecular complexes.

Scaling up the system size of these experiments with more qubits and more realistic parameters, some of which require entering the USC regime, may lead to actual quantum simulations of biological complexes and quantum chemistry.

\section{Physics of the ultrastrong coupling regime}\label{sec:5}

In this section, we review some of the intrinsic physics occurring in the USC regime, and what kind of applications have been proposed for ultrastrongly coupled systems. First, we present several instances in which novel quantum optical phenomena are possible in the USC regime and how they could be useful for quantum information processing purposes. We continue with an important application in quantum computing as is the generation of ultrafast quantum gates. The section closes with a description of how dissipative systems must be treated in the USC regime.  

\subsection{Quantum optics}\label{sec:5A}

The achievement of ultrastrong couplings in any physical platform opens up the possibility to study counterintuitive phenomena appearing in the Rabi model which is not present in the more familiar JC model~\cite{RidolfoetAl12PRL,StassietAl13PRL,felicetti2014,garziano2014,garziano2015,ma2015three}. Beyond the instances described in this section, concepts appearing in other branches of physics are also being studied in the USC regime, such as symmetry breaking and Higgs mechanism~\cite{garziano2014} and approaches relating to Feynman diagrams~\cite{1367-2630-19-5-053010}. 

\subsubsection{Two atoms excited by a single photon}

A particular instance is the case of a photon which excites two atoms at the same time in a reversible manner~\cite{garziano2016}. In a generalized version of the Rabi model, two two-level atoms interact with a single mode of a cavity [see also Eq.~(\ref{eq:Hfq})], given by the Hamiltonian 
\begin{multline}
\label{eq:garziano}
\hat{\cal H}=\frac{\hbar\Omega}{2} \sum_{i} \hat{\sigma}_{z}^{(i)} + \hbar\omega \hat{a}^{\dagger}\hat{a} \\ +\hbar g \left( \hat{a}+\hat{a}^{\dagger}\right) \sum_{i} \left[ \cos{\left( \theta \right)} \hat{\sigma}^{(i)}_{x}+ \sin{\left( \theta \right)} \hat{\sigma}^{(i)}_{z} \right].
\end{multline} 
As shown in Figs.~\ref{fig:garziano_levels} and~\ref{fig:garziano_spectrum}, a mixing exists in third-order perturbation theory between states $|g,g,1\rangle$ and $|e,e,0\rangle$ due to the counterrotating terms. At the resonance point where the frequency of the cavity is twice the frequency of each atom, the effective Hamiltonian is given by $\hat{\cal H}_{\rm eff} = - \hbar\Omega_{\rm eff} \left( |e,e,0\rangle \langle g,g,1| + \text{h.c.} \right)$, where the maximum coupling is achieved when 
\begin{equation}
\frac{\Omega_{\rm eff} }{ \Omega} \bigg|_{\theta=\cos^{-1}{\sqrt{2/3}}} = \frac{16}{9\sqrt{2}} \left( \frac{g}{\Omega}\right)^{3}.
\end{equation}
Analogous work \cite{Kockum2017} studied the frequency conversion in a system of two cavities coupled to the same atom in the USC regime. Moreover, it has been shown \cite{kockum2017deterministic} that other processes similar to the ones described in this section find interesting applications for nonlinear optics. Also, an analogous process to the one just described \cite{stassi2017quantum} can result in a single photon exciting multiple atoms. Furthermore, processes that do not conserve the excitation number can also be used for generating entanglement between photons \cite{macri2018simple}.

\begin{figure}[!hbt]
\centering
\includegraphics[width = 7cm]{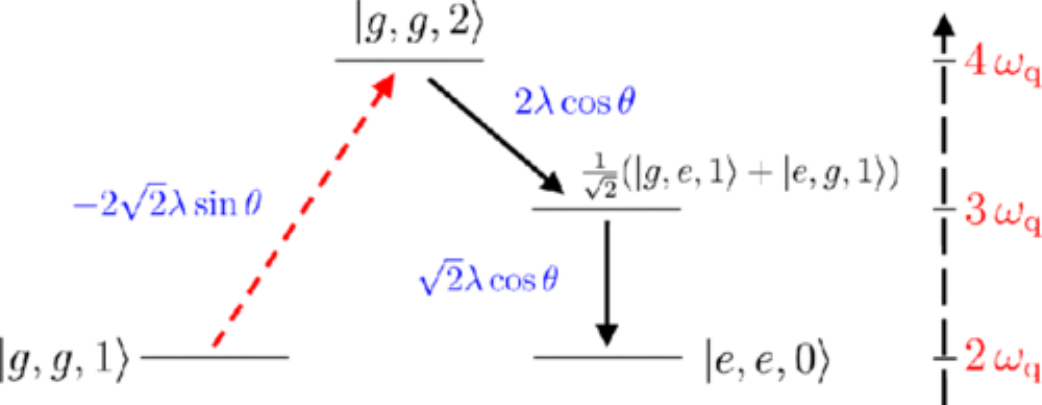}
\caption{\label{fig:garziano_levels} (Color online) Multiatom excitation with a single photon. As a result of USC physics, two or more atoms in an optical cavity can absorb a single photon. A sketch of the process is shown giving the main contribution to the effective coupling between the bare states $|g,g,1\rangle$ and $|e,e,0\rangle$, via intermediate virtual transitions. The coupling $\lambda$ corresponds to $g$ in the main text. The initial state $|g,g,1\rangle$ transitions to virtual intermediate excited states which would not conserve the total energy. At the end of the process, the final state $|e,e,0\rangle$ is excited, preserving the total system energy. Here the processes which do not conserve the excitation number are represented by an arrowed dashed line. Each path includes three virtual transitions involving out-of-resonance intermediate states. Only the process is displayed that gives the main contribution to the effective coupling between the bare states $|g,g,1\rangle$ and $|e,e,0\rangle$. Higher-order processes, depending on the atom-field interaction strength, can also contribute. The transition matrix elements are also shown. From~\cite{garziano2016}.}
\end{figure}

\begin{figure}[!hbt]
\centering
\includegraphics[width = 5cm]{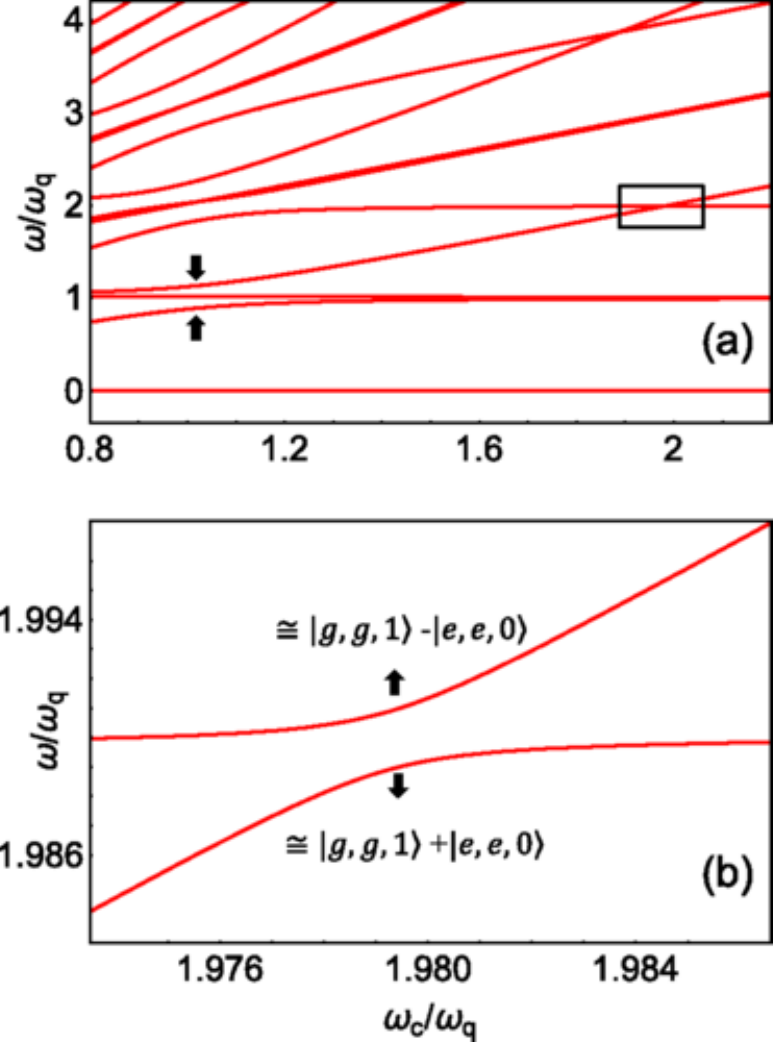}
\caption{\label{fig:garziano_spectrum} (Color online) (a) Frequency differences $\omega_{i,o} = \omega_{i} - \omega_{o}$ for the lowest-energy eigenstates of Eq.~(\ref{eq:garziano}) as a function of the resonator frequency $\omega_{c}/\omega_{q}$. $\omega_c$ and $\omega_q$ correspond to $\omega$ and $\Omega$ of the main text, respectively. Starting from the lowest excited states of the spectrum, a large anticrossing around $\omega_{c}/\omega_{q} =1$ can be observed, corresponding to the standard vacuum Rabi splitting. Here we consider a normalized coupling rate $g/\omega_{q}=0.1$ between the resonator and each of the two qubits. The particular case $\theta = \pi/6$ is shown. The arrows indicate the ordinary vacuum Rabi splitting arising from the coupling between the states $|g,g,1\rangle$ and $(|g,e,0\rangle + |e,g,0 \rangle)/\sqrt{2}$. (b) Enlarged view of the spectral region delimited by a square in (a), where the third and fourth levels display an apparent crossing. The enlarged view shows a clear avoided-level crossing. The level splitting originates from the hybridization of the states $|g,g,1\rangle$ and $|e,e,0\rangle$ due to the presence of counterrotating terms in the system Hamiltonian. The resulting states are well approximated by $(|g,g,1\rangle \pm |e,e,0\rangle)/\sqrt{2}$. This splitting is not present in the RWA, where the coherent coupling between states of a different number of excitations is not allowed. From~\cite{garziano2016}.}
\end{figure}

\subsubsection{Ancilla qubit spectroscopy}

Given the extreme parameters required to reach the USC regime, there is an intrinsic difficulty in performing direct spectroscopic measurements of the system as well as observing its dynamics. By contrast, several proposals were put forward to use a second qubit, known as an ancilla qubit, coupled to the USC system to extract some of its properties \cite{lolli2015, felicetti2015b, garziano2014}. In a separate proposal of a qubit-cavity system in the USC regime \cite{andersen2016}, higher-order modes of the cavity were suggested as an ancillary system to extract information of the cavity mode which is ultrastrongly coupled to the qubit via the cross-Kerr interaction which exists between any pair of modes due to the nonlinearity induced by the qubit.
In all cases, the ancilla-system coupling strength is in the strong coupling regime. In this configuration, the spectrum of the ancilla qubit or cavity contains information on the eigenstates of the USC system. Therefore, the ancillary system can be used as a probe of the many properties of the otherwise inaccessible ultrastrongly coupled system. There exist other proposals for nondemolition detection of USC ground-state properties, e.g., measuring the virtual radiation pressure exerted by the photons in the ground state on a mechanical mirror in an opto-mechanical system \cite{cirio2017amplified}.

In the particular configuration studied by \cite{lolli2015}, the Hamiltonian that describes the dynamics of the system is given by 
\begin{equation}
\label{eq:lolli}
\hat{\cal H} = \hat{\cal H}_{S} + \frac{\Omega_{\rm an}}{2} \hat{\sigma}_{z}^{(\rm an)} + g_{\rm an} (\hat{a}+\hat{a}^{\dagger}) \hat{\sigma}_{x}^{(\rm an)} + \Omega_{d} \cos{\left( \omega_{d} t\right)} \hat{\sigma}_{x}^{(\rm an)},
\end{equation} 
where $\Omega_{\rm an}$ is the natural frequency of the ancillary qubit, $g_{\rm an}$ is the coupling of the ancilla qubit to a single mode of the cavity, $\Omega_{d}$ and $\omega_{d}$ characterize the periodic driving of the ancilla qubit with a classical field, and $\hat{\cal H}_{S}$ is the Hamiltonian of the ultrastrongly coupled system [Eq.~(\ref{QRH})] the ancilla qubit is probing. In the particular work of Lolli \emph{et al.}, the ultrastrongly coupled system consists of a single cavity mode coupled to an ensemble of identical two-level systems with a collective coupling well in the USC regime. Several instances were studied corresponding to the Dicke, Tavis-Cummings \cite{tavis1968exact}, and Hopfield \cite{Hopfield58PR} models, whose respective Hamiltonians are
\begin{equation}
\label{eq:DTH}
\begin{split}
&\hat{\cal H}_{\text{Dicke}}= \omega \hat{a}^{\dagger}\hat{a} + \Omega \hat{J}_{z} + \frac{g}{\sqrt{N}} \left( \hat{a} + \hat{a}^{\dagger} \right) \left( \hat{J}_{+} + \hat{J}_{-}\right), \\
&\hat{\cal H}_{\text{TC}}= \omega \hat{a}^{\dagger}\hat{a} + \Omega \hat{J}_{z} + \frac{g}{\sqrt{N}} \left( \hat{a} \hat{J}_{+} + \hat{a}^{\dagger} \hat{J}_{-} \right), \\
&\hat{\cal H}_{\text{Hopfield}}= \hat{\cal H}_{Dicke} + \frac{g^{2}}{\Omega} \left( \hat{a} + \hat{a}^{\dagger}\right)^{2}.
\end{split}
\end{equation}
$\omega$ is the frequency of the single-mode cavity, $\Omega$ corresponds to the transition frequency of the $N$ identical two-level atoms, $\lambda$ describes the collective coupling, and the collective operators are given by $\hat{J}_{z} = (1/2) \sum_{i} \hat{\sigma}_{z}^{(i)}$ and $\hat{J}_{\pm}=\sum_{i} \hat{\sigma}_{\pm}^{(i)}$. 
\begin{figure}[!hbt]
\centering
   \includegraphics[width=202pt]{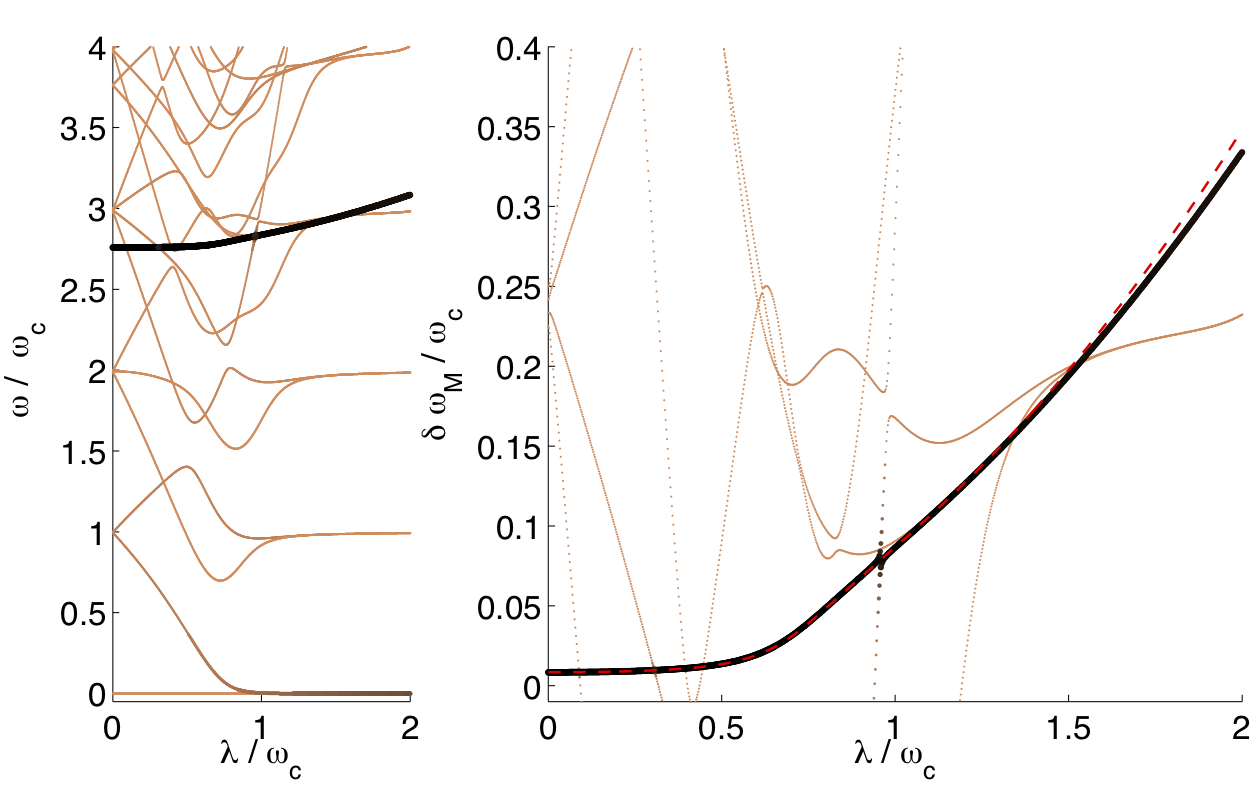}
   \\
   \includegraphics[width=202pt]{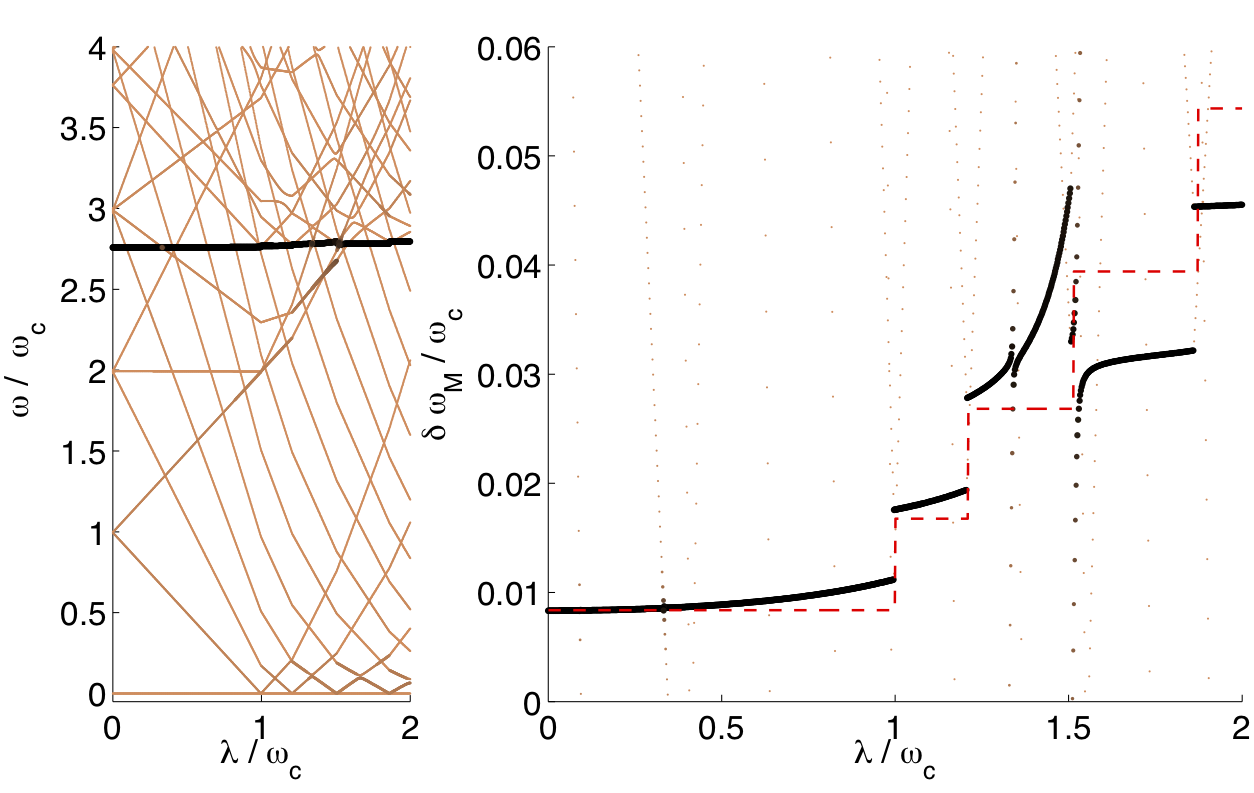}
   \\
     \includegraphics[width=202pt]{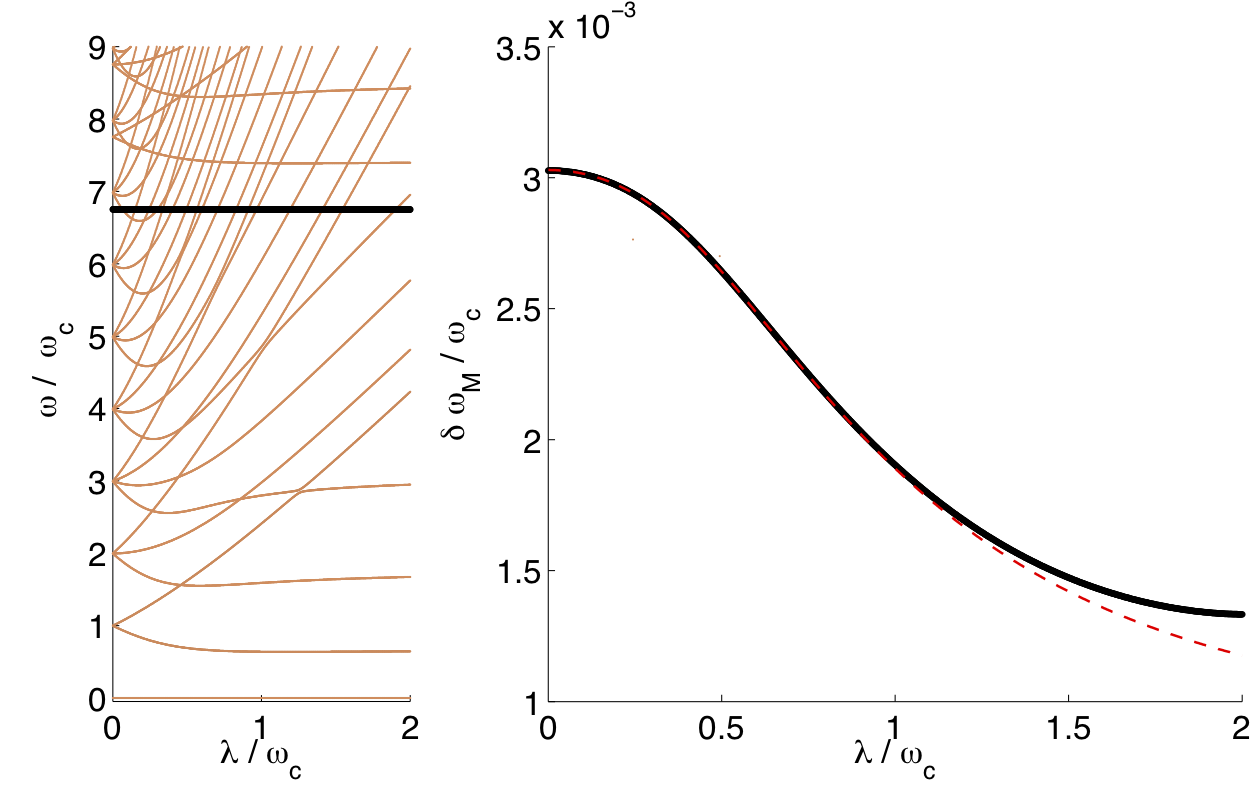}
\caption{\label{fig:lolli} (Color online) Ancilla qubit spectroscopy of ultrastrongly coupled systems. The top, middle, and bottom panels correspond to Eq.~(\ref{eq:lolli}) with the system $S$ being the Dicke, Tavis-Cummings, and Hopfield models, respectively [Eq.~(\ref{eq:DTH})]. $\omega_c$ and $\lambda$ correspond to $\omega$ and $g$ of the main text, respectively. Considering the ancilla qubit as the measurement qubit $M$, for finite $g_{M}$ the coupling between $S$ and $M$ creates a mixing between states of the form $|\psi_{S}\rangle \otimes |\psi_{M}\rangle$ and the driving induces transitions from the ground state $|G_{S+M}\rangle$ to excited states. Therefore, the relevant excited states $|l\rangle$ are those having the largest values of $| \langle G_{S+M} | \hat{\sigma}^{(M)}_{x}| l \rangle |^{2}$. The results show that, due to the off-resonant coupling, there is only one dominant spectroscopically active level (thick black solid line), which has a strong overlap with the state $|G_{S}\rangle \otimes |\uparrow \rangle$. Left panels: excitation energies for the three considered systems $S$ vs the coupling $\lambda$ between the boson field and the $N$ atoms. Right panels: Lamb shift of the ancillary qubit transition as a function of the coupling $\lambda/\omega_c$ of the coupled system $S$ under consideration. The dashed red lines in the right panels depict the shift predicted by the analytic calculation. The agreement between the numerical diagonalization results and the analytical formula [Eq.(\ref{eq:lamb-lolli})] is excellent in the considered range of values for $\lambda / \omega_{c}$, except for points where there are avoided crossings with other levels. From~\cite{lolli2015}.}
\end{figure}

All three models are shown in Fig.~\ref{fig:lolli}. Because of the ancilla-system coupling, a measurable Lamb shift in the frequency of the ancillary qubit appears. Up to second order in perturbation theory in $g_{\rm an}$, this shift can be analytically calculated to be
\begin{equation}\label{eq:lamb-lolli}
\begin{split}
\delta \omega_{\rm an} &\sim g^{2}_{\rm an} \left( \frac{1}{\omega_{\rm an}-\omega} + \frac{1}{\omega_{\rm an}+\omega} \right) \langle \left( \hat{a} + \hat{a}^{\dagger} \right)^{2} \rangle \\
&+ g^{2}_{\rm an} \left( \frac{1}{\left( \omega_{\rm an}-\omega \right)^{2}} - \frac{1}{\left(\omega_{\rm an}+\omega\right)^{2}} \right) \langle \hat{V}^{(S)} \rangle, 
\end{split}
\end{equation}
where $\hat{V}^{(\text{Dicke})} = g N^{-1/2} \left(\hat{a} + \hat{a}^{\dagger} \right) \hat{J}_{x}$, $\hat{V}^{(\text{TC})} = g N^{-1/2} \left( \hat{a}^{\dagger} \hat{J}_{-} + \hat{a} \hat{J}_{+} \right)$, and $\hat{V}^{(\text{Hopfield})} = \hat{V}^{(\text{Dicke})} + 2 g^{2}\Omega^{-1} \left( \hat{a} + \hat{a}^{\dagger}\right)^{2}$. As is explicit from the equations, the shift depends on the ground-state photon population $\langle \hat{a}^{\dagger}\hat{a}\rangle$, on the anomalous expectation value $\langle \hat{a}^{\dagger 2} + \hat{a}^{2} \rangle$, and on the correlations between the cavity and the $N$ two-level systems. Figure~\ref{fig:lolli} shows the Lamb shift for the three discussed models. 

\subsubsection{Optomechanics in the USC regime}

Solid-state nanoelectromechanical resonators have been considered as a mediator of the interactions between qubits~\cite{PhysRevA.70.052315}. Ultrastrongly coupled optomechanical setups have been proposed to prepare quantum states of motion \cite{garziano2015single}. In the same scenario, another type of quantum states which can be obtained as a consequence of ultrastrong interactions are NOON states which are entangled states that represent a superposition of $N$ excitations in one mode with zero excitations in a second mode, and viceversa~\cite{macri2016}. It has been shown that the preparation of NOON states in ultrastrongly coupled optomechanical systems is possible following a completely controlled and deterministic procedure. The setup consists of two identical, optically coupled optomechanical systems which can be modeled by the photonic modes of the optical cavities and the phononic modes from the mechanical oscillators (see the description of the setup in Fig.~\ref{fig:macri}). The dynamics of each independent optomechanical subsystem are characterized by the Hamiltonian 
\begin{equation}
\hat{\cal H}_{0}^{(i)} = \hbar\omega_{R} \hat{a}^{\dagger}_{i} \hat{a}_{i} + \hbar\omega_{M} \hat{b}^{\dagger}_{i} \hat{b}_{i} + \hbar g_{M} \hat{a}^{\dagger}_{i} \hat{a}_{i} \left( \hat{b}_{i} + \hat{b}_{i}^{\dagger} \right),
\end{equation} 
in the local Fock basis $|n_{i},m_{i}\rangle$, where the integers $n_{i}$ and $m_{i}$ represent the number of photons and vibrational excitations in the $i$th optomechanical system. The preparation of mechanical entangled NOON states requires two interacting optical cavities with an interaction Hamiltonian $\hat{\cal H}_{I}=\hbar g_{R}\left( \hat{a}_{1}^{\dagger} \hat{a}_{2} + \hat{a}_{1} \hat{a}^{\dagger}_{2} \right)$. Starting in the ground state of the system that contains no photons or phonons in either system, one of the optical resonators is excited with an external $\pi$ pulse resonant with the transition $|0_{1},0_{1};0_{2},0_{2}\rangle \leftrightarrow |1_{1},m_{1};0_{2},0_{2}\rangle$. Then the system freely evolves with the interaction Hamiltonian undergoing Rabi oscillations. The time-dependent quantum state is then given by $|\Psi(t) \rangle = \cos{\left(g_{R}t\right)} |1_{1},m_{1};0_{2},0_{2}\rangle - i \sin{\left(g_{R}t\right)} |0_{1},0_{1};1_{2},m_{2}\rangle$. A second resonant pulse with the transition $|1_{i},m_{i}\rangle \leftrightarrow |0_{i},N_{i}\rangle$ will produce the desired NOON state,
\begin{equation}
|\Psi\rangle = \alpha |0_{1},N_{1};0_{2},0_{2}\rangle -i \beta |0_{1},0_{1};0_{2},N_{2}\rangle.
\end{equation}

\begin{figure}[!hbt]
\centering
\includegraphics[width = 7cm]{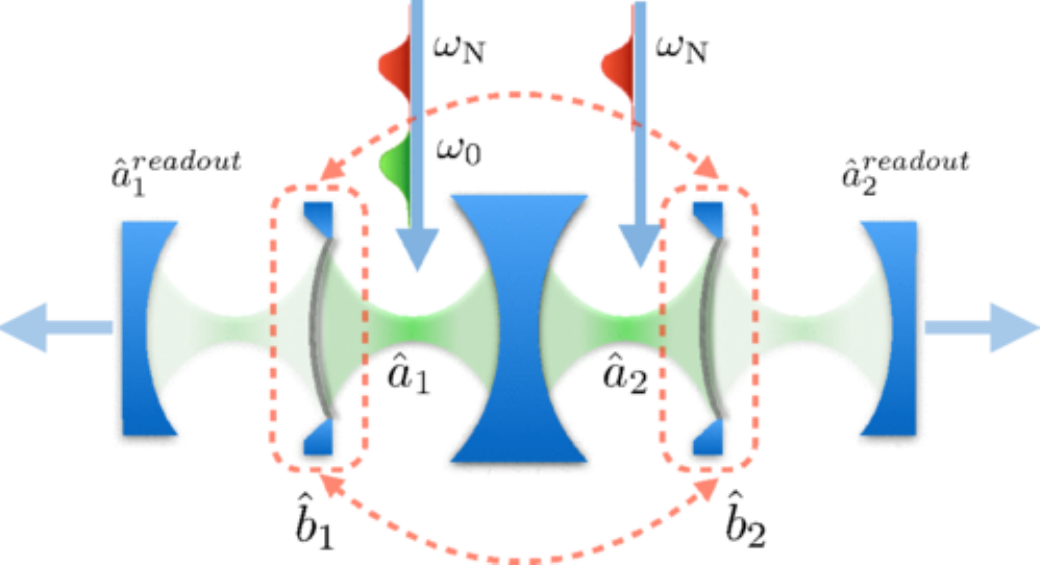}
\caption{\label{fig:macri} (Color online) Optomechanical USC. Two identical coupled optomechanical systems, with frequency $\omega_{M}$, are parametrically coupled with a single-mode optical resonator or cavity, which can be driven by external optical pulses with specific central frequencies. One cavity mirror can be added to the end of both the optomechanical systems for optical readout. From~\cite{macri2016}.}
\end{figure}

It is worth mentioning that further developments and applications of the USC and DSC regimes to coupled mechanical systems are expected, given that the physical conditions are not necessarily equivalent to those of coupled electromagnetic oscillators~\cite{Sudhir2012}. 

Other work in ultrastrongly coupled oscillator systems, including optomechanics, have investigated the influence of $A^2$ terms and their possible detection in a real experiment \cite{Tufarelli2015, Rossil2017}. A recent coupled oscillator experiment in a superconducting circuit \cite{Fedortchenko2017} observed simultaneous single-mode and two-mode squeezing of the radiated field below vacuum fluctuations \cite{Markovic2018}.

\subsection{Quantum computation}\label{sec:5B}

Being able to tune the coupling strength in a light-matter system from strong to the ultrastrong regime allows one to observe and propose new strategies and protocols in quantum information processing, such as remote entanglement applications~\cite{PhysRevLett.120.093601,PhysRevLett.120.093602}. In this section, we discuss the possibility to achieve ultrafast quantum computation, protected qubits to store quantum information, and to manipulate and prepare a desired quantum state.

\subsubsection{Ultrafast quantum computation}

Ultrafast two-qubit gates have been considered as one potential application~\cite{kyaw2015,wang2012,kyaw2015prb,wang2016} of the USC regime in quantum computation~\cite{Romero2012}. In the original proposal \cite{Romero2012}, a two-qubit Hamiltonian was considered
\begin{equation}
\hat{\cal H}=\sum_{i} \frac{\hbar \Omega_{i}}{2} \hat{\sigma}_{z}^{(i)} + \hbar \omega \hat{a}^{\dagger}\hat{a} - \sum_{i} \hbar g_{i} \hat{\sigma}_{z}^{(i)} \left( \hat{a} + \hat{a}^{\dagger} \right),
\end{equation}
with switchable longitudinal couplings $g_{i}$ (see the circuit diagram of the experimental proposal in Fig.~\ref{fig:romero}). Based on a four-step sequential displacement of the cavity $\hat{\mathcal{D}}\left( \beta \hat{\sigma}_{z} \right)= \exp{ \left[ \left( \beta \hat{a}^{\dagger} - \beta^{*} \hat{a} \right) \hat{\sigma}_{z} \right] }$, using $\hat{\mathcal{D}}\left(\alpha \right) \hat{\mathcal{D}}\left(\beta \right) = e^{i \rm{Im}\left( \alpha \beta^{*} \right)} \hat{\mathcal{D}} \left( \alpha + \beta \right)$, the two-qubit gate was shown to be proportional to a CPHASE quantum gate
\begin{equation}
\hat{\mathcal{U}} \propto \exp{ \left[ 4 i \frac{g_{1}g_{2}}{\omega^{2}}\sin{\left(\omega t_{1} \right)} \hat{\sigma}_{z}^{(1)} \hat{\sigma}_{z}^{(2)}\right] }, 
\end{equation}
where the fidelity of the gate can reach $99\%$ in the nanosecond time scale for realistic circuit QED technology.

This protocol relies on being able to switch fluxes in the qubit local bias lines faster than the coupling rate $g$, which implies subnanosecond pulses. Implementing these short pulses comes at a technological cost. First, the entire system bandwidth should be able to transmit the pulses without distortion which would slow down the fast edge. Second, pulse generators able to synthesize picosecond pulses do exist, albeit at a cost which would not easily lead to controlling a large number of qubits. Further technological developments of fast pulse generators are necessary before this technology can be implemented in a scalable way, beyond a two-qubit proof of principle.

\begin{figure}[!hbt]
\centering
\includegraphics[width = 7cm]{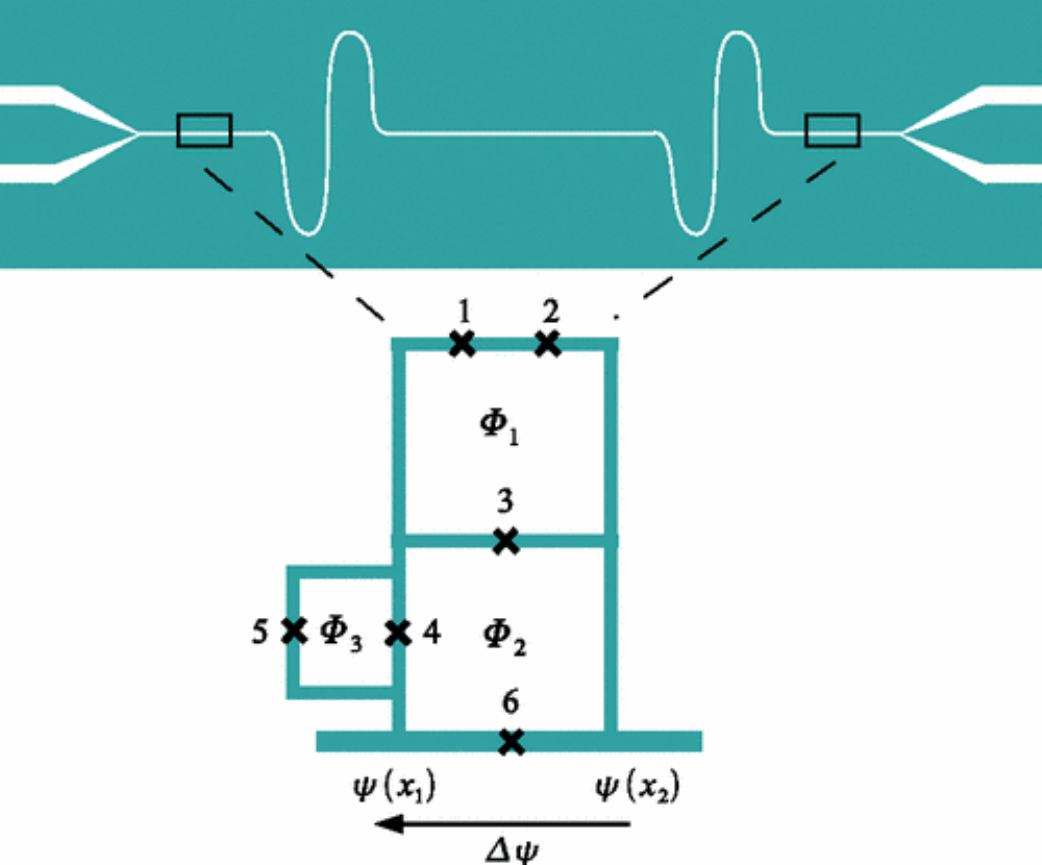}
\caption{\label{fig:romero} (Color online) Ultrafast two-qubit gates. Circuit schematic to realize ultrafast two-qubit controlled phase gates between two flux qubits galvanically coupled to a single-mode transmission line resonator. The bottom image shows a six Josephson-junction circuit coupled galvanically to a resonator. The flux qubit is defined by three Josephson junctions in the upper loop threaded by external flux $\Phi_1$. Two additional loops allow a tunable and a switchable qubit-resonator coupling by controlling fluxes $\Phi_2, and \Phi_3$. The coupling is defined by the phase drop $\Delta\psi$ across the shared junction. From~\cite{Romero2012}.}
\end{figure}

\subsubsection{Protected qubits}

Another important example where the USC regime may become relevant in quantum computation is in the encoding of protected qubits~\cite{nataf2011}. Nataf and Ciuti considered the case of multiple qubits coupled to the same resonator mode
\begin{equation}
\hat{\cal H}/\hbar = \omega \hat{a}^{\dagger}\hat{a} + \frac{\Omega}{2} \sum_{j} \hat{\sigma}_{z}^{(j)} + i \frac{g}{\sqrt{N}} \left( \hat{a} - \hat{a}^{\dagger}\right) \sum_{j} \hat{\sigma}_{x}^{(j)}. 
\end{equation}
Here $N$ is the total number of qubits coupled to the resonator. It turns out that when the collective coupling of all qubits reaches very large values, the two quasidegenerated lowest states of the Hamiltonian become
\begin{equation}
\begin{split}
&| \Psi_{G} \rangle \sim \frac{1}{\sqrt{2}} \left[ | \alpha \rangle_{c}  | + \rangle^{\otimes N} + \left(-1 \right)^{N} |- \alpha \rangle_{c}  | - \rangle^{\otimes N} \right], \\
&| \Psi_{E} \rangle \sim \frac{1}{\sqrt{2}} \left[ | \alpha \rangle_{c}  | + \rangle^{\otimes N} - \left(-1 \right)^{N} |- \alpha \rangle_{c}  | - \rangle^{\otimes N} \right],
\end{split}
\end{equation}
with $|\pm\rangle$ being eigenstates of $\hat{\sigma}_{x}$. Both these states are weakly coupled to each other as they belong to a different parity chain \cite{casanova2010}. This doublet $\{ | \Psi_{G} \rangle , | \Psi_{E} \rangle\}$ therefore  forms a robust qubit, with an energy difference $\delta \sim \Omega\exp{\left( -2g^{2}\omega^{-2} N \right)}$. The analysis of the coherence times is shown in Fig.~\ref{fig:nataf}. Clearly, for increasing coupling strengths, and also for increasing number of qubits, the decoherence rate decreases yielding a more protected qubit, up to a certain value of the coupling where the decoherence rate saturates. In a different work, a proposal \cite{stassi2018long} analyzed a protected quantum memory in the DSC regime.
\begin{figure}[!hbt]
\centering
\includegraphics[width = 7cm]{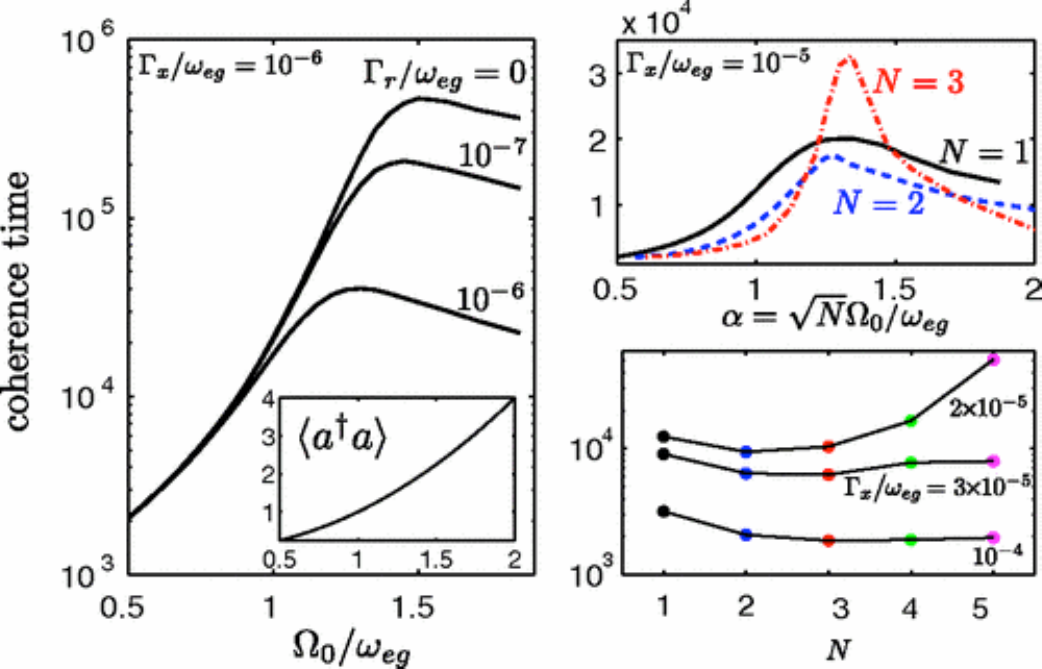}
\caption{\label{fig:nataf} (Color online) Protected quantum computation in the USC regime. $\Omega_0$ and $\omega_{eg}$ correspond to $g$ and $\Omega$ of the main text, respectively. To investigate the robustness of the coherence between the two quasidegenerate vacua $|\Psi_{G}\rangle$ and $|\Psi_{E}\rangle$, they study the nonunitary dynamics of the initially prepared pure state $|\Psi_{0}\rangle = \cos{\theta} |\Psi_{E}\rangle + \sin{\theta} e^{i\phi} |\Psi_{G} \rangle$ in the presence of anisotropic qubit dissipation rates  $\Gamma_{y}, ~ \Gamma_{z} \gg \Gamma_{x}$ and for several cavity loss rates. The simulations proved that the coherence time increased while increasing the normalized vacuum Rabi frequency $g / \Omega$. In fact, the coherence time was exponentially enhanced before reaching a saturation value. Left-hand panel: Coherence time vs the normalized vacuum Rabi frequency for one atom. Inset: Number of photons plotted vs the normalized vacuum Rabi frequency. Top right-hand panel: Coherence time for $N=1,2,$ and $3$ atoms. Bottom right-hand panel: Maximum coherence time as a function of the number of atoms. From~\cite{nataf2011}.}
\end{figure}

\subsubsection{State preparation: Qubit-resonator entangled states}

The eigenstates of a system in the USC regime result in many-body qubit-resonator entangled quantum states~\cite{Ashhab2010, felicetti2015b,garziano2016}. Certain quantum information processing protocols may require the generation of this type of states, an example being cat-state-based quantum error correction \cite{Ofek2016}. For instance, a paradigmatic multipartite entangled state, the $N$-qubit GHZ state, results from a system of superconducting qubits coupled to a transmission line resonator~\cite{wang2010}. In this system, the Hamiltonian in the interaction picture reads $\hat{\cal H}_{I}(t) = \hbar g \sum_{i} \left( \hat{a}^{\dagger} e^{i \omega t} + \hat{a} e^{-i \omega t} \right) \hat{\sigma}_{x}^{(i)}$. For particular periods $T_{n}=2 \pi n / \omega$ commensurate with the cavity frequency $\omega$, the time evolution operator in the Schr\"odinger picture takes the form
\begin{equation}
\hat{U}(T_{n}) \propto \exp{\left[ -i  \theta (n) \sum_{i \neq j} \hat{\sigma}_{x}^{(i)} \hat{\sigma}_{x}^{(j)} \right]},
\end{equation}
with $\theta (n)=g^{2}/\omega^{2} 2 \pi n$. Hence, starting from a product state $| \Psi(0) \rangle = \otimes_{i=1}^{N} |-\rangle^{(i)}_{z}$, where $\hat{\sigma}^{(i)}_{z} |-\rangle^{(i)}_{z} =- |-\rangle^{(i)}_{z}$, the system evolves into a GHZ state of the form
\begin{equation}
|\Psi (T_{\rm min}) \rangle = \frac{1}{\sqrt{2}} \left( \otimes_{i=1}^{N} |-\rangle^{(i)}_{z} + e^{i \pi (N+1)/2} \otimes_{i=1}^{N} |+\rangle^{(i)}_{z} \right),
\end{equation}
for the minimum preparation time given by $T_{\rm min}=\pi \omega/8 g^{2}$~\cite{wang2010}.

\subsection{Dissipation in the ultrastrong coupling regime}\label{sec:5C}

Dissipation, decay or decoherence rates are natural scales that appear in various platforms of quantum information processing due to the coupling of qubits to any external degrees of freedom. The first study of dissipation in the USC regime \cite{DeLiberato2009} used the second-order time-convolutionless projection operator method. In later work, an equivalent method was found by projecting the master equation in the dressed-state basis~\cite{Beaudoin2011}. Using either technique, modifications of the standard quantum optics master equation were obtained which do not display unphysical effects when the USC regime is reached. 

Here, we follow the master equation projection method \cite{Beaudoin2011} to obtain a suitable description of the system dynamics in the dissipative QRM, valid in the Bloch-Siegert regime (perturbative USC). The standard (Lindblad) form of the master equation at zero temperature $T=0$ is given by
\begin{equation}\label{eq:Lindblad}
\frac{d\hat{\rho}}{dt} = -i[\hat{\mathcal{H}}, \hat{\rho}] + \mathcal{L}\hat{\rho},
\end{equation}
where $\hat{\rho}$ is the density matrix of the whole system. The Lindbladian $\mathcal{L}$ in the standard form is defined as
\begin{equation}\label{eq:Lind}
\mathcal{L}\hat{\rho} = \kappa\mathcal{D}[\hat{a}]\hat{\rho} + \gamma_{\rm ge}\mathcal{D}[\hat{\sigma}_-]\hat{\rho} + \frac{\gamma_{\phi}}{2}\mathcal{D}[\hat{\sigma}_z]\hat{\rho}.
\end{equation}
Here $\kappa, \gamma_{\rm ge}, and \gamma_{\phi}$ are the cavity decay, qubit decay, and qubit dephasing rates, respectively. The superoperator $\mathcal{D}[\mathcal{\hat{O}}]$ is defined as $\mathcal{D}[\mathcal{\hat{O}}]\hat{\rho} = (1/2)(2\mathcal{\hat{O}}\hat{\rho}\mathcal{\hat{O}}^{\dag} - \hat{\rho}\mathcal{\hat{O}}^{\dag}\mathcal{\hat{O}} - \mathcal{\hat{O}}^{\dag}\mathcal{\hat{O}}\hat{\rho})$. Equation~(\ref{eq:Lindblad}) assumes that the ground state of the qubit $|g\rangle$ plus the vacuum of the cavity $|0\rangle$ is the ground state of the whole system $|g0\rangle$. However, in the QRM the ground state is a superposition of different states of both subsystems, it is a superposition of multiple photon number states entangled with the qubit states (see Sec.~\ref{sec:2}). Therefore, the master equation needs to be modified in such a way that it damps any initial state toward the actual ground state $\widetilde{|g0\rangle}$. In Fig.~\ref{fig:ME} it is possible to observe the detrimental effect of not using the proper form of the master equation, which results in a fictitious heating rate. 
\begin{figure}[!hbt]
\centering
\includegraphics[width = 0.9\columnwidth]{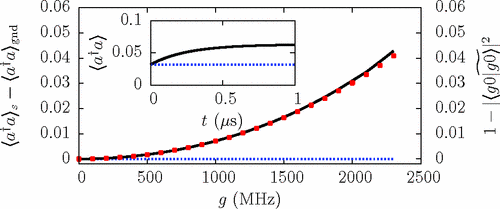}
\caption{\label{fig:ME} Excess in the mean photon number due to relaxation in the steady state of the ultrastrongly coupled qubit-resonator system \cite{Beaudoin2011}. Initially, the system is in its true ground state $\widetilde{|g0\rangle}$, but, under the standard master equation~(\ref{eq:Lindblad}), relaxation unphysically excites the system even at $T=0$. The black line, which corresponds to the left axis, represents the number of additional photons introduced in the steady state by dissipation. The red dots, associated with the right axis, designate 1 minus the fidelity of the Rabi ground state $\widetilde{|g0\rangle}$ to the vacuum state $|g0\rangle$. The parameters used are $\Omega/2\pi = \omega/2\pi = 6~\rm{GHz}, \kappa/2\pi = \gamma/2\pi = 0.1$ MHz, and no pure dephasing. $\kappa$ and $\gamma$ are the resonator and qubit energy damping rates, respectively. Inset: Mean photon number as a function of time for the system starting in its ground state with $g/2\pi=2~$GHz. In both the main plot and the inset, the blue dashed line indicates results for the fidelity and the photon number as obtained with the master equation given by the Lindbladian in Eq.~(\ref{eq:LQRM}).}
\end{figure}

To obtain a master equation that takes into account the actual eigenvalues of the QRM, we first move to the frame that diagonalizes the quantum Rabi Hamiltonian [Eq.~(\ref{QRH})] for both the system and the system-bath Hamiltonians. Under experimentally reasonable approximations\footnote{Neglecting high-frequency terms, the resulting expressions involve transitions $|j\rangle\leftrightarrow|k\rangle$ between eigenstates at a rate that depends on the noise spectral density at frequency $\Delta_{kj} = \omega_k - \omega_j$. If their linewidth is small enough, these transitions can be treated as due to independent baths. As a result, these independent baths can each be treated in the Markov approximation.}, the correct form of the Lindbladian at zero temperature $T=0$ reads
\begin{multline}\label{eq:LQRM}
\mathcal{L}_{\rm QRM}\circ = \mathcal{D}\left[\sum_j\Phi^j|j\rangle\langle j|\right]\circ + \sum_{j,k\ne j}\Gamma_{\phi}^{jk}\mathcal{D}[|j\rangle\langle k|]\circ \\+ \sum_{j,k>j}(\Gamma_{\kappa}^{jk}+\Gamma_{\gamma}^{jk})\mathcal{D}[|j\rangle\langle k|]\circ.
\end{multline}
$|k\rangle$ and $|j\rangle$ are eigenstates of the QRM. The circle $\circ$ represents the operator on which the Lindbladian is acting on. The first two terms in Eq.~(\ref{eq:LQRM}) are the contributions
from the bath that caused only dephasing in the standard master equation [last term in Eq.~(\ref{eq:Lind})]. Here this $\hat{\sigma}_z$ bath causes dephasing in the eigenstate basis with
\begin{equation}
\Phi_j = \sqrt{\frac{\gamma_{\phi}(0)}{2}}\sigma_z^{jj},
\end{equation}
where $\gamma_{\phi}(\omega)$ is the dephasing rate corresponding to noise at frequency $\omega$ due to the noise spectral density $\sigma_z^{jk} = \langle j|\hat{\sigma}_z|k\rangle$. The fact that $\hat{\sigma}_z$ is not diagonal in the system eigenbasis causes undesired transitions at rate
\begin{equation}
\Gamma_{\phi}^{jk} = \frac{\gamma_{\phi}(\Delta_{jk})}{2}|\sigma_z^{jk}|^2.
\end{equation}
This noise will be significant only if the power spectral density of dephasing noise at frequency $\Delta_{jk}$ is significant. This is the case away from the sweet spot in superconducting qubits. The longitudinal noise along $\sigma_z$ may stimulate transitions between the QRM eigenstates $|j\rangle$, leading to dephasing-induced generation of photons and qubit excitations, a phenomenon linked to the dynamical Casimir effect. 

The last two terms in Eq.~(\ref{eq:LQRM}) are the contributions from the resonator and qubit own baths that caused relaxation in the standard master equation. These baths now cause transitions between eigenstates at rates
\begin{align}
\Gamma_{\kappa}^{jk} &= \kappa(\Delta_{jk})|X_{jk}|^2,\\
\Gamma_{\gamma}^{jk} &= \gamma(\Delta_{jk})|\sigma_x^{jk}|^2,
\end{align}
where
\begin{align}
X_{jk} &= \langle j| \hat{X} |k \rangle,\\
\sigma_x^{jk} &= \langle j|\hat{\sigma}_x |k\rangle.
\end{align}
The rates $\kappa(\omega)$ and $\gamma(\omega)$ are proportional to noise spectra from the resonator and qubit baths, respectively. $\hat{X}$ is the cavity quadrature $\hat{X}=\hat{a}^{\dag}+\hat{a}$. The Lindbladian in Eq.~(\ref{eq:LQRM}) correctly predicts the system evolution of the QRM under the presence of dissipation and dephasing baths. This is illustrated by the dashed blue line in Fig.~\ref{fig:ME}. The new decay rates have specific selection rules due to the parity of the eigenstates in the quantum Rabi Hamiltonian. A direct consequence of the modification of the emission rates is the appearance of an asymmetry in the vacuum Rabi splitting when qubit and resonator are resonant. The spectrum of the system could be used in this way to probe dephasing noise \cite{Beaudoin2011}. A more general treatment has been used to describe open systems in the USC regime at finite temperatures \cite{Settineri2018}.

With the corrected version of the master equation, it was demonstrated \cite{DeLiberato2009} that a harmonic modulation of the qubit-cavity interaction strength in the USC regime with a functional form
\begin{equation}
g(t) = g_0 + \Delta g\sin(\omega_{\rm mod}t)
\end{equation}
produces extracavity radiation originated from the spontaneous emission of virtual photons existing in the ground state of an ultrastrongly coupled system. Calculating the emitted radiation employing the standard master equation [Eq.~(\ref{eq:Lindblad})] instead produces the unphysical picture of generating radiation even when the drive is very far from the cavity resonance, which clearly violates energy conservation rules (see Fig.~\ref{fig:Deliberato}).
\begin{figure}[!hbt]
\centering
\includegraphics[width = 0.8\columnwidth]{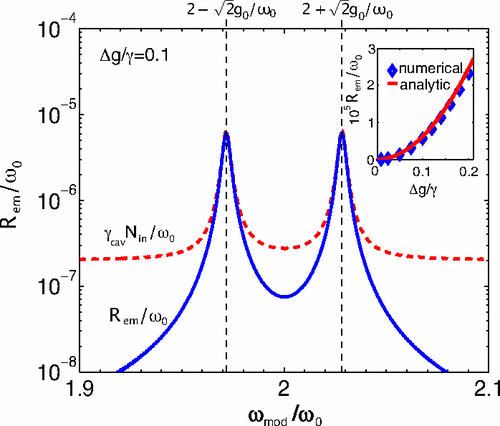}
\caption{\label{fig:Deliberato} (Color online) Extracavity photon emission rate $R_{\rm em}$ (in units of $\omega_0$, the cavity frequency) for a resonant qubit-cavity system as a function of the modulation frequency $\omega_{\rm mod}$ for a modulation amplitude of the vacuum Rabi frequency $\Delta g/\gamma=0.1$, where $\gamma$ is the qubit and cavity emission rate. For comparison, the dashed line shows the extracavity emission rate $\gamma_{\rm cav}N_{\rm in}$ where $N_{\rm in}$ is the steady-state intracavity photon number that would be predicted by the Markovian approximation: note the unphysical prediction of a finite value of the emission even far from resonance. The inset shows the dependence of the photon emission rate on the modulation amplitude, calculated both numerically and analytically. From~\cite{DeLiberato2009}.}
\end{figure}

An important aspect related to dissipation that was just recently addressed~\cite{DeLiberato2017} is the impact of the decay rates on the number of photons in the ground state of a system in the USC regime. The ground state in an ultrastrongly coupled qubit-cavity system is composed of hybridized qubit-cavity states which lead to a nonzero value of the expectation value of the photon number operator, defined as\footnote{We remind the reader that the photon number operator $\hat{N}$ as defined in traditional quantum optics textbooks is not a good quantum number in the USC regime, as it does not commute with the quantum Rabi Hamiltonian $[\mathcal{\hat{H}}_R,\hat{N}]\ne0$. The consequence is a nonstationary value of the population of photonnumber states of the cavity, as shown by \cite{casanova2010}.} $\hat{N}=\langle \hat{a}^{\dag}\hat{a}\rangle$. It is then crucial to understand what is the impact of the qubit and cavity decay rates on the population of photons in the USC ground state. The result is a bit surprising, as it turns out that the USC effects are only quantitatively affected by losses. Thus, USC phenomena such as extracavity emission may be observed in systems with very high losses, even when the usual condition of strong coupling is not satisfied $\gamma>g$.

Another quantum optical phenomenon in open quantum systems that is modified in the USC regime is photon blockade~\cite{RidolfoetAl12PRL}. In the strong coupling condition where the RWA applies, the temporal photon-photon correlation function shows an oscillatory behavior with a frequency given by the Rabi frequency of the externally applied drive. Instead, in the USC regime the frequency is given by the ultrastrong emitter-photon coupling which can be traced back to the presence of two-photon cascade decays induced by counterrotating interaction terms. In order to reach these conclusions, a generalized version of the input-output relations had to be extended to the USC regime. The result is the following relation:
\begin{equation}
\label{eq:i-o}
\hat{a}_{\rm out}(t) = \hat{a}_{\rm in}(t) - i \frac{\epsilon_c}{\sqrt{8\pi^2\hbar\epsilon_0v}}\dot{\hat{P}}^+.
\end{equation}
Here $\epsilon_c$ is a coupling parameter to the environment, $\epsilon_0$ describes the dielectric properties of the output waveguide, and $v$ is the phase velocity. Crucially, $\dot{\hat{P}}^+$ is not proportional to the intracavity field $\hat{a}$ as is usual in quantum optics. Its explicit form is $\dot{\hat{P}}^+ = -i\sum_{j,k>j}\Delta_{kj}P_{jk}|j\rangle\langle k|$, where $\Delta_{jk} = \omega_j - \omega_k$, and $P_{jk} = \langle j|\hat{P}|k\rangle$, with $\hat{P} = -iP_0(\hat{a}-\hat{a}^{\dag})$. Here $|j\rangle$ are the QRM eigenstates. Note that $P^+|0\rangle = 0$, while $a|0\rangle\ne0$. This redefinition of the input-output relations has a direct impact on the output photon number flux, which otherwise would show a finite value even without an externally applied drive. 

Finally, a novel topic that has emerged is that of discrete time crystals, which are out-of-equilibrium dynamical phases recently proposed and observed. The analysis of these systems in the context of open dissipative regimes has also been carried out in terms of the open Dicke model \cite{Gong2018}.

\section{Conclusions and Outlook}\label{sec:6}

The interaction between light and matter can be considered as the essential dialogue that describes and explains most fundamental phenomena in nature, emerging rather late in the history of physics out of stepwise developments in mechanics and optics. With the arrival of atomic physics in the 20th century, after the success of electromagnetism at the end of the 19th century, light-matter models were proposed to account for quantum effects observed in the laboratory, giving rise to the (semiclassical) Rabi model. Along these lines, a final key improvement had to be performed with the quantization of light to produce the full-fledged quantum Rabi model. This review aims at producing a biased overview of light-matter interactions where the ultrastrong and deep strong coupling regimes are necessary for describing the interplay between models and experimental observations. Somehow, we needed the advent of advanced tools in quantum control of atoms and photons, in the wide frame of quantum technologies at the beginning of this 21st century, to produce key experimental results and their corresponding theoretical descriptions in the USC and DSC regimes. Exploring these novel extreme coupling strengths between quantized light and quantized matter is a fundamental task of high scientific relevance, which required conceptual and experimental improvements during the last decade. As frequently happens in the interplay between science and technology, the discovered USC and DSC phenomena may find a variety of applications in quantum simulations, quantum sensing, quantum communication, and quantum computing. Accelerating quantum dynamics should also inspire novel protocols in scalable quantum processing. We believe the study of USC and DSC regimes is still in its infancy and that most advanced discoveries and applications are still waiting to be discovered.

\begin{acknowledgements}
The authors acknowledge Motoaki Bamba, Xinwei Li, and Vladimir E.~Manucharyan for fruitful discussions, and M.~Bajcsy for providing references for the figures in the introduction. J.K. thanks Jeffery Horowitz and Jorge Zepeda for careful proofreading. J.K. acknowledges support from the National Science Foundation through Grant No. DMR-1310138. P. F.-D. acknowledges funding from the Ministry of Economy and Competitiveness, through contracts FIS2017-89860-P and Severo Ochoa SEV-2016-0588 and the support of the Beatriu de Pin\'os postdoctoral programme of the Government of Catalonia's Secretariat for Universities and Research of the Ministry of Economy and Knowledge. L. L., E. R., and E. S. acknowledge funding from MINECO/FEDER FIS2015-69983-P and Basque Government IT986-16, while L. L. is also supported by Ram\'on y Cajal Grant No. RYC-2012-11391.

\end{acknowledgements}

\bibliography{USCbiblio}
\end{document}